\documentclass[aps,prl,reprint,amsmath,amssymb,superscriptaddress, 10pt]{revtex4-2}

\usepackage[usenames,dvipsnames]{xcolor}
\usepackage{amsmath,amssymb}
\usepackage{graphicx}
\usepackage{hyperref}
\usepackage{braket}
\usepackage[normalem]{ulem}
\usepackage{enumerate}
\usepackage{MnSymbol}
\usepackage{esint}
\usepackage{comment}
\newcommand{\Cite}[1]{\mbox{\cite{#1}}}
\usepackage[normalem]{ulem}
\usepackage{siunitx}
\newcommand{\be}{\begin{equation}}
\newcommand{\ee}{\end{equation}}
\def\bes{\begin{subequations}}
\def\esu{\end{subequations}}

\newcommand{\dr}{\text{dr}}

\newcommand{\eff}{\text{eff}}

\newcommand{\abs}[1]{\left\lvert #1 \right\rvert}
\graphicspath{{Figures/}}

\newcommand{\scexp}{{\xi}}
\newcommand{\dd}{{\rm d}}

\newcommand{\qq}{{\rm q}}

\newcommand{\para}[1]{ \noindent\emph{\textbf{#1---}}}

\newcommand{\com}[1]{{}} 

\newcommand{\ch}[1]{{\color{black}  #1}}

\newcommand{\abC}[1]{{\com{[\emph{\color{blue} #1}]}}}

\newcommand{\ps}[1]{{\color{black}  #1}}

\begin{document}

\newcommand{\titleinfo}{Experimental observation of ballistic correlations in integrable turbulence}

\title{\titleinfo}

\author{Elias Charnay} 

\author{Adrien Escoubet} 

\author{François Copie}

\author{Stéphane Randoux}
\affiliation{Univ. Lille, CNRS, UMR 8523, PhLAM – Physique des Lasers, Atomes et Molécules, F-59000 Lille, France}
\author{Thibault Bonnemain} 
\affiliation{Laboratoire de Physique Théorique et Modélisation, CNRS UMR 8089,
CY Cergy Paris Université, 95302 Cergy-Pontoise Cedex, France}
\author{Alvise Bastianello} 
\affiliation{CEREMADE, CNRS, Universit\'e Paris-Dauphine, Universit\'e PSL, 75016 Paris, France}
\author{Pierre Suret}
\affiliation{Univ. Lille, CNRS, UMR 8523, PhLAM – Physique des Lasers, Atomes et Molécules, F-59000 Lille, France}

\begin{abstract}
Unequal-time correlation functions fundamentally characterize emergent statistical properties in complex systems, yet their direct measurement in experiments is challenging.
We report the experimental observation of two-time, ballistic correlations in a photonic platform governed by the focusing nonlinear Schrödinger equation. Using a recirculating optical fiber loop with heterodyne field detection, we acquire the full space-time dynamics of partially coherent optical waves and extract the intensity correlator in stationary states of integrable turbulence. The correlators collapse under ballistic rescaling and quantitatively agree with predictions from Generalized Hydrodynamics (GHD) evaluated using the density of states obtained via inverse scattering analysis of the recorded fields. Our results provide a direct test (without any fitting parameters) of GHD in an integrable waves system. {\bf This work is published in Phys. Rev. Lett. 137, 097202 (2026).}
\end{abstract}

\maketitle

\para{Introduction.}
Correlation functions are central to turbulence
and many-body physics, encoding the spectrum of the constituents of matter~\cite{rickayzen2013} and lying at the foundations of field theory~\cite{Schwinger1951}.
In strongly interacting—or highly nonlinear—systems, correlation functions are intricate because they capture the complex interplay of many excitations, providing a systematic probe of collective phenomena~\cite{spohn2012large,giamarchi2003}.
Seemingly disparate systems, ranging from quantum or classical particles, magnetism, ocean waves, and nonlinear optics to living matter, may look entirely different at microscopic scales, yet they can share the same large-scale \emph{emergent hydrodynamics}~\cite{Kadanoff1963,Spohn2014} built upon conservation laws and symmetries.

In particular, the correlations of conserved charges densities \(\qq(t,{\rm x})\) (whose integral \(\mathcal{Q}=\int \dd {\rm x}\,\qq(t,{\rm x})\) is conserved in time) exhibit remarkable spatiotemporal scalings.
Consider the two-time connected correlations
\be\label{eq_C}
C(\Delta t, \Delta {\rm x})=\big\langle \qq(t,{\rm x})\qq(t_0,{\rm x}_0)\big\rangle-\big\langle \qq(t,{\rm x})\big\rangle\big\langle \qq(t_0,{\rm x}_0)\big\rangle
\ee
where $\Delta t=t-t_0$ and $\Delta{\rm x}={\rm x}-{\rm x}_0$ in homogeneous and stationary states. Averages are taken over intrinsic randomness (e.g., thermal or quantum fluctuations), over space, or over large ensembles of random initial conditions.
Typical systems feature \emph{diffusion} at large scales such as
\(C(\Delta t,\Delta{\rm x})\simeq \Delta t^{-D\scexp}\,\mathcal{C}\!\left(\Delta{\rm x}\,\Delta t^{-\scexp}\right)\), where \(D\) is the system dimensionality and \(\scexp=1/2\). Other scenarios are also possible, for example ballistic scaling \(\scexp=1\)~\cite{Bastianello2022,Bertini2021}, subdiffusion \(\scexp<1/2\)~\cite{Gromov2020,Feldmeier2020,Sanchez2020}, or  superdiffusion \(1/2<\scexp<1\)~\cite{Ljubotina2017,Bulchandani2021,Schuckert2020}. Diffusion naturally follows from Fick’s law, which states that average currents are proportional to the gradient of the charge density; deviations from diffusion indicate exotic hydrodynamics.

In integrable models, Fick’s law breaks down~\cite{Bertini2021,Bastianello2022}. These systems are strongly interacting and possess infinitely many conservation laws,
leading through unconventional ``integrable turbulence"~\cite{Zakharov2009} to relaxation to exotic steady states coined generalized Gibbs ensembles (GGE)~\cite{jaynes1957information,jaynes1957information2,Rigol2007,Calabrese2016}.
The GGE concept formed within quantum many-body physics, and subsequently percolated to integrable PDEs in theory~\cite{DeLuca_2016,Bastianello2018,DelVecchio2020,Koch2022,bonnemain2022generalized, bonnemain2025two} and in experiments~\cite{bastianello2025}. 
The conserved quantities deeply modify the large scale hydrodynamics of integrable models, described by the complementary frameworks of  generalized hydrodynamics (GHD)~\Cite{Alvaredo2016,Bertini2016} and soliton gas (SG) theory~\Cite{zakharov1971, El2005, el2020spectral}.
 Integrable models showcase ballistic correlation functions, as supported by numerical evidence~\cite{Bertini2021} and analytical results~\cite{Spohn1982,Doyon2018,Doyon2022}. However, a direct \emph{experimental} probe of correlation functions has remained a major challenge: access to full spatio-temporal dynamics is often unavailable. This difficulty is especially acute in cold-atom experiments~\cite{Bloch2012} probing quantum dynamics, where measurements are generally destructive~\cite{Schweigler2017,gooding2025}. Despite numerous experimental tests of GHD in cold-atoms \cite{Schemmer2019,Malvania2021,Moller2021,Cataldini2022,Schuttelkopf2024,dubois2024,Yang2024Phantom,horvath2025,zeng2026}, ballistic correlations have never been observed in integrable systems. By contrast, photonics offers a versatile platform to investigate nonlinear integrable partial differential equations (PDEs)~\cite{suret2024soliton}. 

In this Letter, we experimentally probe unequal-time correlation functions in a photonic platform governed by the focusing one-dimensional nonlinear Schrödinger (fNLS) equation—a universal  integrable PDE~\cite{shabat1972exact} with countless applications~\cite{yang2010nonlinear}. In the experiments, initial partially coherent waves (PCWs) first evolve to a stationary state on which we subsequently measure two-point correlation functions of the field's intensity, which is a locally conserved charge density. We show that the two-time correlation of intensity exhibits a striking ballistic scaling as per Eq. \eqref{eq_C} with $D=1$ and $\scexp=1$.

Secondly, we provide a quantitative test of GHD by demonstrating a parameter-free comparison between experimental data and analytical predictions for correlations originally derived in GHD for quantum integrable models~\cite{Doyon2018} and later extended to SG via semiclassical limits of quantum integrability~\cite{DeLuca_2016,Bastianello2018,DelVecchio2020,Koch2022,Koch2023,Bastianello2024}. The fNLS equation features solitons, particle-like excitations with well-defined velocities. The stability of solitons and the elastic nature of their interactions are responsible for the ballistic scaling of the correlation function,  due to solitons that travel with velocity $\Delta x/\Delta t$ and propagate information along their trajectories.
The comparison between theoretical and experimental correlations requires the accurate measurement of the phase-space density of solitons, called density of states (DOS)~\cite{el2020spectral, suret2024soliton}.
Since solitons are infinitely-lived quasiparticles whose decay is prevented by the conservation laws, the DOS is conserved in time and fully characterizes the emergent GGE~\cite{ilievski2015complete, bonnemain2022generalized}.

\para{Experimental setup and model.} Revealing two-time correlations requires access to the full space-time dynamics, which we obtain using an experimental platform based on a recirculating optical fiber loop (ROFL)~\cite{copie2023space,fache2025perturbed,kraych2019statistical}. A schematic of the setup is shown in Fig.~\ref{Fig_setup} and experimental details are provided in the Supplementary Material (SM)~\cite{suppmat}.
At leading order, light propagation in the ROFL is described by the fNLS equation~\cite{kraych2019statistical,copie2023space}
$\partial_z A=-i\,\frac{\beta_2}{2}\,\partial_T^{2}A+i\gamma |A|^{2}A$, 
where \(z\) is the propagation distance, \(T\) is the time in the co-propagating frame at the group velocity of the carrier wave, \(\beta_2=-22~\mathrm{ps}^2\,\mathrm{km}^{-1}\) is the group-velocity-dispersion coefficient, \(\gamma=1.3~\mathrm{W}^{-1}\,\mathrm{km}^{-1}\) is the Kerr nonlinearity coefficient of the fiber, and \(A(z,T)\) is the slowly varying envelope of the electric field~\cite{agrawal2000nonlinear}. Upon introducing the dimensionless variables \(\psi= A/\sqrt{P_0}\), \(t=\gamma P_0 z/2\), and \(x=\sqrt{\gamma P_0/|\beta_2|}\,T\), with \(P_0=\langle |A|^2\rangle\) the mean optical power, we obtain the fNLS equation in dimensionless form
\be
i \partial_t \psi = -\partial_{ x}^{2}\psi - 2|\psi|^{2}\psi .
\label{eq_nls}
\ee

Previous studies using a ROFL focused on modulation instability arising during the propagation of a single plane-wave initial condition seeded with noise~\cite{fache2025perturbed,kraych2019statistical}. This protocol excites solitons with zero  velocity~\cite{gelash2019bound}.
To highlight ballistic correlation functions, which are propagated by solitons with non-zero relative velocity, we have designed an optical source of PCWs. PCWs are  linear superpositions of many statistically independent Fourier components and are well-described by a Gaussian random process with a finite spectral bandwidth $\Delta \nu$~\cite{Mandel_Wolf_1995, walczak_2015}.

\begin{figure}[t!]
\includegraphics[width=0.5\textwidth]{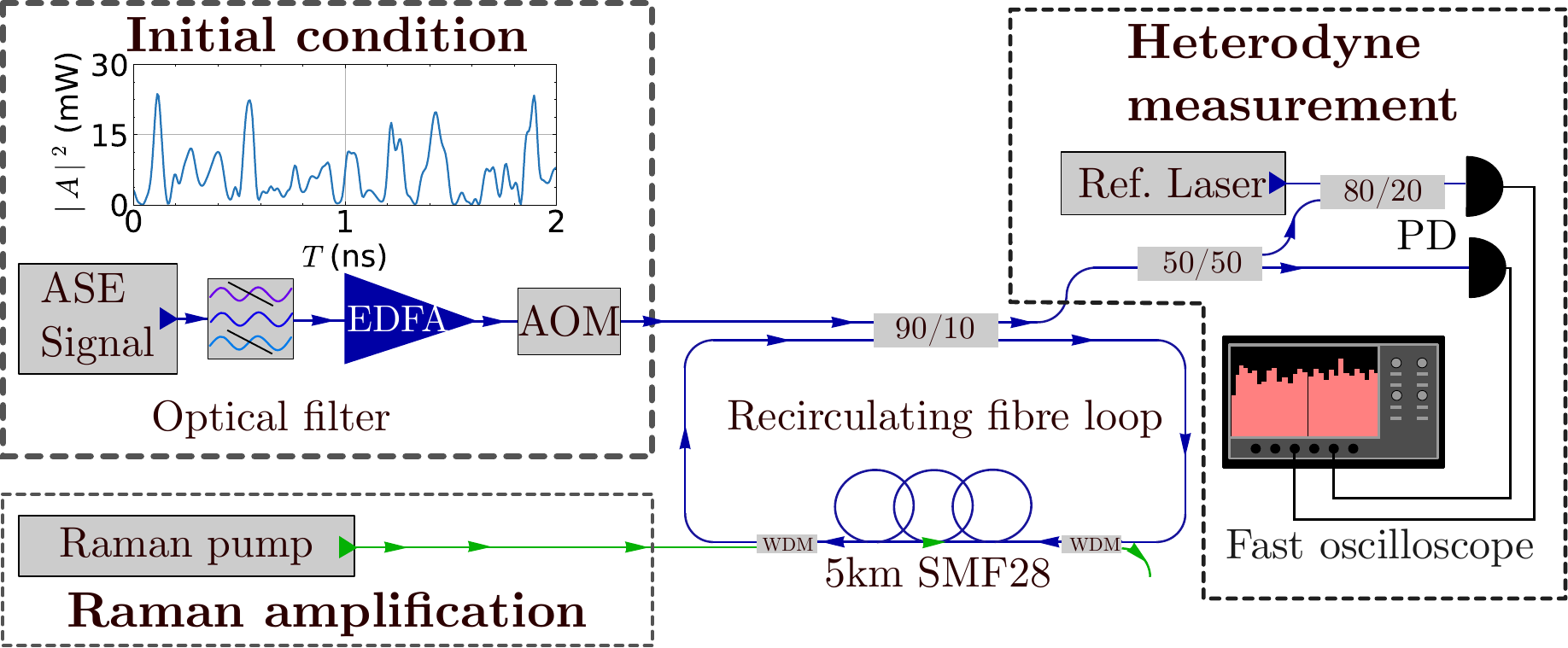}
\caption{\textbf{Experimental setup---} The initial conditions are PCWs generated by an ASE source. The signal is spectrally filtered, amplified by an Erbium Doped fiber Amplifier (EDFA). A \SI{1}{\micro \second}-long section of the signal is sliced by using an Acousto-Optic Modulator (AOM) before being launched into the \SI{5}{km}-long single-mode fiber loop. The signal is recorded at each round trip using both direct and heterodyne detection of the field with photodetectors (PD). Losses are compensated by using a contrapropagating Raman amplifier.}\label{Fig_setup}
\end{figure}
\begin{figure*}[t!]
\includegraphics[width=0.9\textwidth]{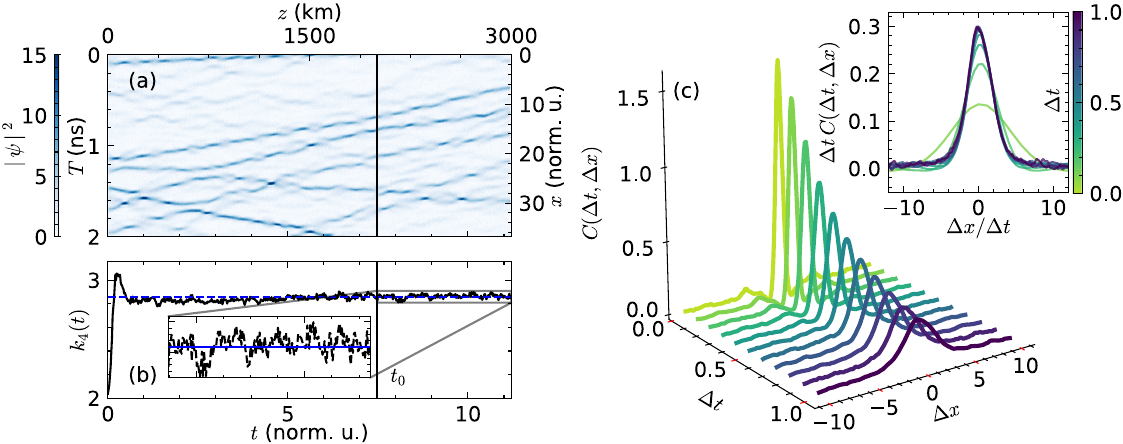}
\caption{\textbf{Ballistic correlation functions in optical fibers (experiments)---} Initial width of the Fourier spectrum of PCWs: $\Delta k = 1.96$.
(a) Spatio-temporal dynamics of the intensity $|\psi(t, x)|^2$ of a typical field configuration. Solitons form at early time and propagate undergoing elastic scattering. $(z, T)$ in physical units correspond to the dimensionless variables $(t,x)$.
(b) Evolution of the kurtosis $\kappa_4(t)$. After a short transient, $\kappa_4$ reaches a stationary value (for $t>5$), hinting at the relaxation of the system to a GGE. (c)  Intensity correlation function $C(\Delta t,\Delta  x)=\langle|\psi(t, x)|^2|\psi(t_0, x_0)|^2\rangle-\langle|\psi(t, x)|^2\rangle\langle|\psi(t_0, x_0)|^2\rangle$ with $\Delta t=t-t_0$ and $\Delta  x= x- x_0$ for $t_0=7.5$, $\Delta t$ is shown in steps of width $0.1$. The inset reports the ballistic rescaling $C(\Delta t, \Delta  x)=\frac{1}{\Delta t}\mathcal{C}(\Delta  x/\Delta t)$, showing the collapse of all the curves for $\Delta t\ge 0.25$. 
}\label{Fig_kurtosis_correl}
\end{figure*}
More precisely, in the experiments reported here, the initial conditions are made of long wavetrains of PCWs generated from a filtered amplified spontaneous emission (ASE) source at a central wavelength $\sim$ \SI{1554}{nm}. Using a tunable spectral filter, we produce a Fourier spectrum with width  \(\Delta \nu \simeq \SI{5}{GHz}\) (see~\cite{suppmat}). The total flat pulse duration is about \SI{1}{\micro \second}, which is much longer than the fluctuation timescale of \(\sim \SI{100}{ps}\) (see inset in Fig.\,\ref{Fig_setup}).

The dimensionless coordinate \(x\) and time \(t\) correspond, respectively, to the physical time \(T\) and the experimental propagation distance \(z\);
accordingly, the physical frequency $\nu$ maps to the dimensionless wavenumber $k$. PCWs have delta-correlated Fourier components \(\langle \tilde{\psi}^{*}(k')\tilde{\psi}(k)\rangle = n(k)\,\delta(k-k')\), leading to Gaussian-distributed field $\psi(t=0, x)$~\cite{walczak_2015}. 
The spectrum $n(k)$ is bell-shaped  (see SM~\cite{suppmat}) with half width at half maximum $\Delta k = 2\pi\Delta \nu\,/\sqrt{\gamma P_0/|\beta_2|}$ which regulates the strength of the nonlinearity:  increasing \(\Delta k\) enhances the relative contribution of \(\partial_x^2\psi\) compared with \(|\psi|^2\psi\) in Eq. \eqref{eq_nls}. In the experiments, \(\Delta \nu\) is fixed, and two values of the mean optical power, \(P_0= \SI{5.85}{mW}\) and \(\SI{2.84}{mW}\), are used to tune $\Delta k$, corresponding to \(\Delta k=1.96\) and \(2.81\).

The initial field is injected into a $\sim 5\mathrm{km}$ single-mode fiber loop  (round-trip time $25~\mu\mathrm{s}$); a 90/10 coupler recirculates \(90\%\) of the power, while \(10\%\) is tapped each round trip and measured with ultrafast photodetectors and an oscilloscope (\SI{32}{GHz} combined bandwidth). Losses induced by fiber propagation and by the per-round-trip extraction are compensated using Raman amplification.

Periodic extraction at each round trip enables stroboscopic monitoring of the wave-field evolution over successive \(5~\mathrm{km}\) propagation intervals. Numerical processing of the temporal signals allows us to construct spatio-temporal dynamics of the intensity $|\psi(t,x)|^2$ over hundreds of round trips within the fiber loop (see Fig.~\ref{Fig_kurtosis_correl}(a)). The space-time diagram reveals that solitons quickly emerge during the propagation of PCWs and collide elastically. In what follows, all the experimental results are reported in the dimensionless units, and averaged over $70$ experimental runs of width $x\in [-500,500]$. We also spatially average across the whole sample, thanks to translational invariance of the GGE.

\para{Measurement of ballistic correlations}  It is well established that, starting from initial PCWs, a system governed by the fNLS equation \eqref{eq_nls} evolves to a statistically stationary state of integrable turbulence~\Cite{Zakharov2009,agafontsev2021extreme} described by a GGE~\cite{Koch2022}. We measure the unequal-time correlation function of the intensity on this stationary GGE, as in Eq.~\eqref{eq_C} with ${\rm q}(t,x)=|\psi(t,x)|^2$ .

To estimate the time \(t=t_{0}\) at which the stationary state is reached we monitor the kurtosis $\kappa_{4}(t)=\langle |\psi(t, x)|^{4}\rangle/\langle |\psi(t, x)|^{2}\rangle^{2}$~\cite{agafontsev2021extreme} shown in Fig.~\ref{Fig_kurtosis_correl}(b). For PCWs, the kurtosis evolves from $\kappa_{4}=2$ toward a fixed value $2\le \kappa_{4}\le 4$ which increases with the strength of nonlinearity~\cite{agafontsev2021extreme}. Stationarity is reached for  times $t\gtrsim 5$, and below we evaluate correlations for $t\ge t_{0}$ with $t_{0}=7.5$.
For $\Delta k=1.96$ and $\Delta k=2.81$, the kurtosis relaxes to $\kappa_4=2.85$ and $\kappa_4=2.50$ respectively.
The experimental intensity correlator \(C(\Delta t,\Delta x)\) exhibits a decrease in amplitude and a concomitant increase in spatial width with increasing \(\Delta t\) (see  Fig.~\ref{Fig_kurtosis_correl}(c)). Our first key experimental result is showing the remarkable collapse of this two-point correlator under {\it ballistic rescaling} $C(\Delta t,\Delta{ x})\simeq \Delta t^{-1}\,\mathcal{C}\!\left(\Delta  x\,\Delta t^{-1}\right)$ for $\Delta t = t-t_0\ge 0.25$ (inset  in Fig.~\ref{Fig_kurtosis_correl}(c)). This ballistic behavior of nonlinear waves is a hallmark of a nearly integrable regime in our experiments.

\para{Comparison with GHD predictions.}
 The comparison between experiments and theory  requires the accurate measurement of the DOS of the nonlinear modes characterizing the GGE. In principle, the whole-line inverse scattering (IST) suggests fNLS features both radiative and solitonic modes \cite{faddeev2007hamiltonian}. However, solitons alone provide a complete description of GGEs \cite{Koch2022,Koch2023,Bastianello2024} and radiation is describable as a collective effect of solitons~\cite{jenkins2024approximation}.
Solitons are labeled by a complex spectral parameter $\lambda$, whose real part $\Re(\lambda)\in(-\infty,\infty)$ is proportional to the free velocity of the soliton  $v(\lambda)=-4\,\Re(\lambda)$, 
and its imaginary part $\Im(\lambda)\in(0,+\infty)$ gives the maximum field's intensity $4\Im(\lambda)$ of the solitonic profile.
The DOS $\rho(\lambda)$ describes the spatial density of solitons with spectral parameter $\lambda$~\cite{el2020spectral}. Given a field configuration $\psi(t,x)$, the associated DOS can be extracted within the IST formalism computing the discrete eigenvalues $\lambda$ of the Zakharov-Shabat problem~\cite{shabat1972exact}, see also SM~\cite{suppmat}. These eigenvalues are constants of motion and $\rho(\lambda)$ (space-averaged) is independent of $t$. 

Solitons collide elastically: the only effect of interactions is a displacement of their positions, or \emph{scattering shift} \(\Delta\), relative to their free trajectories~\cite{zakharov1973interaction,copie2023space}. In a soliton gas, at leading order on large scales, solitons can be regarded as particles moving with an \emph{effective velocity} renormalized by collisions with other solitons. On large scales, an inhomogeneous system is described by a space-time generalization of the DOS $\rho_{t, x}(\lambda)$ which obeys the fundamental kinetic equation~\cite{zakharov1971,El2005, Alvaredo2016,Bertini2016} 
\be\label{eq_ghd}
\partial_t \rho_{t, x}(\lambda)+\partial_{  x} \!\left[v_{t, x}^{\mathrm{eff}}(\lambda)\,\rho_{t, x}(\lambda)\right]=0\,,
\ee
where \(v^{\mathrm{eff}}_{t,x}(\lambda)\) is the effective velocity, determined self-consistently from the integral equation
\be\label{eq_veff}
v_{t,  x}^{\mathrm{eff}}(\lambda)= v(\lambda)
+ \int_{\Gamma} \dd \mu \;\Delta(\lambda,\mu)\,\rho_{t, x}(\mu)\,\big[v_{t,x}^{\mathrm{eff}}(\lambda)-v_{t, x}^{\mathrm{eff}}(\mu)\big]\,,
\ee
thereby making Eq.~\eqref{eq_ghd} highly nonlinear. Here, \(\Delta(\lambda,\mu)=\frac{1}{\Im(\lambda)}\log\!\left|\frac{\mu-\bar\lambda}{\mu-\lambda}\right|\) is the scattering shift and \(\Gamma\) is the domain of the upper half-plane where \(\mu\) takes values.
Equations~\eqref{eq_ghd} and \eqref{eq_veff} have been derived independently for the fNLS equation within SG (finite-gap) theory~\cite{el2020spectral} and GHD~\cite{Koch2022}, and it is supported by extensive numerical~\cite{congy2021soliton,congy2025riemann} and experimental~\cite{suret2023soliton,fache2024interaction} benchmarks. 

Equation~\eqref{eq_ghd} does not directly provide access to the correlations in Eq.~\eqref{eq_C}, as it is insensitive to fluctuations in the soliton densities, which are instead captured by the GGE. 
The GGE is an invariant measure under fNLS' dynamics given by $e^{-\sum_j \beta_j \mathcal{Q}_j[\psi]}$, where $\{\beta_j\}$ are suitable Lagrange multipliers and the summation is over all the infinitely many conserved charges $\mathcal{Q}_j[\psi]$. GGEs generalize conventional Gibbs ensembles \(\propto e^{-\beta \mathcal{H}[\psi]}\) for nonintegrable systems with only one charge and Lagrange multiplier (the Hamiltonian \(\mathcal{H}[\psi]\) and inverse temperature \(\beta\) respectively). As it is well known from statistical physics~\cite{huang2009}, the average charge $\langle \mathcal{Q}_i[\psi]\rangle_{\{\beta_j\}}$ as a function of $\{\beta_j\}$ also determines equal-time fluctuations via  $-\partial_{\beta_{i'}}\langle\mathcal{Q}_i[\psi]\rangle_{\{\beta_j\}_j} = \langle \mathcal{Q}_{i'}[\psi]\mathcal{Q}_i[\psi]\rangle-\langle \mathcal{Q}_{i'}[\psi]\rangle\langle\mathcal{Q}_i[\psi]\rangle$.
\begin{figure}[t!]
\includegraphics[width=0.48\textwidth]{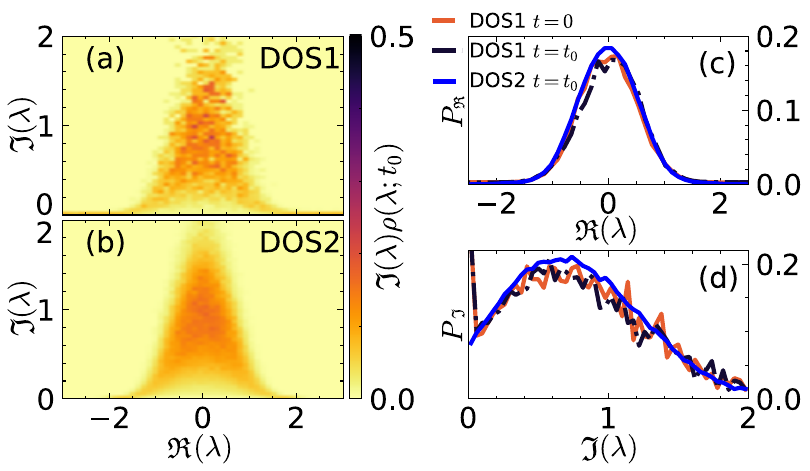}
\caption{\textbf{Density of states (DOS) measurement.}
Panels (a,b): The density of states DOS1 and DOS2 in the complex $\lambda$-plane at $t=t_0=7.5$ (initial spectral width of PCWs $\Delta k=1.96$)~\cite{suppmat}. Since $\rho(\lambda)$ diverges as $\lambda\to\pm\infty+i0$, we plot $\Im(\lambda)\rho(\lambda)$, which is finite~\cite{Koch2022}.
Panels (c,d): DOS marginals $P_{\Re}=\int \dd\Im(\lambda)\,\Im(\lambda)\rho(\lambda)$ and $P_{\Im}=\int \dd\Re(\lambda)\,\Im(\lambda)\rho(\lambda)$. Marginals for DOS1 at $t=0$ (orange, solid) and $t=7.5$ (black, dotted) demonstrate DOS conservation, and they are compared with the marginal of DOS2 shown at $t=7.5$ (blue line).}
\label{Fig_DOS}
\end{figure}

The method of \emph{hydrodynamic projections}~\cite{spohn2012large,Doyon2018,Doyon2022} combines GGEs' fluctuations in linear response with Eq.~\eqref{eq_ghd} to access unequal-time correlation functions. 

The idea is to compute $\langle{\rm q}_{i'}(t, x)\rangle$ propagating from an initial localized and inhomogeneous perturbation of $\beta_i$ in the GGE.
The key point is now connecting the DOS with the Lagrange multipliers $\{\beta_j\}_j$. In classical PDEs, this remained an open problem for a long time \cite{Chung1990}, until its recent solution through semiclassical limits of quantum integrability~\cite{Koch2022,Koch2023,Bastianello2024}. 
The SM~\cite{suppmat} summarizes the hydrodynamic projections' approach for fNLS, yielding a compact expression for the intensity correlations \( |\psi|^{2} \):
\be\label{eq_GHD_C}
C(\Delta t,\Delta  x)\simeq \frac{1}{\Delta t}\int_\Gamma \dd\lambda\, \delta\!\left(\tfrac{\Delta  x}{\Delta t}-v^\eff(\lambda)\right)\,\rho(\lambda)\,[h_0^\dr(\lambda)]^{2}\,,
\ee
where $h_0^\dr(\lambda)=4\,\Im(\lambda)[1-\int \dd\mu\, \Delta(\lambda,\mu)\,\rho(\mu)]$.
This equation has a natural interpretation: solitons are stable and propagate ballistically with their effective velocity along the ray \(v^\eff(\lambda)=\Delta  x/\Delta t\), carrying information and correlations. Their contribution, equal to \(4\,\Im(\lambda)\) in the absence of surrounding solitons, is renormalized by interactions to \(h_0^\dr(\lambda)\).
\begin{figure}[t!] 
\includegraphics[width=0.85\linewidth]{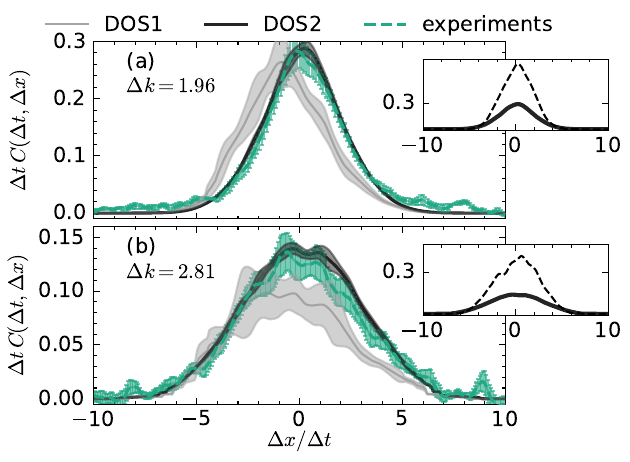}
\caption{\textbf{Comparison between experimental and theoretical correlations---} Rescaled ballistic correlations from Eq.~(\ref{eq_GHD_C}) (DOS1 : gray lines; DOS2 : black lines). Experimental correlations for $\Delta t =1$ (solid green line). The shaded area represents the root mean square error. (a) Strong nonlinearity $\Delta k=1.96$ (b) Weak nonlinearity $\Delta k=2.81$. Insets: the correlation for free particles, computed without dressing or effective velocities in Eq.~\ref{eq_GHD_C} from DOS2 (dashed black lines)}
    \label{fig_comparison}
\end{figure}

The comparison between the experimentally measured correlations and Eq.~\eqref{eq_GHD_C} requires the determination of the DOS from the full field (phase and amplitude),  measured by heterodyne detection~\cite{fache2025perturbed,suppmat} (see Fig.~\ref{Fig_setup}). Since the numerical evaluation of Eq.~(\ref{eq_GHD_C}) is highly sensitive to tiny errors in the DOS measurement, 
we investigated different strategies to retrieve $\rho(\lambda)$ from the experimental data (see~\cite{suppmat} for details). In the first strategy (DOS1), the field is measured directly at $t=t_0$. In the second, more indirect, strategy (DOS2): (i) we compute the Fourier spectrum of the field measured at $t=0$, (ii) we generate numerically configurations of PCWs by assuming random phases for the Fourier components, (iii) we propagate the field up to $t=t_0$ by integrating numerically the fNLSE, taking into account the residual losses of the ROFL. As the Fourier spectrum broadens during propagation~\cite{suppmat}, the field measured at $t=0$ used for DOS2 is much less distorted by the finite detection bandwidth than the one measured at time $t=t_0$ (DOS1). Moreover, the DOS2 strategy enables the numerical generation of a larger number of random realizations, reducing the statistical uncertainty.

In Fig.~\ref{Fig_DOS}(a-b) we show DOS1 and DOS2 in the complex \(\lambda\)-plane. Figs.~\ref{Fig_DOS}(c,d) report the marginals of the DOS computed at two different times, demonstrating the conservation of the DOS during propagation and the fact that our system operates very close to integrability.

In Fig.~\ref{fig_comparison}, we compare the ballistic rescaling of correlators measured directly in the experiments with those computed from Eq.~\eqref{eq_GHD_C} for the two strengths of the nonlinearity and both DOS1 and DOS2.
Remarkably, the experimental results (green lines) and the theoretical prediction from DOS2 (black line), computed without fitting parameters, agree with good accuracy in both cases. Predictions from DOS1 show  deviations, interpreted as systematic imperfections due to the finite bandwidth of detection and asymmetric transfer function for broad Fourier spectra~\cite{suppmat}. Alternative theoretical approaches cannot describe the experimentally measured ballistic correlations. The dynamics is markedly non-linear and featuring sharp structures, ruling out Bogoliubov theory~\cite{Kovrizhin2001} or phase-density approaches~\cite{kivshar2003optical}. Conventional hydrodynamics based on few conservation laws predicts sub-ballistic broadening governed by Kardar--Parisi--Zhang dynamics~\cite{Spohn2014}.

A further phenomenological approach is to treat solitons as if they were free particles, by replacing in Eq.~\eqref{eq_GHD_C} $v^{\mathrm{eff}}(\lambda)$ with the free soliton velocity $v(\lambda)$ and \(h_0^{\mathrm{dr}}\) with the “free charge” \(4\,\Im(\lambda)\). The free particle prediction (dashed lines in the insets of Fig.~\ref{fig_comparison}, shown only for DOS2) is markedly different from GHD, showing the latter is the only theoretical framework to quantitatively reproduce the observed correlators.

\para{Discussion.} Unequal-time correlations offer deep insight into nonlinear and many-body dynamics, encoding symmetries, conservation laws, and emergent hydrodynamics. We have experimentally measured ballistic correlation functions in stationary states of the fNLS equation (an archetypal integrable PDE) realized in a ROFL. Our measurements quantitatively confirm that the exact predictions of GHD~\cite{Doyon2018,Doyon2022,Koch2022} (originally developed for quantum systems) describe stationary properties of integrable turbulence. 
Our results represent a unique  probe of the statistical mechanics and hydrodynamics of integrable models, while two-time correlations remain notoriously difficult to access in quantum experiments~\cite{Schweigler2017,gooding2025}. Our approach opens the way to experimental studies of fundamental open problems such as the transient approach to ballistic scaling and the onset of diffusion from convection~\cite{DeNardis2018},  multipoint correlators~\cite{Doyon2018},  large deviations of charge fluctuations~\cite{Doyon2020,Doyon2023} or the emergence of long-range correlations~\cite{Doyon2023L,hubner2025diffusive}. Our work also contributes to testing the robustness of IST analysis of complex experimental fields, which may find applications in nonlinear optical telecommunications~\cite{prilepsky2014nonlinear}.

\para{Data availability} The data that support the findings of this article are not publicly available upon publication because it is not technically feasible and/or the cost of preparing, depositing, and hosting the data would be prohibitive within the terms of this research project. The data are available from the authors upon reasonable request.

\para{Acknowledgments} The authors thank Benjamin Doyon
for fruitful discussions. E. C. acknowledges the Student
Workshop on Integrability where part of this work has been completed, and thanks Loïc Fache for help with the
heterodyne detection method. The authors are grateful to
the Centre d’Etudes et de Recherche Lasers et Application
(CERLA) for technical support. This work was partially
supported by the Agence Nationale de la Recherche
through the StormWave (No. ANR-21-CE30-0009) and
SOGOOD (No. ANR-21-CE30-0061) projects, the
Ministry of Higher Education and Research, Hauts de
France Council, and European Regional Development
Fund (ERDF) through the Contrat de Projets Etat-
R´egion (CPER Photonics for Society P4S). The authors
acknowledge the support of the CDP C2EMPI, as well as
the French State under the France-2030 program, the
University of Lille, the Initiative of Excellence of the
University of Lille, the European Metropolis of Lille for
their funding and support of the R-CDP-24-004-C2EMPI
project and the WILL-CHAIRES-25-003-GHyPhot project
(No. ANR-21-IDES-0006). A. B. is supported by the ERC
StG No. 101220476–NonErgHydro.

\nocite{takahashi2005, congy2024}

%

\clearpage
\onecolumngrid
\clearpage
\centerline{\large \bf Supplementary Material}
\centerline{\large \bf Experimental observation of ballistic correlations in integrable turbulence}

\setcounter{equation}{0}  
\setcounter{figure}{0}
\setcounter{page}{1}
\setcounter{section}{0}    
\renewcommand\thesection{\arabic{section}}    
\renewcommand\thesubsection{\arabic{subsection}}    
\renewcommand{\thetable}{S\arabic{table}}
\renewcommand{\theequation}{S\arabic{equation}}
\renewcommand{\thefigure}{S\arabic{figure}}
\setcounter{secnumdepth}{2}  

\tableofcontents

\section{Detailed experimental setup and DOS measurement}

\begin{figure*}[h]
\includegraphics[width=0.95\textwidth]{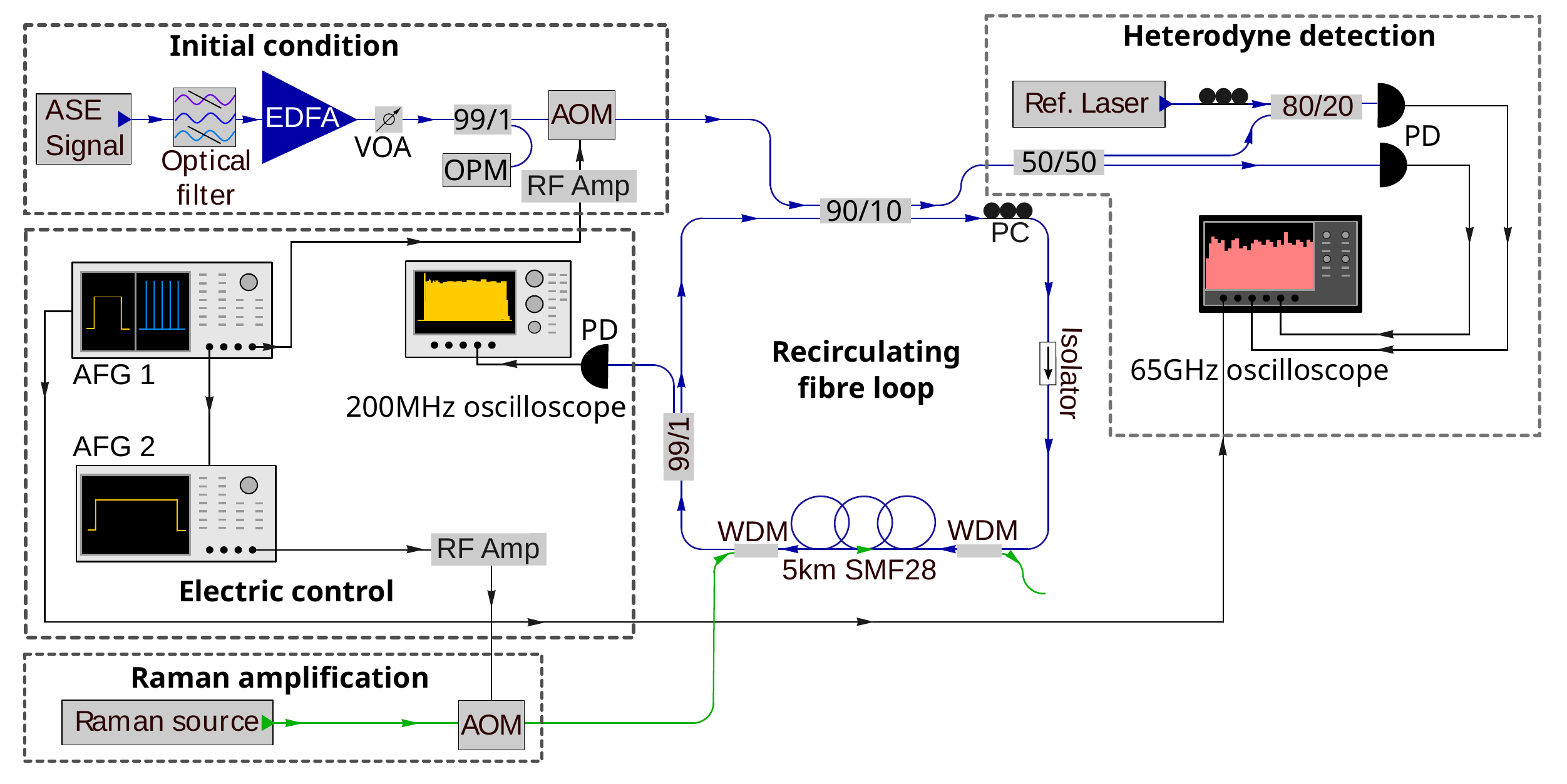}
\label{Fig_scheme_full}
\caption{Detailed experimental setup. ASE: Amplified Spontaneous Emission. EDFA: Erbrium Doped Fiber Amplifier. VOA: Variable Optic Attenuator. OPM: Optical Power Meter. AOM: Acousto-Optic Modulator. RF Amp: Radio-frequency amplifier. AFG: Arbitrary Function Generator. WDM: Wave Division Multiplexer. PD: Photodetector. }
\end{figure*}

\subsection{Experimental setup}

In this work, we use an experimental platform based on a recirculating optical fiber loop (ROFL)~\cite{SMcopie2023space,SMfache2025perturbed}. In contrast to previous studies on modulation instability seeded by a single-frequency laser, we employ partially coherent waves (PCWs) generated by an amplified spontaneous emission (ASE) source. Our experimental setup, with technical details, is shown in Fig.~\ref{Fig_scheme_full}.\\

\textbf{In brief}, the initial conditions consist of PCWs produced by a filtered ASE source at a central wavelength of approximately \SI{1554}{nm}. A tunable spectral filter sets a full width at half maximum (FWHM) of \(\Delta\nu= \SI{10}{GHz}\). A temporal window of \SI{1}{\micro \second} is carved by an acousto–optic modulator (AOM), and the resulting signal is subsequently amplified with an erbium-doped fiber amplifier (EDFA). This standard initial condition corresponds to the linear superposition of many statistically independent Fourier components and is therefore well described by a Gaussian random process.

The initial field is injected into a fiber loop consisting of approximately \SI{5}{km} of single-mode fiber (SMF), closed via a 90/10 coupler. The coupler is configured to recirculate \(90\%\) of the optical power. The optical signal propagates clockwise; at each round trip, \(10\%\) of the circulating power is extracted and sent to photodetectors (PDs) connected to a high-speed oscilloscope. Periodic extraction at every round trip enables stroboscopic monitoring of the wave–field evolution over successive \SI{5}{km} propagation intervals. Losses from fiber propagation and per-round-trip extraction are compensated using Raman amplification.\\

\textbf{More precisely}, the ASE source (BKtel TBS-C220-FCAPC) delivers \(\sim\SI{20}{mW}\) of average optical power, which is spectrally filtered (EXFO XTM-50) to retain \(\sim\SI{10}{GHz}\) around \(\lambda=\SI{1554.43}{nm}\). The field is then amplified twice by erbium-doped fiber amplifiers: first by a preamplifier to \(\sim\SI{5}{mW}\), and then by a two-stage EDFA to a mean power of \(\sim\SI{1}{W}\). The random waves are gated by an AOM (VSF MT110-IR25-Fio) into \(\sim\SI{1}{\micro \second}\) windows before injection into the ROFL.

Ten percent of the optical power enters the fiber loop. The polarization is adjusted by a manual polarization controller (PC) before propagation through \SI{5}{km} of standard telecom fiber (SMF-28) with group-velocity dispersion \(\beta_{2}=\SI{-22}{ps^2/km}\) and Kerr coefficient \(\gamma=\SI{1.3}{W^{-1}km^{-1}}\). The light emitted by a Raman fiber laser (CRFL, Keopsys) at \(\SI{1455}{nm}\) counterpropagates in the fiber loop and serves as the pump for Raman amplification. This amplification compensates fiber absorption and fiber coupler losses. An AOM gates the Raman pump into \(\SI{20}{ms}\) pulses, thereby setting the experiment duration.

The mean optical power is monitored by diverting \(1\%\) of the circulating light to a slow PD (DET08 FC) and a \SI{200}{MHz} oscilloscope (Tektronix 3 Series).

Fast detection is provided by two high-speed PDs (Finisar XPDV2120R) connected to the “slow” and “fast” channels of a Teledyne LeCroy LabMaster MCM-ZI oscilloscope (up to \(\SI{160}{GS/s}\)), with bandwidths of \(\SI{32}{GHz}\) and \(\SI{65}{GHz}\), respectively. Half of the signal is sent to the first PD for direct intensity measurements, providing accurate power calibration and faithful reconstruction of the dynamics. On the second PD, the signal is mixed with a continuous-wave reference (APEX AP35520) at \(\SI{1554.76}{nm}\), enabling a heterodyne measurement. This produces a frequency-shifted replica at \(\delta\nu\approx\SI{41.5}{GHz}\), set by the beat between the signal’s central frequency and the reference; we numerically extract the components around \(\delta\nu\) as in Ref.~\cite{SMfache2025perturbed}.

Electronic synchronization is required between the initial condition generation, the Raman amplification and the heterodyne recording. It is assured by two AFGs, the first (AFG Tektronix 31000 series) pilots the initial condition and the detection, and triggers the second one (AFG Tektronix 1062) for the Raman amplification.

\subsection{Generation and measurement of Partially coherent waves}

The filtered ASE source generates the initial random partially coherent waves (PCWs). These correspond to a linear superposition of many \emph{delta-correlated} Fourier modes,
\(\langle \tilde{\psi}^*(k')\tilde{\psi}(k)\rangle = n(k)\,\delta(k-k')\),
such that, in real space, the real and imaginary parts are Gaussian distributed. PCWs constitute a benchmark for extreme-wave statistics and are well known in integrable turbulence~\cite{SMwalczak_2015}. The kurtosis—a metric quantifying the propensity for extreme events—starts at 2 and asymptotically converges through propagation to a value between 2 and 4, depending on the strength of the nonlinearity~\cite{SMagafontsev2021extreme}.\\

\noindent{\bf Heterodyne measurement}\\
\\
The experimental signal is mixed with a stable reference laser, producing a frequency shift \(\delta \nu \approx \SI{41.5}{GHz}\).
Following the procedure described in~\cite{SMfache2025perturbed}, the full field (phase and amplitude) is retrieved from the recorded traces.
To improve measurement accuracy, we correct the signal using the transfer function of the detection chain (photodiode and oscilloscope).

\begin{figure}[h]
    \centering
    \includegraphics[width=0.8\linewidth]{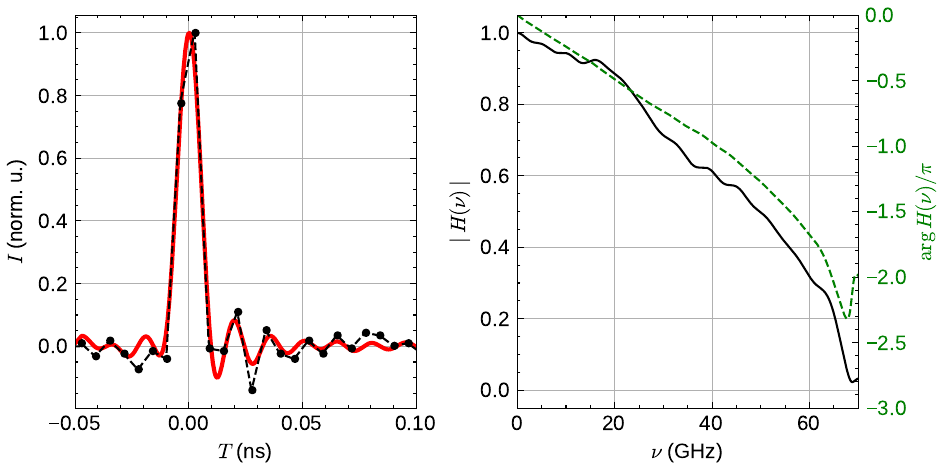}
    \caption{On the left: \ch{impulse response of a single signal (black dashed lines with dots),} and interpolated averaged signals (red line). On the right: reconstructed transfer function from taking the FFT of the red signal and using Bode's relations, amplitude (black line) and phase (green dashed line).}
    \label{fig:SM_transfer_function}
\end{figure}

The impulse response of the detection system is measured with a mode-locked femtosecond laser (Pritel).
To eliminate timing jitter, we fit a Gaussian to each interpolated trace to locate the maximum, recenter the waveform at \(T=0\), and average over \(10^{4}\) recorded pulses.
A fast Fourier transform (FFT) then yields the magnitude of the transfer function, \(|H(\nu)|\).
Using the Bode gain–phase relation, we reconstruct the phase,
\begin{equation}
    \arg H(\omega)
    = \frac{1}{\pi}\,\mathcal{P}\!\int_{-\infty}^{+\infty}\dd\omega'
      \frac{\ln |H(\omega')|}{\omega - \omega'}\,,
\end{equation}
which provides the full transfer function \(H(\omega)\).
In the demodulation procedure of the experimental signals, we divide the spectrum by \(H(\omega)\), thereby partially correcting the bandwidth-induced asymmetries (see below Fig~\ref{fig:SM_transfer_function} for the impulse response and reconstructed transfer function and Fig~\ref{fig:spectrum} for the influence of transfer function on the Fourier spectrum).\\

\noindent{\bf Spectrum}\\

\begin{figure}[b!]
    \centering
    \includegraphics[width=0.5\linewidth]{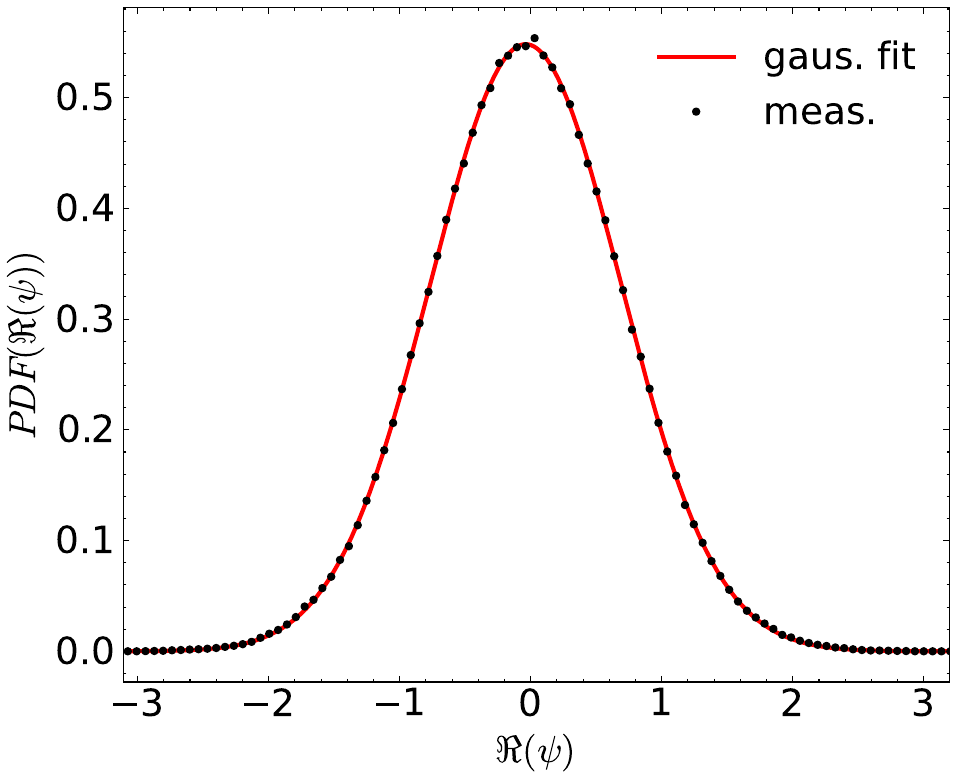}
    \caption{\abC{Add caption}\ch{Experimental measurements of the initial probability density functions (PDF) of the field: PDF of the real part (in black) with a gaussian fit (in red)}}
    \label{fig:SM_gaussianity}
\end{figure}

\begin{figure}[h!]
\centering
\includegraphics[width=0.8\linewidth]{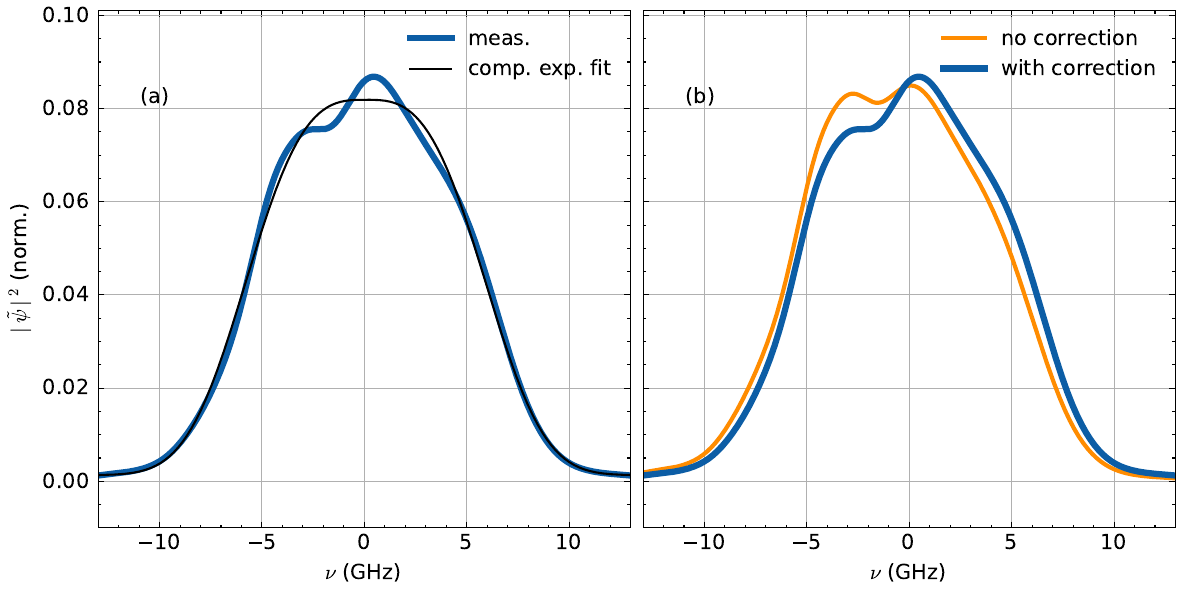}
\caption{Experimental measurement of the filtered-ASE Fourier spectrum. Panel (a) : smoothed average of the measurements (blue),  compressed-exponential fit to the averaged spectrum (black) of the form \ch{$n(\nu) \propto \exp{\left[-\ln{2} \abs{\frac{\nu}{\Delta \nu}}^p\right]}$ }where \(p=3.08\) and \ch{$\Delta \nu = \SI{5.80}{GHz}$}. Panel (b) : comparison between the spectra with (in blue) and without (in orange) bandwidth correction. All displayed spectra are normalized by their mass $\int \dd\nu\abs{\tilde\psi}^2 $.}
\label{fig:spectrum}
\end{figure}

\noindent With our heterodyne detection, we record a hundred of time-domain traces of the optical field’s phase and amplitude dynamics. \ch{First, we check that the initial field is gaussianly distributed. In Fig.~\ref{fig:SM_gaussianity} we show that the probability distribution of the real part of the field $\psi(x)$ follows a suitably-normalized Gaussian distribution. The imaginary part (not shown) is analogously distributed. Then, we move on to characterize the Fourier spectrum. } We compute the complex-field spectrum using standard FFTs and then average and smooth the spectra, both with and without the transfer-function correction (see above). Owing to the optical filtering, the measured Fourier spectrum decays more sharply than a Gaussian. An empirical compressed-exponential fit of the form \ch{$n(\nu) \propto \exp{\left[-\ln{2} \abs{\frac{\nu}{\Delta \nu}}^p\right]}$ yields $p=3.08$ and $\Delta\nu = \SI{5.80}{GHz}$. } 
\ch{We checked with numerical simulations that a more precise interpolation of the spectrum obtained combining the compressed-exponential with high-order polynomials $(a+b\nu+c\nu^2+d\nu^3+e\nu^4)\exp{\left[-\lvert(\nu-\nu_0)/\Delta\nu\rvert^p\right]}$ does not alter the measured correlation functions within the errorbars. Hence, we use the simpler compressed-exponential fit for simplicity. }
The parameters \(\Delta \nu\) and the initial power \(P_{0}\) can then be recast into the dimensionless spectral width—i.e., an inverse measure of the nonlinearity strength—via
\(\Delta k = 2\pi\,\Delta \nu \sqrt{\dfrac{|\beta_{2}|}{\gamma P_{0}}}\).

\begin{figure}[h!]
    \centering
    \includegraphics[width=0.5\linewidth]{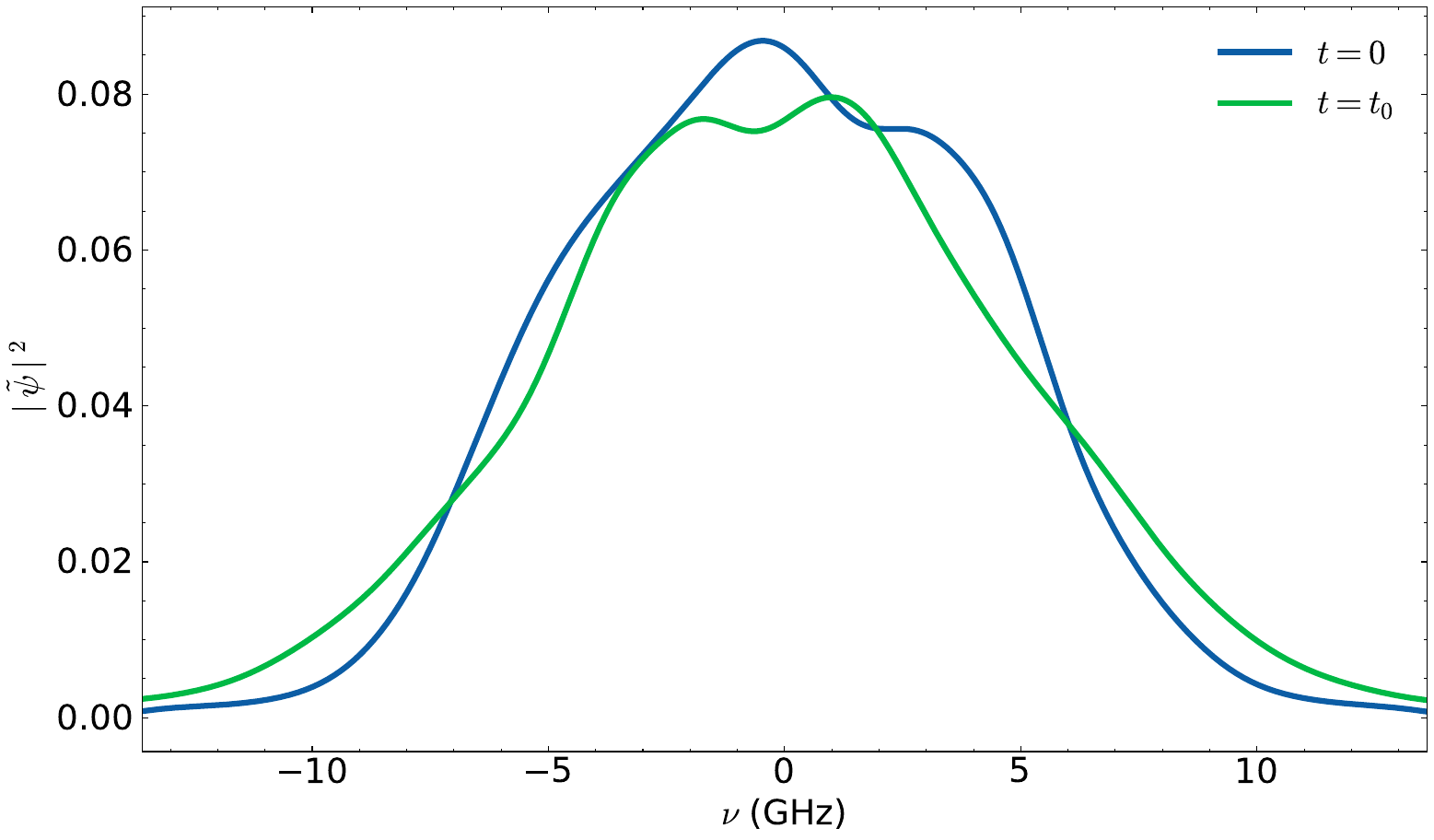}
    \caption{\ps{Experimental measurement of the Fourier spectrum of the initial field ($t=0$) and after nonlinear propagation ($t=t_0$). The spectral width increases during propagation.}}
    \label{fig:specBroad}
\end{figure}

\ps{Finally, it is important to note that the Fourier spectrum is broader after nonlinear propagation ($t=t_0$; see Fig.~\ref{fig:specBroad}). We use this property to improve the accuracy of the DOS determination by adopting the DOS2 strategy rather than the DOS1 strategy, as described below.}

\subsection{Measurement of the DOS}

\noindent{\bf Direct scattering transform}\\

\noindent We compute the DST by numerically solving the Zakharov–Shabat problem~\cite{SMyang2010nonlinear}. Introducing an auxiliary two-component complex field \(Y(x)\), we consider the eigenvalue equation
\be\label{eq_ZS}
\begin{pmatrix}
-\partial_x & \psi(x)\\
\psi^*(x) & \partial_x
\end{pmatrix} Y(x)= i\lambda\, Y(x)\, .
\ee
Eigenvalues \(\lambda\) with positive imaginary part are the rapidities of the solitons supported by the field configuration \(\psi(x)\). To obtain the DOS, we solve the above eigenvalue problem for each field realization \(\psi\) and build a histogram of the resulting eigenvalues, which converges to the DOS. A cutoff is applied to discard eigenvalues with very small imaginary parts (\(\lesssim 10^{-10}\)) from the histogram. For the numerical solution of Eq.~\eqref{eq_ZS}, \(Y(x)\) is represented on the same lattice as \(\psi(x)\), while the derivative operator \(\partial_x\) is discretized using the Fourier collocation method~\cite{SMyang2010nonlinear}: \(\partial_x Y(x)\) is Fourier transformed, \(\partial_x Y(x)\to -i k\,\tilde{Y}(k)\), and then sampled on the lattice-allowed Fourier modes \(\{k_j\}\) as \(-i k_j \tilde{Y}(k_j)\).
The DST is highly sensitive to fluctuations of the field and therefore requires very accurate field measurements. In our work, we employed two complementary strategies to reconstruct the DOS from the experimental data (see below).\\

\noindent{\bf Direct computation of the DOS from the experimentally measured field (DOS1)}\\

\noindent The first strategy (DOS1) to compute the DOS is to apply the direct scattering transform to the measured complex field \(\psi(t,x)\) using the Fourier collocation method at $t=t_0$ (see above and Ref.~\cite{SMyang2010nonlinear}) over \(70\) realizations. The resulting eigenvalue distributions are then averaged over all realizations, and the normalized probability density in the complex plane is divided by the observation-window length (see, e.g., Ref.~\cite{SMgelash2019bound}). 

\begin{figure}[h!]
    \centering
    \includegraphics[width=0.8\linewidth]{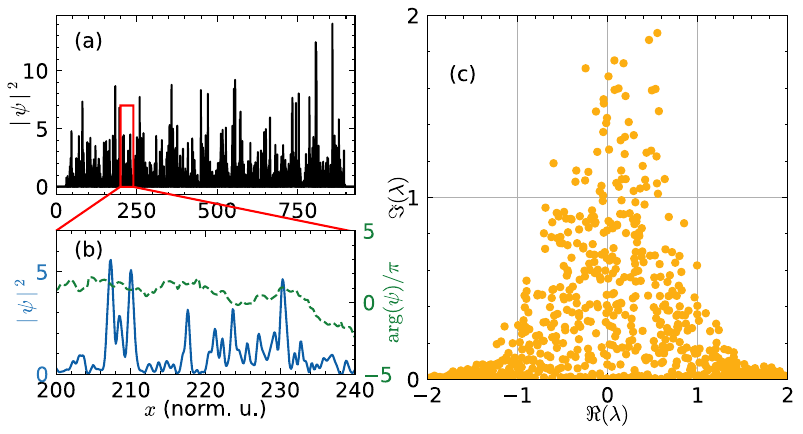}
    \caption{\textbf{DOS1 measurement for a single experimental realization at $t=t_0$.}
\ch{{\bf (a)} Example of a full experimental sample with spatial width $L=1000$ in dimensionless units, corresponding to $\SI{57}{ns}$ in physical units. The field is measured at $t=t_0$, and the initial Fourier spectral width is \(\Delta k = 1.96\).
{\bf (b)} Zoom of the same sample showing the reconstructed intensity (blue) and phase (green). The temporal width of the zoomed window is $\SI{2.15}{ns}$.
}{\bf (c)} IST spectrum extracted from the measured field. The IST is computed over the full temporal window shown in panel~{\bf (a)}.}
    \label{fig:field_IST}
\end{figure}

More precisely, the number of solitons having an eigenvalue $\lambda$ such that $\Re(\lambda)\in [\Re(\lambda),\Re(\lambda)+\Delta \Re(\lambda)]$ and $\Im(\lambda)\in [\Im(\lambda),\Im(\lambda)+\Delta \Im(\lambda)]$ at time $t$ in the space window $[x, x+\dd x]$ is $\dd N_S(x, t) = \rho(x, t;\lambda) \, \dd\Re(\lambda)\,\dd\Im(\lambda) \dd x$. Integrating over \(x\) and using statistical homogeneity, the DOS computed from a single realization is
\[
\rho(\lambda)=\frac{N_{S}(\lambda)}{\Delta\Re(\lambda)\,\Delta\Im(\lambda)\,L}\,,
\]
where \(L\) is the window length and \(N_{S}(\lambda)\) is the number of eigenvalues falling in the bin
\([\Re(\lambda),\,\Re(\lambda)+\Delta\Re(\lambda)]\times[\Im(\lambda),\,\Im(\lambda)+\Delta\Im(\lambda)]\). \ch{See Fig.~\ref{fig:field_IST} for one example of IST spectrum computed from one experimental realization.} 

The final DOS1 is obtained by averaging \(\rho(\lambda)\) over $n$ realizations \ch{(see Fig. 3.a of the main Letter)}. Indeed, to reduce array sizes and computation time, each window on which the IST spectrum is computed is limited to 8192 points. More precisely, from the $n_{exp}=5$ experimental realizations of \(\SI{1}{\micro s}\) record we select a central \(\SI{798}{ns}\) segment and partition it into $n_w=14$ fields of 8192 points each, of width $\SI{57}{ns}$. Such reduced windows correspond each to a dimensionless width of $L \simeq 1060$ for $P_0= \SI{5.85}{mW}$ and to $L \simeq 738$ for $P_0 = \SI{2.84}{mW}$. Each field is multiplied by a \ch{super}-Gaussian $\exp(-(x/l)^{50})$, enforcing zero-boundary conditions compatible with Fourier collocation. An example of a typical signal (at \(t=0\)) and its IST spectrum is shown in Fig.~\ref{fig:field_IST}. \ps{In the manuscript, the displayed DOS1 is computed from the averaging} over the $n_{exp}\times n_w=70$ windows of observation.\\

\noindent{\bf Computation of the DOS from the experimentally measured Fourier spectrum (DOS2)}\\

\noindent \ch{The second strategy (DOS2) consists in computing the IST spectrum using the Fourier collocation method on signals numerically generated from the Fourier spectrum measured at $t=0$. This Fourier spectrum is obtained by applying an FFT to the complex field measured by heterodyne detection (see above). We then assign random phases uniformly distributed between $0$ and $2\pi$ to the Fourier components. Finally, the field is propagated numerically until $t=t_0$ by integrating the fNLS equation including residual losses of the ROFL. Since the Fourier spectrum at $t=t_0$ is broader than that at $t=0$, this second strategy (DOS2) provides a more accurate determination of the DOS. It therefore appears that the DOS computed using this second strategy is less sensitive to the asymmetric transfer function of our detection apparatus.}

\ch{In the DOS2 procedure, the random phases do not aim at reproducing the phases of a specific experimental realization, but at sampling the statistical ensemble of initial PCWs generated by the source; the phase correlations at later times are then generated by the nonlinear fNLSE evolution.}

\begin{figure}[h!]
    \centering
    \includegraphics[width=0.8\linewidth]{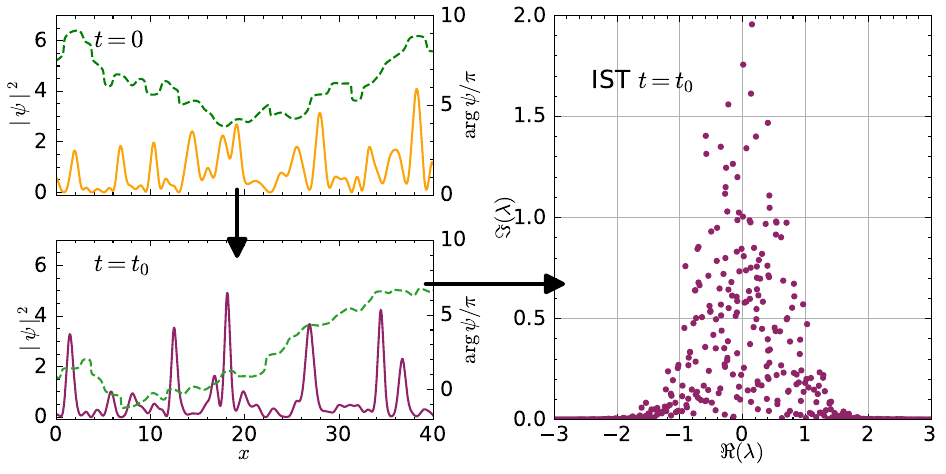}
    \caption{\ch{\textbf{DOS2 measurement for a single realization.}
Numerical realizations of partially coherent waves are generated from the experimentally measured Fourier spectrum at $t=0$ (see Fig.~\ref{fig:spectrum}), assuming random phases for the Fourier components.
{\bf Top left panel:} Zoom of one realization at $t=0$; the total sample length is $L=400$. Orange solid line: intensity. Dashed green line: phase.
{\bf Bottom left panel:} Field after propagation up to $t=t_0$, obtained by numerical integration of the fNLS equation including the residual losses of the ROFL. Purple solid line: intensity. Dashed green line: phase.
{\bf Right panel:} IST spectrum of the realization computed at $t=t_0$ over the full temporal window.}}
    \label{fig:num_IST}
\end{figure}

More precisely, we use the compressed-exponential fit (i.e. $n(\nu)=\nu_0\exp[-\ln{2}\abs{\nu/\Delta \nu}^p]$ where $p$, $n_0$ and $\Delta \nu$ are fitted parameters) of the spectrum reported above and generate many realizations of partially coherent waves by sampling independent random Fourier phases. The IST spectrum of each realization is then computed (see examples in Fig.~\ref{fig:num_IST}) and averaged (see Fig 3.b of the main Letter). We use \(n=1000\) realizations of length \(L=400\) to obtain the numerical DOS2.\\

\ch{In the manuscript,} we compare the DOS obtained with the first strategy (DOS1: direct IST on the experimental field) and the second strategy (DOS2: numerical realizations generated from the measured Fourier spectrum) in \ch{Fig.3 of the main Letter}. DOS1 and DOS2 are broadly consistent; in particular, the weighted marginals \(P_{\Re}\) and \(P_{\Im}\) nearly coincide. However, DOS1 is not perfectly even in \(\Re(\lambda)\) because the detection transfer function is not flat around \(\nu \sim 40~\mathrm{GHz}\) (see Fig.~\ref{fig:SM_transfer_function}). Moreover, since the second strategy permits generating many more samples, DOS2 is smoother than DOS1 and allows for a finer binning. Our extensive analysis shows that using DOS2 improves the agreement between the experimental correlations and the theoretical predictions. 

Note that small residual losses are present at short times in the ROFL. In the strongly nonlinear case (\(\Delta k=1.96\)), losses are below \(2\%\) and can be neglected. In the weaker nonlinearity (\(\Delta k=2.81\)), about \(8\%\) losses are observed during the transient of Raman amplification (\(t<1/2\)) before the system stabilizes into a lossless regime. These low losses slightly affect the measured DOS. To mitigate their impact on DOS2 in the weakly nonlinear case, we propagate the measured initial field from \(t=0\) to \(t=t_{0}\) using numerical simulations of the fNLS that include residual losses (see~\cite{SMfache2025perturbed}); DOS2 is then computed at time \(t_{0}\).

\section{Thermodynamics and soliton gases in the focusing Non-Linear-Schroedinger}

This section provides a short overview of the key concepts at the roots of the analytical expression for the two-point correlators, Eq. (1) in the main text. To this end, we first review the methodology that in quantum integrability is referred to as thermodynamic Bethe ansatz (TBA) \cite{SMtakahashi2005} which allows to connect the notion of a GGE and the DOS. Then, we merge these concepts with hydrodynamics and present a short derivation of Eq. (5) based on Ref. \cite{SMDoyon2018}.

\subsection{Thermodynamic Bethe ansatz}
\label{subsec_TBA}

In integrable models, thermodynamics can be computed by treating the excitations --i.e. the solitons in the fNLS-- as if they were extended particles with an effective length given by the scattering shift, but otherwise non-interacting. This program in quantum integrability is fulfilled within TBA, which has been recently carried to classical integrable PDEs through semiclassical limits of quantum integrability: for the fNLS, this has been carried out in Ref. \cite{SMKoch2022} which we follow here.
Before delving into the details, a short comment on the notation is due: in Ref. \cite{SMKoch2022}, a parametrization of the rapidities and of the DOS more natural from the quantum perspective has been used, which we found it more convenient for the scope to this section. Hence, as a first step we will clarify the notation's dictionary.
The complex spectral parameter used in soliton gases is split into its real and imaginary part as follows
\be\label{eq_rap_dictionary}
\lambda=-\frac{1}{2}\theta+\frac{i}{4}s\, ,
\ee
where $\lambda$ is called the ``rapidity" of the soliton, and $s$ labels solitons of different amplitude.
Hence, the upper complex plane characterizing the spectral parameter of solitons is now parametrized as $\theta\in (-\infty,\infty)$ and $s\in(0,\infty)$. The use of the new variables reflects into a different normalization for the DOS: we use $\bar{\rho}_s(\theta)$ for this new parametrization, the correspondence with the SG's DOS is obtained by keeping into account the Jacobian of the change of variables $ \bar{\rho}_s(\theta)=\frac{1}{8}\rho(\lambda)\Big|_{\lambda=-\frac{1}{2}\theta+\frac{i}{4}s}$. 
Two further remarks are useful:
\begin{itemize}
\item In SG, or equivalently GHD, the DOS is a local object varying in space and time. Section \ref{subsec_TBA} is concerned with homogeneous and time-independent ensembles, therefore we drop the --in this case-- trivial space-time dependency of the DOS, which will be later restored introducing hydrodynamics in Section \ref{subsec_hydrocorr}.
\item We notice that, in principle, the (integrated in space) DOS can be defined for a given field configuration without invoking any notion of averaging. However, since within this section we focus on the statistical properties of GGEs, we always refer to the averaged DOS, equivalently obtained either by averaging over the initial conditions, or on the proper GGE.
\end{itemize}

Integrable models feature infinitely many extensive conserved quantities of the form $\mathcal{Q}_i=\int \dd x\, {\rm q}_i(x)$, with ${\rm q}_i(x)$ being localized around position $x$. For example, in the fNLS, for the intensity one has ${\rm q}_i(x)\to|\psi(x)|^2$, for the momentum ${\rm q}_i(x)\to \frac{1}{2}\left(i \psi\partial_x\psi^*-i\psi^*\partial_x\psi\right)$, and for the energy ${\rm q}_i(x)\to |\partial_x\psi|^2-|\psi|^4$. Other conserved quantities have more intricated dependence on the fields, but they can be systematically generated through the inverse scattering method \cite{SMfaddeev2007hamiltonian}.
In GGEs, the average density of charges is directly connected with the DOS as
\be\label{eq_ch_exp}
\langle {\rm q}_i(x)\rangle=\int_{-\infty}^{\infty} \dd\theta\int_0^\infty\dd s\, h_{i;s}(\theta)\rho_s(\theta)\, ,
\ee
where $h_{i;s}(\theta)$ is called the charge-eigenvalue associated to the conserved quantities. For the intensity one has $h_{i;s}(\theta)\to n_s\equiv s$, for the momentum one has $h_{i;s}(\theta)\to p_s(\theta)\equiv s\theta$, and for the energy $h_{i;s}(\theta)\to \epsilon_s(\theta)\equiv s\theta^2-\frac{1}{12}s^3$. It should be noticed that these charge eigenvalues are indeed the intensity, momentum, and energy carried by a soliton through the correspondence \eqref{eq_rap_dictionary}.
A key quantity in SG is the scattering shift between solitons: it is convenient to introduce a scattering kernel $\bar{\varphi}_{s,s'}(\theta-\theta')$ defined as
\be\label{eq_ghd_kernel}
\bar{\varphi}_{s,s'}(\theta-\theta')=2\log\frac{(s+s')^2+4(\theta-\theta')^2}{(s-s')^2+4(\theta-\theta')^2}
\ee
Notice the immediate correspondence between the DOS scattering kernel defined in the main text and Eq. \eqref{eq_ghd_kernel} as $\varphi(\lambda,\mu)=\partial_\theta p_s(\theta)\, \bar{\varphi}_{s,s'}(\theta-\theta')\Big|^{\lambda=-\frac{1}{2}\theta+\frac{i}{4}s}_{\mu=-\frac{1}{2}\theta'+\frac{i}{4}s'}$. Indeed, in TBA and GHD, the connection between the scattering kernel and the scattering shift is generally obtained by multiplying for the momentum derivative \cite{SMzakharov1971,SMBertini2016,SMAlvaredo2016}.
It is also useful to introduce the  filling function $\vartheta_s(\theta)$ and the dressing operation.
The filling function is defined as
\be\label{eq_ghd_fill}
\frac{1}{\vartheta_s(\theta)}\bar{\rho}_s(\theta)=\frac{\partial_\theta p_s(\theta)}{2\pi}-\int_{-\infty}^{+\infty}\dd \theta' \int_0^{+\infty}\dd s' \bar{\varphi}_{s,s'}(\theta-\theta')\bar{\rho}_{s'}(\theta')
\ee

Then, for any given test function $\tau_s(\theta)$, one defines the dressing $\tau_s(\lambda)\to \tau_s^\dr(\lambda)$ through the solution of the linear integral equation
\be\label{eq_ghd_dr}
\tau^\dr_s(\theta)=\tau_s(\theta)-\int_{-\infty}^{+\infty}\dd \theta' \int_0^{+\infty}\dd s'\, \frac{1}{2\pi} \bar{\varphi}_{s,s'}(\theta-\theta')\vartheta_{s'}(\theta')\tau^\dr_{s'}(\theta')\, .
\ee
Notice that, by direct comparison of Eqs. \eqref{eq_ghd_fill} and \eqref{eq_ghd_dr}, one has $\rho_s(\theta)=\frac{1}{2\pi}(\partial_\theta p_s)^\dr \bar{\rho}_s(\theta)$.
A less trivial, but useful, identity relates the effective velocity as in Eq. (4) to the dressing operation. For notation consistency, we define $\bar{v}^\eff_s(\theta)\equiv v^\eff(\lambda)\Big|_{\lambda=-\frac{1}{2}\theta+\frac{i}{4}s}$, and by manipulating the integral equations defining the dressing operation (see Ref. \cite{SMAlvaredo2016}) one can show $\bar{v}^\eff_s(\theta)=\tfrac{(\partial_\theta \epsilon_s(\theta))^\dr}{(\partial_\theta p_s(\theta))^\dr}$.
We can now finally focus on the GGE. To this end, it is usual to formally define the GGE partition function as a path integral of the GGE's measure over possible field configurations $\mathcal{Z}\equiv \int \mathcal{D}\psi\, e^{-\sum_i \beta_i \mathcal{Q}_i[\psi]}$. Path-integrals are notoriously hard to be rigorously defined and we will not delve into this difficult matter, treating the definition of $\mathcal{Z}$ as a formal expression. Notice that  $\mathcal{Z}$ is well-defined in integrable models on a lattice.
Of particular interest is the free energy $\mathcal{F}\equiv -\log\mathcal{Z}$: within the TBA's framework \cite{SMtakahashi2005}, the free energy is written in terms of the DOS as 
\be\label{eq_free_energy}
\mathcal{F}=L\int_{-\infty}^{+\infty}\dd\theta\int_0^{+\infty}\dd s\,\left\{\sum_i \beta_i \, h_{i;s}(\theta)\bar{\rho}_s(\theta)-\frac{(\partial_\theta p_s(\theta))^\dr}{2\pi} S_s(\vartheta_s(\theta))\right\}\, .
\ee
Above, we introduced the system's size $L$ for extensivity, and we readily recognize the sum of the average local charges weigthed with the generalized temperatures $\langle\sum_i \beta_i \mathcal{Q}_i \rangle=L\int_{-\infty}^{+\infty}\dd\theta\int_0^{+\infty}\dd s\,\sum_i \beta_i \, h_{i;s}(\theta)\bar{\rho}_s(\theta)$. The second term in Eq. \eqref{eq_free_energy} is an entropic term and it accounts for the fact that there are several physical inequivalent soliton's configurations resulting in the same DOS. For example, one could consider two different ensambles of solitons with equal rapidities, but different soliton's positions, which are clearly two different field's configurations. The entropy accounts for such a degeneracy: the function $S_s(X)$ will be specified later, for the time being it is useful to proceed in full generality.

The DOS associated with a given GGE is obtained by asking $\bar{\rho}_s(\theta)$ to \emph{minimize the free energy}. Looking for the saddle point $\tfrac{\delta \mathcal{F}}{\delta \bar{\rho}_s(\theta)}=0$ (which it can be shown to be unique, and a minimum \cite{SMtakahashi2005}), one reaches the non-linear integral equations
\be\label{eq_TBA_eq}
S_s'(\vartheta_s(\theta))=\sum_i \beta_{i} h_{i;s}(\theta)- \int_{-\infty}^{+\infty}\dd\theta'\int_0^{+\infty}\dd s'\,\frac{1}{2\pi} \bar{\varphi}_{s,s'}(\theta-\theta') \left[\vartheta_{s'}(\theta') S_{s'}'(\vartheta_{s'}(\theta'))-S_{s'}(\vartheta_{s'}(\theta'))\right]\, ,
\ee
where $S_s'(X)$ is the derivative of $S_s(X)$ with respect to its argument $S_s'(X)\equiv \tfrac{\dd S_s(X)}{\dd X}$. Eq. \eqref{eq_TBA_eq} is often referred to as the \emph{TBA equation}: once one obtains the filling function $\vartheta_s(\theta)$ solving Eq. \eqref{eq_TBA_eq}, through Eq. \eqref{eq_ghd_fill} one reaches the DOS, and finally with Eq. \eqref{eq_ch_exp} one can compute the expectation values of the conserved quantities.
In thermodynamics, it is canonical to retrieve the expectation value of the energy --in our case, a generic conserved quantity-- by deriving the free energy with respect to the (generalized) inverse temperature $\partial_{\beta_i}\mathcal{F}=\langle\mathcal{Q}_i\rangle$: by manipulating the integral equations, it is possible to show the consistency with the simple expression Eq. \eqref{eq_ch_exp}.
As outlined in the main text, through second derivatives one obtains charge-charge connected correlation functions 
\be \label{eq_ddF}-\partial_{\beta_i}\partial_{\beta_j}\mathcal{F}=-\partial_{\beta_i} \langle \mathcal{Q}_j\rangle=\langle \mathcal{Q}_i \mathcal{Q}_j\rangle-\langle \mathcal{Q}_i\rangle \langle\mathcal{Q}_j\rangle
\ee
which results in the expression \cite{SMtakahashi2005}
\be\label{eq_TBA_twopoint}
\langle \mathcal{Q}_i \mathcal{Q}_j\rangle-\langle \mathcal{Q}_i\rangle \langle\mathcal{Q}_j\rangle=L \int_{-\infty}^{+\infty}\dd\theta\int_0^{+\infty}\dd s\, \bar{\rho}_s(\theta) \left[-\frac{1}{\vartheta_s(\theta)S_s''(\vartheta_s(\theta))}\right] h_{i;s}^\dr(\theta) h_{j;s}^\dr(\theta)\, .
\ee
Above, $S_s''(X)\equiv \tfrac{\dd^2 S}{\dd X^2}$. We will present a short derivation of this expression in Section \ref{subsec_hydrocorr}, by considering the more general case of two-time correlation functions directly.
Notice that, even though Eq. \eqref{eq_TBA_eq} is used to compute the $\beta_i-$derivatives leading to Eq. \eqref{eq_TBA_twopoint}, the final result is entirely expressed in terms of the DOS, the filling fraction and the dressing operation. Therefore, in practice, there is no need to determine the $\beta_i$ and solve the difficult non-linear TBA equations \eqref{eq_TBA_eq}, and one has to solve the much easier dressing equation \eqref{eq_ghd_dr}.

To be able to evaluate the above expressions in practice, one has first to assign the right entropy weight $S_s(X)$. A plausible ansatz would be using a Maxwell-Boltzmann distribution with entropy $S_s(X)\stackrel{?}{=}S_\text{MB}(X)\equiv X(1-\log X)$, but it turns out to be not correct.

Recently, this problem has been solved through semiclassical limits of quantum integrability. In quantum integrable models, there are no ambiguities in which excitations should be included in the thermodynamics and their entropy contribution \cite{SMtakahashi2005}. Therefore, the strategy is first to consider thermodynamics of the quantum version of the NLS (known as Lieb-Liniger model), and retrieve thermodynamics by taking a proper semiclassical limit. For the fNLS this has been done in Ref. \cite{SMKoch2022}.
The final result is that \emph{GGEs can be fully described by solitons} without having to explicitly account for radiation, but \emph{solitons have unconventional entropy}
\be\label{eq_entropy}
S_s(X)= X(1-\log X)-X \log s^2-s^{-2}\, .
\ee
Which is similar to the Maxwell-Boltzmann entropy, with the addition of two terms.
It is possible to show that radiative modes can be understood as a collective effect of solitons of small amplitude and large width.
It should be noted that in the original reference on the fNLS \cite{SMKoch2022} the explicit expression for $S_s(X)$ is omitted and only the TBA equations \eqref{eq_TBA_eq} are reported, obtained from the semiclassical limit of the quantum Lieb-Liniger model. From that equations, the entropy $S_s(x)$ can be identified by direct comparison with the general form Eq. \eqref{eq_TBA_eq}, and leading to Eq. \eqref{eq_entropy}.
Notice that, as long as only the charge-charge correlator is considered, using the correct entropy Eq. \eqref{eq_entropy} of the Maxwell-Boltzmann (incorrect) ansatz $S_\text{SM}$ is equivalent, as in Eq. \eqref{eq_TBA_twopoint} $S_s(X)$ contributes only through its second derivative and $\tfrac{\dd^2 S_\text{MB}(X)}{\dd X^2}=\tfrac{\dd^2 S_s(X)}{\dd X^2}$.

Before moving to hydrodynamics, we quickly mention the analytical formula for the kurtosis, which is an important quantity in determining the emergenge of a GGE. Indeed, the conserved charges are time-independent by definition, and therefore --even though they are easy to compute, see Eq. \eqref{eq_ch_exp}-- they are not a good indicator to see relaxation to a GGE.
Instead, as we discuss in the main text, the kurtosis undergoes non-trivial evolution in our nonequilibrium protocol before reaching a stationary value described by a GGE.
On the GGE, the kurtosis can be computed as \cite{SMKoch2022}
\be\label{eq_kurt_Alvise}
\langle |\psi|^4\rangle=\int_{-\infty}^{+\infty}\dd\theta\int_0^{+\infty}\dd s\,\left[-\bar{\rho}_s(\theta)\frac{1}{6}s^3+\frac{2 s\theta}{2\pi} \vartheta_s(\theta) f_s^\dr(\theta)\right]\, ,
\ee
where $f_s(\theta)=\int \dd s' \int \dd\theta'\, (\theta'-\theta)\varphi_{s,s'}(\theta-\theta')\bar{\rho}_{s'}(\theta')$, \ch{and then $f^\dr_s(\theta)$ is computed accordingly to Eq.~\eqref{eq_ghd_dr}.}
Equation~\eqref{eq_kurt_Alvise} has been derived in Ref. \cite{SMKoch2022} from the semiclassical limit of available results available in the quantum Lieb-Liniger.
Shortly after, another expression for the kurtosis has been found within purely classical methods within the soliton gas picture \cite{SMCongy2024}
\be\label{eq_kurt_Thib}
\langle |\psi|^4\rangle=- 16 \operatorname{Im}\left[\frac{2}{3} \overline{\lambda^3}+\frac{1}{4} \overline{\lambda^2 v^\text{eff}(\lambda)}\right],\;\text{with} \, ,
\ee
\ch{where, for any function $h(\lambda)$, we define $\overline{h(\lambda)}=\int \dd\lambda\,h(\lambda) \rho(\lambda)$.}
The two expressions are indeed equivalent, as it will be discussed in a forthcoming work.

\subsection{Hydrodynamic correlation functions}
\label{subsec_hydrocorr}

Armed with the methodology presented in Section \ref{subsec_TBA}, we now merge the ability of GGEs to capture charge-charge equal-time correlations \eqref{eq_TBA_twopoint} with the kinetic, or hydrodynamic, equation for soliton gases, and eventually access the two-time correlators.
This methodology is framed within \emph{hydrodynamic projections} \cite{SMspohn2012large}, applied to integrable systems \cite{SMDoyon2018,SMDoyon2022}: the idea is to regard $\langle \qq(t,x)\qq(t=0,x=0)\rangle$ as the evolution of the average of $\qq(t,x)$ from an inhomogeneous initial state, obtained by locally perturbing the GGE with a small shift of the associated $\beta_i\to \beta_i +\delta\beta_i(x)$.
In this Section, the DOS becomes again a space-time dependent quantity $\bar{\rho}_s(\theta)\to \bar{\rho}_{t,x;s}(\theta)$: we can likewise associate to the DOS a space-time dependent filling fraction $\vartheta_s(\theta)\to \vartheta_{t,x;s}(\theta)$ and, for the initial conditions of time evolution, a weakly inhomogeneous GGE as outlined above.
Mirroring the computation of equal-time charge-charge correlations from deriving the average charges in $\beta_i$ as in Eq. \eqref{eq_ddF}, we can use
\be
\langle {\rm q}_i(t,x){\rm q}_j(0,0)\rangle-\langle {\rm q}_i(t,x)\rangle\langle {\rm q}_j(0,0)\rangle=-\frac{\delta}{\delta \beta_i(0)} \langle {\rm q}_j(t,x)\rangle\, ,
\ee
where $\langle {\rm q}_j(t,x)\rangle=\int_{-\infty}^{+\infty}\dd\theta\int_0^{+\infty}\dd s\, h_{j;s}(\theta)\rho_{t,x;s}(
\theta)$. 
To take the derivative, we consider a small deformation of the DOS on a homogeneous background $\rho_{t,x;s}\to\rho_{s}+\delta \rho_{t,x;s}$, obtaining $\delta\langle {\rm q}_j(t,x)\rangle=\int_{-\infty}^{+\infty}\dd\theta\int_0^{+\infty}\dd s\, h_{j;s}(\theta)\delta\rho_{t,x;s}(
\theta)$.
It is convenient using a parametrization in terms of the filling fraction. We rewrite $\rho_{t,x;s}(\theta)=\frac{1}{2\pi}(\partial_\theta p_s(\theta))^\dr \vartheta_{t,x;s}(\theta)$, and by using Eq. \eqref{eq_ghd_dr} we can readily find
\be
\delta \rho_{t;x,s}+\vartheta_{s}(
\theta)\int_{-\infty}^{+\infty}\dd \theta' \int_0^{+\infty}\dd s' \bar{\varphi}_{s,s'}(\theta-\theta')\delta \rho_{t;x,s}(\theta')=\frac{(\partial_\theta p_s)^\dr}{2\pi}\delta \vartheta_{t,x;s}(\theta)\, .
\ee
Plugging this equation in the expression for $\delta \langle {\rm q}_j(t,x)\rangle$ and noticing a dressing operation, we reach the compact expression
\be\label{eq_S15}
\delta\langle {\rm q}_j(t,x)\rangle=\int_{-\infty}^{+\infty}\dd\theta\int_0^{+\infty}\dd s\, h^\dr_{j;s}(\theta)\frac{(\partial_\theta p_s)^\dr}{2\pi}\delta \vartheta_{t,x;s}(\theta)\, .
\ee
As a next step, we connect $\delta \vartheta_{t,s;x}(\theta)$ with the inhomogeneous initial conditions by using the kinetic equation for soliton gases $\partial_t \bar{\rho}_{t,x;s}(\theta)+\partial_x[\bar{v}^\eff_{t,x;s}(\theta)\bar{\rho}_{t,x;s}(\theta)]=0$ which can be rewritten in terms of the filling fraction as $\partial_t \vartheta_{t,x;s}(\theta)+\bar{v}^\eff_{t,x;s}(\theta)\partial_x\vartheta_{t,x;s}(\theta)=0$ \cite{SMAlvaredo2016,SMBertini2016}. Expanding the latter for small inhomogeneities of the filling function, we get the simple linear equation $\partial_t \delta\vartheta_{t,x;s}(\theta)+\bar{v}^\eff_{s}(\theta)\partial_x\delta \vartheta_{t,x;s}(\theta)=0$, where the effective velocity is computed on the homogeneous and stationary background. These linearized equations can be easily solved
\be\label{eq_S16}
\delta \vartheta_{t,x;s}(\theta)=\delta \vartheta_{0,x-t v^\eff_s(\theta);s}(\theta)=\int \dd y\, \delta(x-t\bar{v}_s^\eff(\theta)-y)\delta\vartheta_{0,y;s}\, ,
\ee
where $\delta(x-t\bar{v}_s^\eff(\theta)-y)$ is a Dirac $\delta-$function. As a final step, one has to link the small perturbation of the initial filling function $\delta\vartheta_{0,y;s}$ with the perturbation in the generalized inverse temperature. To do so, one perturbs Eq. \eqref{eq_TBA_eq} finding

\be
S_s''(\vartheta_{s}(\theta))\delta\vartheta_{0,y;s}(\theta)=\sum_i \delta\beta_{i}(y) h_{i;s}(\theta)- \int_{-\infty}^{+\infty}\dd\theta'\int_0^{+\infty}\dd s'\,\frac{1}{2\pi} \bar{\varphi}_{s,s'}(\theta-\theta')\vartheta_{s'}(\theta') S_{s'}''(\vartheta_s'(\theta')) \delta \vartheta_{0,y;s'}(\theta)\, .
\ee
By direct comparison with the dressing equation Eq. \eqref{eq_ghd_dr}, one recognizes $S_s''(\vartheta_{s}(\theta))\delta\vartheta_{0,y;s}(\theta)=\sum_i \delta \beta_i(y) h^\dr_{i;s}(\theta)$. Plugging this last identity in Eq. \eqref{eq_S16}, and then in Eq. \eqref{eq_S15}, we finally reach the sought result
\be\label{eq_ch_ch_ghd}
\langle {\rm q}_i(t,x){\rm q}_j(0,0)\rangle-\langle {\rm q}_i(t,x)\rangle\langle {\rm q}_j(0,0)\rangle= \int_{-\infty}^{+\infty}\dd\theta\int_0^{+\infty}\dd s\, \bar{\rho}_s(\theta) \left[-\frac{1}{\vartheta_s(\theta)S_s''(\vartheta_s(\theta))}\right] h_{i;s}^\dr(\theta) h_{j;s}^\dr(\theta)\delta(x-t \bar{v}^\eff_s(
\theta))\, .
\ee

By rewriting this final result in the conventional notation of soliton gases, and specializing it to the correlations of the intensity $q_i(t,x)\to |\psi(t,x)|^2$, one reaches Eq. (5) of the main text. Notice that, if one integrates Eq. \eqref{eq_ch_ch_ghd} over $x$, Eq. \eqref{eq_TBA_twopoint} (a part from an overall $L$ factor coming from translational invariance) is recovered, regardless the value of $t$. Indeed, this must be the case since the total charges $\mathcal{Q}_i$ are integral of motion.

\subsection{Numerical computation of the DOS and of the hydrodynamic correlation functions}
\label{subsec_num_TBA}

\begin{figure*}[h!]
\includegraphics[width=0.99\textwidth]{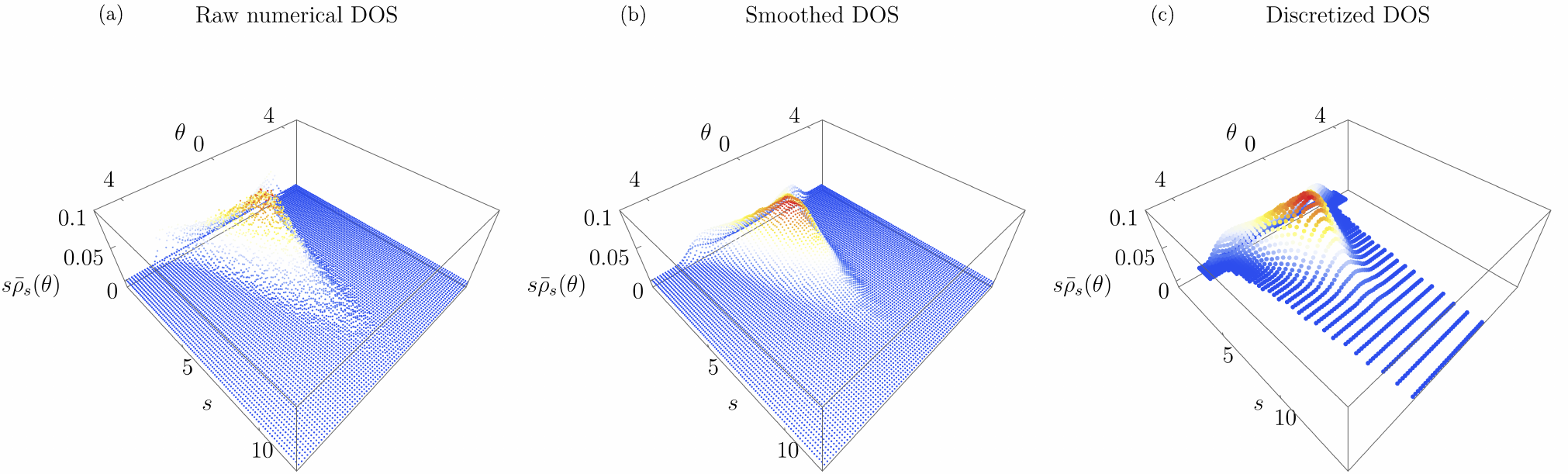}
\label{Fig_S_DOS}
\caption{\textbf{Numerical DOS and its discretization.} Panel (a): numerical DOS extracted from the Zakharov-Shabbat problem, we plot $s\bar{\rho}_s(\theta)$. We simulate coherent waves associated with the strong-nonlinear regime, the field is discretized on $2^9$ points with lattice $\dd x=0.1$, and we show the histrogram built on $1000$ field configurations.
Panel (b): statistical fluctuatins of the raw DOS are suppressed by a convolution with a narrow gaussian, in this case with a width of $\sim 1.5$ phase-space's bins.
Panel (c): the DOS obtained in panel (b) is interpolated, and from this the discretized DOS used in the integral equations is obtained. We choose a non-monotonous discretization denser at small values of $s$, and truncate regions where the DOS is vanishing. Here, we use about $1800$ points in discretizing the integral equations.}
\end{figure*}

We briefly overview the numerical method used to conveniently discretizing the integral equations presented in the previous sections having the DOS as an input.
 Extra care should be put into discretizing the integral equations presented in Section \ref{subsec_TBA} and \ref{subsec_hydrocorr} as the filling fraction, for every fixed value of $\theta$, develops a singularity for small $s$ as $\lim_{s\to 0}s^2\vartheta_{s}(\theta)\Big|_{\theta \text {fixed}}=1$ \cite{SMKoch2022}. At the same time, for every test function $\tau_s(\theta)$ that vanishes linearly in $s$ for $s\to 0$ (eg. the intensity, momentum and energy of the solitons), the associated dressed function $\tau^\dr_s(
\theta)$ \emph{vanishes quadratically $\sim s^2$} as $s\to 0$ \cite{SMKoch2022}. The small $s$ limits of the filling fraction and the dressing respectively makes the equations difficult to be naively discretized. To this end, we use the same discretization presented in Ref. \cite{SMKoch2022} where the DOS, and the filling functions are first reparametrized to account for their small $s$ behavior, and a non-uniform grid (denser for small $s$) is used. The scattering kernel $\bar{\varphi}_{s,s'}(\theta-\theta')$ approaches a Dirac $\delta$ in the rapidities for $s,s'\to 0$, therefore a n\"aive discretization with a midpoint rule is not optimal, and it is greatly improved by integrating $\bar{\varphi}_{s,s'}(\theta-\theta')$ over the discretization's plaquettes. 
The details of the discretization are carefully discussed in Ref. \cite{SMKoch2022}.
These precautions are enough to solve the TBA equations \eqref{eq_TBA_eq} and from then the dressing equations, and the correlations. However, in the case where the DOS is numerically sampled from the fields rather than obtained from Eq. \eqref{eq_TBA_eq}, one should additionally consider that the statistical noise in the histrogram of the DOS is a further challenge.
This problem is solved in two steps: \emph{i)} reparametrize the DOS such that it is a function as smooth as possible, and \emph{ii)} smoothen the DOS through the convolution with a narrow gaussian, in such a way fluctuations are smoothed on a length scale larger than the bins' size of the histrogram. At the end, we verify the computed correlation functions are stable upon changing the smoothening of point \emph{ii)}.
To address point \emph{i)}, it is useful to know the asymptotic behavior of $\bar{\rho}_s(
\theta)$ at large $\theta$. At large rapidities, the support of $\bar{\rho}_s(\theta)$ focuses on $s$ nearby $0$: from the asymptotic analysis of the TBA equations, one can obtain the following asymptotic behavior
\be\label{eq_rho_asymp}
\bar{\rho}_s(\theta)\simeq A_\theta \frac{(A_\theta s)^2}{4 \sinh^2(A_\theta s/2)}\frac{(s A_\theta)\coth(sA_\theta/2)-2}{(sA_\theta)^2}\, .
\ee
where $A_\theta^{-1}=\int_0^\infty \dd s\, s\bar{\rho}_s(\theta)$.
From Eq. \eqref{eq_rho_asymp}, we notice that $\rho_{s=0}(
\theta)\sim A_\theta$: at large rapidities, $A_\theta$ is expected to grow and thus $\rho_s(\theta)$ grows as well, as it is shown in the numerical data of Fig. \ref{Fig_S_DOS}.
In contrast, by plotting the histogram of $s\bar{\rho}_s(\theta)$, we see the its support is squeezed at smal values of $s$ at the tails, but it does not diverge. As a last step, we smoothen $s\bar{\rho}_s(\theta)$ as anticipated at point \emph{ii)}: the smoothed function is then discretized and used as an input for the integral equations ultimately leading to Eq. \eqref{eq_ch_ch_ghd}.


\begin{thebibliography}{80}%
\makeatletter
\providecommand \@ifxundefined [1]{%
 \@ifx{#1\undefined}
}%
\providecommand \@ifnum [1]{%
 \ifnum #1\expandafter \@firstoftwo
 \else \expandafter \@secondoftwo
 \fi
}%
\providecommand \@ifx [1]{%
 \ifx #1\expandafter \@firstoftwo
 \else \expandafter \@secondoftwo
 \fi
}%
\providecommand \natexlab [1]{#1}%
\providecommand \enquote  [1]{``#1''}%
\providecommand \bibnamefont  [1]{#1}%
\providecommand \bibfnamefont [1]{#1}%
\providecommand \citenamefont [1]{#1}%
\providecommand \href@noop [0]{\@secondoftwo}%
\providecommand \href [0]{\begingroup \@sanitize@url \@href}%
\providecommand \@href[1]{\@@startlink{#1}\@@href}%
\providecommand \@@href[1]{\endgroup#1\@@endlink}%
\providecommand \@sanitize@url [0]{\catcode `\\12\catcode `\$12\catcode
  `\&12\catcode `\#12\catcode `\^12\catcode `\_12\catcode `\%12\relax}%
\providecommand \@@startlink[1]{}%
\providecommand \@@endlink[0]{}%
\providecommand \url  [0]{\begingroup\@sanitize@url \@url }%
\providecommand \@url [1]{\endgroup\@href {#1}{\urlprefix }}%
\providecommand \urlprefix  [0]{URL }%
\providecommand \Eprint [0]{\href }%
\providecommand \doibase [0]{https://doi.org/}%
\providecommand \selectlanguage [0]{\@gobble}%
\providecommand \bibinfo  [0]{\@secondoftwo}%
\providecommand \bibfield  [0]{\@secondoftwo}%
\providecommand \translation [1]{[#1]}%
\providecommand \BibitemOpen [0]{}%
\providecommand \bibitemStop [0]{}%
\providecommand \bibitemNoStop [0]{.\EOS\space}%
\providecommand \EOS [0]{\spacefactor3000\relax}%
\providecommand \BibitemShut  [1]{\csname bibitem#1\endcsname}%
\let\auto@bib@innerbib\@empty
\bibitem [{\citenamefont {Rickayzen}(2013)}]{rickayzen2013}%
  \BibitemOpen
  \bibfield  {author} {\bibinfo {author} {\bibfnamefont {G.}~\bibnamefont
  {Rickayzen}},\ }\href@noop {} {\emph {\bibinfo {title} {Green's functions and
  condensed matter}}}\ (\bibinfo  {publisher} {Courier Corporation},\ \bibinfo
  {year} {2013})\BibitemShut {NoStop}%
\bibitem [{\citenamefont {Schwinger}(1951)}]{Schwinger1951}%
  \BibitemOpen
  \bibfield  {author} {\bibinfo {author} {\bibfnamefont {J.}~\bibnamefont
  {Schwinger}},\ }\bibfield  {title} {\bibinfo {title} {On the green’s
  functions of quantized fields. i},\ }\href
  {https://doi.org/10.1073/pnas.37.7.452} {\bibfield  {journal} {\bibinfo
  {journal} {Proceedings of the National Academy of Sciences}\ }\textbf
  {\bibinfo {volume} {37}},\ \bibinfo {pages} {452} (\bibinfo {year} {1951})},\
  \Eprint
  {https://arxiv.org/abs/https://www.pnas.org/doi/pdf/10.1073/pnas.37.7.452}
  {https://www.pnas.org/doi/pdf/10.1073/pnas.37.7.452} \BibitemShut {NoStop}%
\bibitem [{\citenamefont {Spohn}(2012)}]{spohn2012large}%
  \BibitemOpen
  \bibfield  {author} {\bibinfo {author} {\bibfnamefont {H.}~\bibnamefont
  {Spohn}},\ }\href@noop {} {\emph {\bibinfo {title} {Large scale dynamics of
  interacting particles}}}\ (\bibinfo  {publisher} {Springer Science \&
  Business Media},\ \bibinfo {year} {2012})\BibitemShut {NoStop}%
\bibitem [{\citenamefont {Giamarchi}(2003)}]{giamarchi2003}%
  \BibitemOpen
  \bibfield  {author} {\bibinfo {author} {\bibfnamefont {T.}~\bibnamefont
  {Giamarchi}},\ }\href@noop {} {\emph {\bibinfo {title} {Quantum physics in
  one dimension}}},\ Vol.\ \bibinfo {volume} {121}\ (\bibinfo  {publisher}
  {Clarendon press},\ \bibinfo {year} {2003})\BibitemShut {NoStop}%
\bibitem [{\citenamefont {Kadanoff}\ and\ \citenamefont
  {Martin}(1963)}]{Kadanoff1963}%
  \BibitemOpen
  \bibfield  {author} {\bibinfo {author} {\bibfnamefont {L.~P.}\ \bibnamefont
  {Kadanoff}}\ and\ \bibinfo {author} {\bibfnamefont {P.~C.}\ \bibnamefont
  {Martin}},\ }\bibfield  {title} {\bibinfo {title} {Hydrodynamic equations and
  correlation functions},\ }\href
  {https://doi.org/https://doi.org/10.1016/0003-4916(63)90078-2} {\bibfield
  {journal} {\bibinfo  {journal} {Annals of Physics}\ }\textbf {\bibinfo
  {volume} {24}},\ \bibinfo {pages} {419} (\bibinfo {year} {1963})}\BibitemShut
  {NoStop}%
\bibitem [{\citenamefont {Spohn}(2014)}]{Spohn2014}%
  \BibitemOpen
  \bibfield  {author} {\bibinfo {author} {\bibfnamefont {H.}~\bibnamefont
  {Spohn}},\ }\bibfield  {title} {\bibinfo {title} {Nonlinear fluctuating
  hydrodynamics for anharmonic chains},\ }\href
  {https://doi.org/10.1007/s10955-014-0933-y} {\bibfield  {journal} {\bibinfo
  {journal} {Journal of Statistical Physics}\ }\textbf {\bibinfo {volume}
  {154}},\ \bibinfo {pages} {1191} (\bibinfo {year} {2014})}\BibitemShut
  {NoStop}%
\bibitem [{\citenamefont {Bastianello}\ \emph {et~al.}(2022)\citenamefont
  {Bastianello}, \citenamefont {Bertini}, \citenamefont {Doyon},\ and\
  \citenamefont {Vasseur}}]{Bastianello2022}%
  \BibitemOpen
  \bibfield  {author} {\bibinfo {author} {\bibfnamefont {A.}~\bibnamefont
  {Bastianello}}, \bibinfo {author} {\bibfnamefont {B.}~\bibnamefont
  {Bertini}}, \bibinfo {author} {\bibfnamefont {B.}~\bibnamefont {Doyon}},\
  and\ \bibinfo {author} {\bibfnamefont {R.}~\bibnamefont {Vasseur}},\
  }\bibfield  {title} {\bibinfo {title} {Introduction to the special issue on
  emergent hydrodynamics in integrable many-body systems},\ }\href
  {https://doi.org/10.1088/1742-5468/ac3e6a} {\bibfield  {journal} {\bibinfo
  {journal} {Journal of Statistical Mechanics: Theory and Experiment}\ }\textbf
  {\bibinfo {volume} {2022}},\ \bibinfo {pages} {014001} (\bibinfo {year}
  {2022})}\BibitemShut {NoStop}%
\bibitem [{\citenamefont {Bertini}\ \emph {et~al.}(2021)\citenamefont
  {Bertini}, \citenamefont {Heidrich-Meisner}, \citenamefont {Karrasch},
  \citenamefont {Prosen}, \citenamefont {Steinigeweg},\ and\ \citenamefont
  {\ifmmode \check{Z}\else \v{Z}\fi{}nidari\ifmmode~\check{c}\else
  \v{c}\fi{}}}]{Bertini2021}%
  \BibitemOpen
  \bibfield  {author} {\bibinfo {author} {\bibfnamefont {B.}~\bibnamefont
  {Bertini}}, \bibinfo {author} {\bibfnamefont {F.}~\bibnamefont
  {Heidrich-Meisner}}, \bibinfo {author} {\bibfnamefont {C.}~\bibnamefont
  {Karrasch}}, \bibinfo {author} {\bibfnamefont {T.}~\bibnamefont {Prosen}},
  \bibinfo {author} {\bibfnamefont {R.}~\bibnamefont {Steinigeweg}},\ and\
  \bibinfo {author} {\bibfnamefont {M.}~\bibnamefont {\ifmmode \check{Z}\else
  \v{Z}\fi{}nidari\ifmmode~\check{c}\else \v{c}\fi{}}},\ }\bibfield  {title}
  {\bibinfo {title} {Finite-temperature transport in one-dimensional quantum
  lattice models},\ }\href {https://doi.org/10.1103/RevModPhys.93.025003}
  {\bibfield  {journal} {\bibinfo  {journal} {Rev. Mod. Phys.}\ }\textbf
  {\bibinfo {volume} {93}},\ \bibinfo {pages} {025003} (\bibinfo {year}
  {2021})}\BibitemShut {NoStop}%
\bibitem [{\citenamefont {Gromov}\ \emph {et~al.}(2020)\citenamefont {Gromov},
  \citenamefont {Lucas},\ and\ \citenamefont {Nandkishore}}]{Gromov2020}%
  \BibitemOpen
  \bibfield  {author} {\bibinfo {author} {\bibfnamefont {A.}~\bibnamefont
  {Gromov}}, \bibinfo {author} {\bibfnamefont {A.}~\bibnamefont {Lucas}},\ and\
  \bibinfo {author} {\bibfnamefont {R.~M.}\ \bibnamefont {Nandkishore}},\
  }\bibfield  {title} {\bibinfo {title} {Fracton hydrodynamics},\ }\href
  {https://doi.org/10.1103/PhysRevResearch.2.033124} {\bibfield  {journal}
  {\bibinfo  {journal} {Phys. Rev. Res.}\ }\textbf {\bibinfo {volume} {2}},\
  \bibinfo {pages} {033124} (\bibinfo {year} {2020})}\BibitemShut {NoStop}%
\bibitem [{\citenamefont {Feldmeier}\ \emph {et~al.}(2020)\citenamefont
  {Feldmeier}, \citenamefont {Sala}, \citenamefont {De~Tomasi}, \citenamefont
  {Pollmann},\ and\ \citenamefont {Knap}}]{Feldmeier2020}%
  \BibitemOpen
  \bibfield  {author} {\bibinfo {author} {\bibfnamefont {J.}~\bibnamefont
  {Feldmeier}}, \bibinfo {author} {\bibfnamefont {P.}~\bibnamefont {Sala}},
  \bibinfo {author} {\bibfnamefont {G.}~\bibnamefont {De~Tomasi}}, \bibinfo
  {author} {\bibfnamefont {F.}~\bibnamefont {Pollmann}},\ and\ \bibinfo
  {author} {\bibfnamefont {M.}~\bibnamefont {Knap}},\ }\bibfield  {title}
  {\bibinfo {title} {Anomalous diffusion in dipole- and
  higher-moment-conserving systems},\ }\href
  {https://doi.org/10.1103/PhysRevLett.125.245303} {\bibfield  {journal}
  {\bibinfo  {journal} {Phys. Rev. Lett.}\ }\textbf {\bibinfo {volume} {125}},\
  \bibinfo {pages} {245303} (\bibinfo {year} {2020})}\BibitemShut {NoStop}%
\bibitem [{\citenamefont {Guardado-Sanchez}\ \emph {et~al.}(2020)\citenamefont
  {Guardado-Sanchez}, \citenamefont {Morningstar}, \citenamefont {Spar},
  \citenamefont {Brown}, \citenamefont {Huse},\ and\ \citenamefont
  {Bakr}}]{Sanchez2020}%
  \BibitemOpen
  \bibfield  {author} {\bibinfo {author} {\bibfnamefont {E.}~\bibnamefont
  {Guardado-Sanchez}}, \bibinfo {author} {\bibfnamefont {A.}~\bibnamefont
  {Morningstar}}, \bibinfo {author} {\bibfnamefont {B.~M.}\ \bibnamefont
  {Spar}}, \bibinfo {author} {\bibfnamefont {P.~T.}\ \bibnamefont {Brown}},
  \bibinfo {author} {\bibfnamefont {D.~A.}\ \bibnamefont {Huse}},\ and\
  \bibinfo {author} {\bibfnamefont {W.~S.}\ \bibnamefont {Bakr}},\ }\bibfield
  {title} {\bibinfo {title} {Subdiffusion and heat transport in a tilted
  two-dimensional fermi-hubbard system},\ }\href
  {https://doi.org/10.1103/PhysRevX.10.011042} {\bibfield  {journal} {\bibinfo
  {journal} {Phys. Rev. X}\ }\textbf {\bibinfo {volume} {10}},\ \bibinfo
  {pages} {011042} (\bibinfo {year} {2020})}\BibitemShut {NoStop}%
\bibitem [{\citenamefont {Ljubotina}\ \emph {et~al.}(2017)\citenamefont
  {Ljubotina}, \citenamefont {{\v{Z}}nidari{\v{c}}},\ and\ \citenamefont
  {Prosen}}]{Ljubotina2017}%
  \BibitemOpen
  \bibfield  {author} {\bibinfo {author} {\bibfnamefont {M.}~\bibnamefont
  {Ljubotina}}, \bibinfo {author} {\bibfnamefont {M.}~\bibnamefont
  {{\v{Z}}nidari{\v{c}}}},\ and\ \bibinfo {author} {\bibfnamefont
  {T.}~\bibnamefont {Prosen}},\ }\bibfield  {title} {\bibinfo {title} {Spin
  diffusion from an inhomogeneous quench in an integrable system},\ }\href
  {https://doi.org/10.1038/ncomms16117} {\bibfield  {journal} {\bibinfo
  {journal} {Nature Communications}\ }\textbf {\bibinfo {volume} {8}},\
  \bibinfo {pages} {16117} (\bibinfo {year} {2017})}\BibitemShut {NoStop}%
\bibitem [{\citenamefont {Bulchandani}\ \emph {et~al.}(2021)\citenamefont
  {Bulchandani}, \citenamefont {Gopalakrishnan},\ and\ \citenamefont
  {Ilievski}}]{Bulchandani2021}%
  \BibitemOpen
  \bibfield  {author} {\bibinfo {author} {\bibfnamefont {V.~B.}\ \bibnamefont
  {Bulchandani}}, \bibinfo {author} {\bibfnamefont {S.}~\bibnamefont
  {Gopalakrishnan}},\ and\ \bibinfo {author} {\bibfnamefont {E.}~\bibnamefont
  {Ilievski}},\ }\bibfield  {title} {\bibinfo {title} {Superdiffusion in spin
  chains},\ }\href {https://doi.org/10.1088/1742-5468/ac12c7} {\bibfield
  {journal} {\bibinfo  {journal} {Journal of Statistical Mechanics: Theory and
  Experiment}\ }\textbf {\bibinfo {volume} {2021}},\ \bibinfo {pages} {084001}
  (\bibinfo {year} {2021})}\BibitemShut {NoStop}%
\bibitem [{\citenamefont {Schuckert}\ \emph {et~al.}(2020)\citenamefont
  {Schuckert}, \citenamefont {Lovas},\ and\ \citenamefont
  {Knap}}]{Schuckert2020}%
  \BibitemOpen
  \bibfield  {author} {\bibinfo {author} {\bibfnamefont {A.}~\bibnamefont
  {Schuckert}}, \bibinfo {author} {\bibfnamefont {I.}~\bibnamefont {Lovas}},\
  and\ \bibinfo {author} {\bibfnamefont {M.}~\bibnamefont {Knap}},\ }\bibfield
  {title} {\bibinfo {title} {Nonlocal emergent hydrodynamics in a long-range
  quantum spin system},\ }\href {https://doi.org/10.1103/PhysRevB.101.020416}
  {\bibfield  {journal} {\bibinfo  {journal} {Phys. Rev. B}\ }\textbf {\bibinfo
  {volume} {101}},\ \bibinfo {pages} {020416} (\bibinfo {year}
  {2020})}\BibitemShut {NoStop}%
\bibitem [{\citenamefont {Zakharov}(2009)}]{Zakharov2009}%
  \BibitemOpen
  \bibfield  {author} {\bibinfo {author} {\bibfnamefont {V.~E.}\ \bibnamefont
  {Zakharov}},\ }\bibfield  {title} {\bibinfo {title} {Turbulence in integrable
  systems},\ }\href
  {https://doi.org/https://doi.org/10.1111/j.1467-9590.2009.00430.x} {\bibfield
   {journal} {\bibinfo  {journal} {Studies in Applied Mathematics}\ }\textbf
  {\bibinfo {volume} {122}},\ \bibinfo {pages} {219} (\bibinfo {year}
  {2009})},\ \Eprint
  {https://arxiv.org/abs/https://onlinelibrary.wiley.com/doi/pdf/10.1111/j.1467-9590.2009.00430.x}
  {https://onlinelibrary.wiley.com/doi/pdf/10.1111/j.1467-9590.2009.00430.x}
  \BibitemShut {NoStop}%
\bibitem [{\citenamefont {Jaynes}(1957{\natexlab{a}})}]{jaynes1957information}%
  \BibitemOpen
  \bibfield  {author} {\bibinfo {author} {\bibfnamefont {E.~T.}\ \bibnamefont
  {Jaynes}},\ }\bibfield  {title} {\bibinfo {title} {Information theory and
  statistical mechanics},\ }\href@noop {} {\bibfield  {journal} {\bibinfo
  {journal} {Physical review}\ }\textbf {\bibinfo {volume} {106}},\ \bibinfo
  {pages} {620} (\bibinfo {year} {1957}{\natexlab{a}})}\BibitemShut {NoStop}%
\bibitem [{\citenamefont
  {Jaynes}(1957{\natexlab{b}})}]{jaynes1957information2}%
  \BibitemOpen
  \bibfield  {author} {\bibinfo {author} {\bibfnamefont {E.~T.}\ \bibnamefont
  {Jaynes}},\ }\bibfield  {title} {\bibinfo {title} {Information theory and
  statistical mechanics. ii},\ }\href@noop {} {\bibfield  {journal} {\bibinfo
  {journal} {Physical review}\ }\textbf {\bibinfo {volume} {108}},\ \bibinfo
  {pages} {171} (\bibinfo {year} {1957}{\natexlab{b}})}\BibitemShut {NoStop}%
\bibitem [{\citenamefont {Rigol}\ \emph {et~al.}(2007)\citenamefont {Rigol},
  \citenamefont {Dunjko}, \citenamefont {Yurovsky},\ and\ \citenamefont
  {Olshanii}}]{Rigol2007}%
  \BibitemOpen
  \bibfield  {author} {\bibinfo {author} {\bibfnamefont {M.}~\bibnamefont
  {Rigol}}, \bibinfo {author} {\bibfnamefont {V.}~\bibnamefont {Dunjko}},
  \bibinfo {author} {\bibfnamefont {V.}~\bibnamefont {Yurovsky}},\ and\
  \bibinfo {author} {\bibfnamefont {M.}~\bibnamefont {Olshanii}},\ }\bibfield
  {title} {\bibinfo {title} {Relaxation in a completely integrable many-body
  quantum system: An ab initio study of the dynamics of the highly excited
  states of 1d lattice hard-core bosons},\ }\href
  {https://doi.org/10.1103/PhysRevLett.98.050405} {\bibfield  {journal}
  {\bibinfo  {journal} {Phys. Rev. Lett.}\ }\textbf {\bibinfo {volume} {98}},\
  \bibinfo {pages} {050405} (\bibinfo {year} {2007})}\BibitemShut {NoStop}%
\bibitem [{\citenamefont {Calabrese}\ \emph {et~al.}(2016)\citenamefont
  {Calabrese}, \citenamefont {Essler},\ and\ \citenamefont
  {Mussardo}}]{Calabrese2016}%
  \BibitemOpen
  \bibfield  {author} {\bibinfo {author} {\bibfnamefont {P.}~\bibnamefont
  {Calabrese}}, \bibinfo {author} {\bibfnamefont {F.~H.~L.}\ \bibnamefont
  {Essler}},\ and\ \bibinfo {author} {\bibfnamefont {G.}~\bibnamefont
  {Mussardo}},\ }\bibfield  {title} {\bibinfo {title} {Introduction to
  ‘quantum integrability in out of equilibrium systems’},\ }\href
  {https://doi.org/10.1088/1742-5468/2016/06/064001} {\bibfield  {journal}
  {\bibinfo  {journal} {Journal of Statistical Mechanics: Theory and
  Experiment}\ }\textbf {\bibinfo {volume} {2016}},\ \bibinfo {pages} {064001}
  (\bibinfo {year} {2016})}\BibitemShut {NoStop}%
\bibitem [{\citenamefont {De~Luca}\ and\ \citenamefont
  {Mussardo}(2016)}]{DeLuca_2016}%
  \BibitemOpen
  \bibfield  {author} {\bibinfo {author} {\bibfnamefont {A.}~\bibnamefont
  {De~Luca}}\ and\ \bibinfo {author} {\bibfnamefont {G.}~\bibnamefont
  {Mussardo}},\ }\bibfield  {title} {\bibinfo {title} {Equilibration properties
  of classical integrable field theories},\ }\href
  {https://doi.org/10.1088/1742-5468/2016/06/064011} {\bibfield  {journal}
  {\bibinfo  {journal} {Journal of Statistical Mechanics: Theory and
  Experiment}\ }\textbf {\bibinfo {volume} {2016}},\ \bibinfo {pages} {064011}
  (\bibinfo {year} {2016})}\BibitemShut {NoStop}%
\bibitem [{\citenamefont {Bastianello}\ \emph {et~al.}(2018)\citenamefont
  {Bastianello}, \citenamefont {Doyon}, \citenamefont {Watts},\ and\
  \citenamefont {Yoshimura}}]{Bastianello2018}%
  \BibitemOpen
  \bibfield  {author} {\bibinfo {author} {\bibfnamefont {A.}~\bibnamefont
  {Bastianello}}, \bibinfo {author} {\bibfnamefont {B.}~\bibnamefont {Doyon}},
  \bibinfo {author} {\bibfnamefont {G.}~\bibnamefont {Watts}},\ and\ \bibinfo
  {author} {\bibfnamefont {T.}~\bibnamefont {Yoshimura}},\ }\bibfield  {title}
  {\bibinfo {title} {{Generalized hydrodynamics of classical integrable field
  theory: the sinh-Gordon model}},\ }\href
  {https://doi.org/10.21468/SciPostPhys.4.6.045} {\bibfield  {journal}
  {\bibinfo  {journal} {SciPost Phys.}\ }\textbf {\bibinfo {volume} {4}},\
  \bibinfo {pages} {045} (\bibinfo {year} {2018})}\BibitemShut {NoStop}%
\bibitem [{\citenamefont {Vecchio}\ \emph {et~al.}(2020)\citenamefont
  {Vecchio}, \citenamefont {Bastianello}, \citenamefont {Luca},\ and\
  \citenamefont {Mussardo}}]{DelVecchio2020}%
  \BibitemOpen
  \bibfield  {author} {\bibinfo {author} {\bibfnamefont {G.~D. V.~D.}\
  \bibnamefont {Vecchio}}, \bibinfo {author} {\bibfnamefont {A.}~\bibnamefont
  {Bastianello}}, \bibinfo {author} {\bibfnamefont {A.~D.}\ \bibnamefont
  {Luca}},\ and\ \bibinfo {author} {\bibfnamefont {G.}~\bibnamefont
  {Mussardo}},\ }\bibfield  {title} {\bibinfo {title} {{Exact
  out-of-equilibrium steady states in the semiclassical limit of the
  interacting Bose gas}},\ }\href
  {https://doi.org/10.21468/SciPostPhys.9.1.002} {\bibfield  {journal}
  {\bibinfo  {journal} {SciPost Phys.}\ }\textbf {\bibinfo {volume} {9}},\
  \bibinfo {pages} {002} (\bibinfo {year} {2020})}\BibitemShut {NoStop}%
\bibitem [{\citenamefont {Koch}\ \emph {et~al.}(2022)\citenamefont {Koch},
  \citenamefont {Caux},\ and\ \citenamefont {Bastianello}}]{Koch2022}%
  \BibitemOpen
  \bibfield  {author} {\bibinfo {author} {\bibfnamefont {R.}~\bibnamefont
  {Koch}}, \bibinfo {author} {\bibfnamefont {J.-S.}\ \bibnamefont {Caux}},\
  and\ \bibinfo {author} {\bibfnamefont {A.}~\bibnamefont {Bastianello}},\
  }\bibfield  {title} {\bibinfo {title} {Generalized hydrodynamics of the
  attractive non-linear schr\"odinger equation},\ }\href
  {https://doi.org/10.1088/1751-8121/ac53c3} {\bibfield  {journal} {\bibinfo
  {journal} {Journal of Physics A: Mathematical and Theoretical}\ }\textbf
  {\bibinfo {volume} {55}},\ \bibinfo {pages} {134001} (\bibinfo {year}
  {2022})}\BibitemShut {NoStop}%
\bibitem [{\citenamefont {Bonnemain}\ \emph {et~al.}(2022)\citenamefont
  {Bonnemain}, \citenamefont {Doyon},\ and\ \citenamefont
  {El}}]{bonnemain2022generalized}%
  \BibitemOpen
  \bibfield  {author} {\bibinfo {author} {\bibfnamefont {T.}~\bibnamefont
  {Bonnemain}}, \bibinfo {author} {\bibfnamefont {B.}~\bibnamefont {Doyon}},\
  and\ \bibinfo {author} {\bibfnamefont {G.}~\bibnamefont {El}},\ }\bibfield
  {title} {\bibinfo {title} {Generalized hydrodynamics of the kdv soliton
  gas},\ }\href {https://doi.org/10.1088/1751-8121/ac8253} {\bibfield
  {journal} {\bibinfo  {journal} {Journal of Physics A: Mathematical and
  Theoretical}\ }\textbf {\bibinfo {volume} {55}},\ \bibinfo {pages} {374004}
  (\bibinfo {year} {2022})}\BibitemShut {NoStop}%
\bibitem [{\citenamefont {Bonnemain}\ \emph {et~al.}(2025)\citenamefont
  {Bonnemain}, \citenamefont {Biondini}, \citenamefont {Doyon}, \citenamefont
  {Roberti},\ and\ \citenamefont {El}}]{bonnemain2025two}%
  \BibitemOpen
  \bibfield  {author} {\bibinfo {author} {\bibfnamefont {T.}~\bibnamefont
  {Bonnemain}}, \bibinfo {author} {\bibfnamefont {G.}~\bibnamefont {Biondini}},
  \bibinfo {author} {\bibfnamefont {B.}~\bibnamefont {Doyon}}, \bibinfo
  {author} {\bibfnamefont {G.}~\bibnamefont {Roberti}},\ and\ \bibinfo {author}
  {\bibfnamefont {G.~A.}\ \bibnamefont {El}},\ }\bibfield  {title} {\bibinfo
  {title} {Two-dimensional stationary soliton gas},\ }\href
  {https://doi.org/10.1103/PhysRevResearch.7.013143} {\bibfield  {journal}
  {\bibinfo  {journal} {Phys. Rev. Res.}\ }\textbf {\bibinfo {volume} {7}},\
  \bibinfo {pages} {013143} (\bibinfo {year} {2025})}\BibitemShut {NoStop}%
\bibitem [{\citenamefont {Bastianello}\ \emph {et~al.}(2026)\citenamefont
  {Bastianello}, \citenamefont {Tikan}, \citenamefont {Copie}, \citenamefont
  {Randoux},\ and\ \citenamefont {Suret}}]{bastianello2025}%
  \BibitemOpen
  \bibfield  {author} {\bibinfo {author} {\bibfnamefont {A.}~\bibnamefont
  {Bastianello}}, \bibinfo {author} {\bibfnamefont {A.}~\bibnamefont {Tikan}},
  \bibinfo {author} {\bibfnamefont {F.}~\bibnamefont {Copie}}, \bibinfo
  {author} {\bibfnamefont {S.}~\bibnamefont {Randoux}},\ and\ \bibinfo {author}
  {\bibfnamefont {P.}~\bibnamefont {Suret}},\ }\bibfield  {title} {\bibinfo
  {title} {Observation of a generalized gibbs ensemble in photonics},\ }\href
  {https://doi.org/10.1103/xz2w-hndp} {\bibfield  {journal} {\bibinfo
  {journal} {Phys. Rev. A}\ }\textbf {\bibinfo {volume} {113}},\ \bibinfo
  {pages} {013514} (\bibinfo {year} {2026})}\BibitemShut {NoStop}%
\bibitem [{\citenamefont {Castro-Alvaredo}\ \emph {et~al.}(2016)\citenamefont
  {Castro-Alvaredo}, \citenamefont {Doyon},\ and\ \citenamefont
  {Yoshimura}}]{Alvaredo2016}%
  \BibitemOpen
  \bibfield  {author} {\bibinfo {author} {\bibfnamefont {O.~A.}\ \bibnamefont
  {Castro-Alvaredo}}, \bibinfo {author} {\bibfnamefont {B.}~\bibnamefont
  {Doyon}},\ and\ \bibinfo {author} {\bibfnamefont {T.}~\bibnamefont
  {Yoshimura}},\ }\bibfield  {title} {\bibinfo {title} {Emergent hydrodynamics
  in integrable quantum systems out of equilibrium},\ }\href
  {https://doi.org/10.1103/PhysRevX.6.041065} {\bibfield  {journal} {\bibinfo
  {journal} {Phys. Rev. X}\ }\textbf {\bibinfo {volume} {6}},\ \bibinfo {pages}
  {041065} (\bibinfo {year} {2016})}\BibitemShut {NoStop}%
\bibitem [{\citenamefont {Bertini}\ \emph {et~al.}(2016)\citenamefont
  {Bertini}, \citenamefont {Collura}, \citenamefont {De~Nardis},\ and\
  \citenamefont {Fagotti}}]{Bertini2016}%
  \BibitemOpen
  \bibfield  {author} {\bibinfo {author} {\bibfnamefont {B.}~\bibnamefont
  {Bertini}}, \bibinfo {author} {\bibfnamefont {M.}~\bibnamefont {Collura}},
  \bibinfo {author} {\bibfnamefont {J.}~\bibnamefont {De~Nardis}},\ and\
  \bibinfo {author} {\bibfnamefont {M.}~\bibnamefont {Fagotti}},\ }\bibfield
  {title} {\bibinfo {title} {Transport in out-of-equilibrium $xxz$ chains:
  Exact profiles of charges and currents},\ }\href
  {https://doi.org/10.1103/PhysRevLett.117.207201} {\bibfield  {journal}
  {\bibinfo  {journal} {Phys. Rev. Lett.}\ }\textbf {\bibinfo {volume} {117}},\
  \bibinfo {pages} {207201} (\bibinfo {year} {2016})}\BibitemShut {NoStop}%
\bibitem [{\citenamefont {Zakharov}(1971)}]{zakharov1971}%
  \BibitemOpen
  \bibfield  {author} {\bibinfo {author} {\bibfnamefont {V.}~\bibnamefont
  {Zakharov}},\ }\bibfield  {title} {\bibinfo {title} {Kinetic equation for
  solitons},\ }\href@noop {} {\bibfield  {journal} {\bibinfo  {journal} {Sov.
  Phys. JETP}\ }\textbf {\bibinfo {volume} {33}},\ \bibinfo {pages} {538}
  (\bibinfo {year} {1971})}\BibitemShut {NoStop}%
\bibitem [{\citenamefont {El}\ and\ \citenamefont {Kamchatnov}(2005)}]{El2005}%
  \BibitemOpen
  \bibfield  {author} {\bibinfo {author} {\bibfnamefont {G.~A.}\ \bibnamefont
  {El}}\ and\ \bibinfo {author} {\bibfnamefont {A.~M.}\ \bibnamefont
  {Kamchatnov}},\ }\bibfield  {title} {\bibinfo {title} {Kinetic equation for a
  dense soliton gas},\ }\href {https://doi.org/10.1103/PhysRevLett.95.204101}
  {\bibfield  {journal} {\bibinfo  {journal} {Phys. Rev. Lett.}\ }\textbf
  {\bibinfo {volume} {95}},\ \bibinfo {pages} {204101} (\bibinfo {year}
  {2005})}\BibitemShut {NoStop}%
\bibitem [{\citenamefont {El}\ and\ \citenamefont
  {Tovbis}(2020)}]{el2020spectral}%
  \BibitemOpen
  \bibfield  {author} {\bibinfo {author} {\bibfnamefont {G.}~\bibnamefont
  {El}}\ and\ \bibinfo {author} {\bibfnamefont {A.}~\bibnamefont {Tovbis}},\
  }\bibfield  {title} {\bibinfo {title} {Spectral theory of soliton and
  breather gases for the focusing nonlinear schr\"odinger equation},\ }\href
  {https://doi.org/10.1103/PhysRevE.101.052207} {\bibfield  {journal} {\bibinfo
   {journal} {Phys. Rev. E}\ }\textbf {\bibinfo {volume} {101}},\ \bibinfo
  {pages} {052207} (\bibinfo {year} {2020})}\BibitemShut {NoStop}%
\bibitem [{\citenamefont {Spohn}(1982)}]{Spohn1982}%
  \BibitemOpen
  \bibfield  {author} {\bibinfo {author} {\bibfnamefont {H.}~\bibnamefont
  {Spohn}},\ }\bibfield  {title} {\bibinfo {title} {Hydrodynamical theory for
  equilibrium time correlation functions of hard rods},\ }\href
  {https://doi.org/https://doi.org/10.1016/0003-4916(82)90292-5} {\bibfield
  {journal} {\bibinfo  {journal} {Annals of Physics}\ }\textbf {\bibinfo
  {volume} {141}},\ \bibinfo {pages} {353} (\bibinfo {year}
  {1982})}\BibitemShut {NoStop}%
\bibitem [{\citenamefont {Doyon}(2018)}]{Doyon2018}%
  \BibitemOpen
  \bibfield  {author} {\bibinfo {author} {\bibfnamefont {B.}~\bibnamefont
  {Doyon}},\ }\bibfield  {title} {\bibinfo {title} {{Exact large-scale
  correlations in integrable systems out of equilibrium}},\ }\href
  {https://doi.org/10.21468/SciPostPhys.5.5.054} {\bibfield  {journal}
  {\bibinfo  {journal} {SciPost Phys.}\ }\textbf {\bibinfo {volume} {5}},\
  \bibinfo {pages} {054} (\bibinfo {year} {2018})}\BibitemShut {NoStop}%
\bibitem [{\citenamefont {Doyon}(2022)}]{Doyon2022}%
  \BibitemOpen
  \bibfield  {author} {\bibinfo {author} {\bibfnamefont {B.}~\bibnamefont
  {Doyon}},\ }\bibfield  {title} {\bibinfo {title} {Hydrodynamic projections
  and the emergence of linearised euler equations in one-dimensional isolated
  systems},\ }\href {https://doi.org/10.1007/s00220-022-04310-3} {\bibfield
  {journal} {\bibinfo  {journal} {Communications in Mathematical Physics}\
  }\textbf {\bibinfo {volume} {391}},\ \bibinfo {pages} {293} (\bibinfo {year}
  {2022})}\BibitemShut {NoStop}%
\bibitem [{\citenamefont {Bloch}\ \emph {et~al.}(2012)\citenamefont {Bloch},
  \citenamefont {Dalibard},\ and\ \citenamefont {Nascimb{\`e}ne}}]{Bloch2012}%
  \BibitemOpen
  \bibfield  {author} {\bibinfo {author} {\bibfnamefont {I.}~\bibnamefont
  {Bloch}}, \bibinfo {author} {\bibfnamefont {J.}~\bibnamefont {Dalibard}},\
  and\ \bibinfo {author} {\bibfnamefont {S.}~\bibnamefont {Nascimb{\`e}ne}},\
  }\bibfield  {title} {\bibinfo {title} {Quantum simulations with ultracold
  quantum gases},\ }\href {https://doi.org/10.1038/nphys2259} {\bibfield
  {journal} {\bibinfo  {journal} {Nature Physics}\ }\textbf {\bibinfo {volume}
  {8}},\ \bibinfo {pages} {267} (\bibinfo {year} {2012})}\BibitemShut {NoStop}%
\bibitem [{\citenamefont {Schweigler}\ \emph {et~al.}(2017)\citenamefont
  {Schweigler}, \citenamefont {Kasper}, \citenamefont {Erne}, \citenamefont
  {Mazets}, \citenamefont {Rauer}, \citenamefont {Cataldini}, \citenamefont
  {Langen}, \citenamefont {Gasenzer}, \citenamefont {Berges},\ and\
  \citenamefont {Schmiedmayer}}]{Schweigler2017}%
  \BibitemOpen
  \bibfield  {author} {\bibinfo {author} {\bibfnamefont {T.}~\bibnamefont
  {Schweigler}}, \bibinfo {author} {\bibfnamefont {V.}~\bibnamefont {Kasper}},
  \bibinfo {author} {\bibfnamefont {S.}~\bibnamefont {Erne}}, \bibinfo {author}
  {\bibfnamefont {I.}~\bibnamefont {Mazets}}, \bibinfo {author} {\bibfnamefont
  {B.}~\bibnamefont {Rauer}}, \bibinfo {author} {\bibfnamefont
  {F.}~\bibnamefont {Cataldini}}, \bibinfo {author} {\bibfnamefont
  {T.}~\bibnamefont {Langen}}, \bibinfo {author} {\bibfnamefont
  {T.}~\bibnamefont {Gasenzer}}, \bibinfo {author} {\bibfnamefont
  {J.}~\bibnamefont {Berges}},\ and\ \bibinfo {author} {\bibfnamefont
  {J.}~\bibnamefont {Schmiedmayer}},\ }\bibfield  {title} {\bibinfo {title}
  {Experimental characterization of a quantum many-body system via higher-order
  correlations},\ }\href {https://doi.org/10.1038/nature22310} {\bibfield
  {journal} {\bibinfo  {journal} {Nature}\ }\textbf {\bibinfo {volume} {545}},\
  \bibinfo {pages} {323} (\bibinfo {year} {2017})}\BibitemShut {NoStop}%
\bibitem [{\citenamefont {Gooding}\ \emph {et~al.}(2025)\citenamefont
  {Gooding}, \citenamefont {Bunney}, \citenamefont {Tajik}, \citenamefont
  {Erne}, \citenamefont {Biermann}, \citenamefont {Schmiedmayer}, \citenamefont
  {Louko}, \citenamefont {Unruh},\ and\ \citenamefont
  {Weinfurtner}}]{gooding2025}%
  \BibitemOpen
  \bibfield  {author} {\bibinfo {author} {\bibfnamefont {C.}~\bibnamefont
  {Gooding}}, \bibinfo {author} {\bibfnamefont {C.~R.~D.}\ \bibnamefont
  {Bunney}}, \bibinfo {author} {\bibfnamefont {S.}~\bibnamefont {Tajik}},
  \bibinfo {author} {\bibfnamefont {S.}~\bibnamefont {Erne}}, \bibinfo {author}
  {\bibfnamefont {S.}~\bibnamefont {Biermann}}, \bibinfo {author}
  {\bibfnamefont {J.}~\bibnamefont {Schmiedmayer}}, \bibinfo {author}
  {\bibfnamefont {J.}~\bibnamefont {Louko}}, \bibinfo {author} {\bibfnamefont
  {W.~G.}\ \bibnamefont {Unruh}},\ and\ \bibinfo {author} {\bibfnamefont
  {S.}~\bibnamefont {Weinfurtner}},\ }\href {https://arxiv.org/abs/2508.01080}
  {\bibinfo {title} {Nondestructive optomechanical detection scheme for
  bose-einstein condensates}} (\bibinfo {year} {2025}),\ \Eprint
  {https://arxiv.org/abs/2508.01080} {arXiv:2508.01080 [cond-mat.quant-gas]}
  \BibitemShut {NoStop}%
\bibitem [{\citenamefont {Schemmer}\ \emph {et~al.}(2019)\citenamefont
  {Schemmer}, \citenamefont {Bouchoule}, \citenamefont {Doyon},\ and\
  \citenamefont {Dubail}}]{Schemmer2019}%
  \BibitemOpen
  \bibfield  {author} {\bibinfo {author} {\bibfnamefont {M.}~\bibnamefont
  {Schemmer}}, \bibinfo {author} {\bibfnamefont {I.}~\bibnamefont {Bouchoule}},
  \bibinfo {author} {\bibfnamefont {B.}~\bibnamefont {Doyon}},\ and\ \bibinfo
  {author} {\bibfnamefont {J.}~\bibnamefont {Dubail}},\ }\bibfield  {title}
  {\bibinfo {title} {Generalized hydrodynamics on an atom chip},\ }\href
  {https://doi.org/10.1103/PhysRevLett.122.090601} {\bibfield  {journal}
  {\bibinfo  {journal} {Phys. Rev. Lett.}\ }\textbf {\bibinfo {volume} {122}},\
  \bibinfo {pages} {090601} (\bibinfo {year} {2019})}\BibitemShut {NoStop}%
\bibitem [{\citenamefont {Malvania}\ \emph {et~al.}(2021)\citenamefont
  {Malvania}, \citenamefont {Zhang}, \citenamefont {Le}, \citenamefont
  {Dubail}, \citenamefont {Rigol},\ and\ \citenamefont {Weiss}}]{Malvania2021}%
  \BibitemOpen
  \bibfield  {author} {\bibinfo {author} {\bibfnamefont {N.}~\bibnamefont
  {Malvania}}, \bibinfo {author} {\bibfnamefont {Y.}~\bibnamefont {Zhang}},
  \bibinfo {author} {\bibfnamefont {Y.}~\bibnamefont {Le}}, \bibinfo {author}
  {\bibfnamefont {J.}~\bibnamefont {Dubail}}, \bibinfo {author} {\bibfnamefont
  {M.}~\bibnamefont {Rigol}},\ and\ \bibinfo {author} {\bibfnamefont {D.~S.}\
  \bibnamefont {Weiss}},\ }\bibfield  {title} {\bibinfo {title} {{Generalized
  hydrodynamics in strongly interacting 1D Bose gases}},\ }\href
  {https://doi.org/10.1126/science.abf0147} {\bibfield  {journal} {\bibinfo
  {journal} {Science}\ }\textbf {\bibinfo {volume} {373}},\ \bibinfo {pages}
  {1129} (\bibinfo {year} {2021})}\BibitemShut {NoStop}%
\bibitem [{\citenamefont {M\o{}ller}\ \emph {et~al.}(2021)\citenamefont
  {M\o{}ller}, \citenamefont {Li}, \citenamefont {Mazets}, \citenamefont
  {Stimming}, \citenamefont {Zhou}, \citenamefont {Zhu}, \citenamefont {Chen},\
  and\ \citenamefont {Schmiedmayer}}]{Moller2021}%
  \BibitemOpen
  \bibfield  {author} {\bibinfo {author} {\bibfnamefont {F.}~\bibnamefont
  {M\o{}ller}}, \bibinfo {author} {\bibfnamefont {C.}~\bibnamefont {Li}},
  \bibinfo {author} {\bibfnamefont {I.}~\bibnamefont {Mazets}}, \bibinfo
  {author} {\bibfnamefont {H.-P.}\ \bibnamefont {Stimming}}, \bibinfo {author}
  {\bibfnamefont {T.}~\bibnamefont {Zhou}}, \bibinfo {author} {\bibfnamefont
  {Z.}~\bibnamefont {Zhu}}, \bibinfo {author} {\bibfnamefont {X.}~\bibnamefont
  {Chen}},\ and\ \bibinfo {author} {\bibfnamefont {J.}~\bibnamefont
  {Schmiedmayer}},\ }\bibfield  {title} {\bibinfo {title} {Extension of the
  generalized hydrodynamics to the dimensional crossover regime},\ }\href
  {https://doi.org/10.1103/PhysRevLett.126.090602} {\bibfield  {journal}
  {\bibinfo  {journal} {Phys. Rev. Lett.}\ }\textbf {\bibinfo {volume} {126}},\
  \bibinfo {pages} {090602} (\bibinfo {year} {2021})}\BibitemShut {NoStop}%
\bibitem [{\citenamefont {Cataldini}\ \emph {et~al.}(2022)\citenamefont
  {Cataldini}, \citenamefont {M\o{}ller}, \citenamefont {Tajik}, \citenamefont
  {Sabino}, \citenamefont {Ji}, \citenamefont {Mazets}, \citenamefont
  {Schweigler}, \citenamefont {Rauer},\ and\ \citenamefont
  {Schmiedmayer}}]{Cataldini2022}%
  \BibitemOpen
  \bibfield  {author} {\bibinfo {author} {\bibfnamefont {F.}~\bibnamefont
  {Cataldini}}, \bibinfo {author} {\bibfnamefont {F.}~\bibnamefont
  {M\o{}ller}}, \bibinfo {author} {\bibfnamefont {M.}~\bibnamefont {Tajik}},
  \bibinfo {author} {\bibfnamefont {J.~a.}\ \bibnamefont {Sabino}}, \bibinfo
  {author} {\bibfnamefont {S.-C.}\ \bibnamefont {Ji}}, \bibinfo {author}
  {\bibfnamefont {I.}~\bibnamefont {Mazets}}, \bibinfo {author} {\bibfnamefont
  {T.}~\bibnamefont {Schweigler}}, \bibinfo {author} {\bibfnamefont
  {B.}~\bibnamefont {Rauer}},\ and\ \bibinfo {author} {\bibfnamefont
  {J.}~\bibnamefont {Schmiedmayer}},\ }\bibfield  {title} {\bibinfo {title}
  {Emergent pauli blocking in a weakly interacting bose gas},\ }\href
  {https://doi.org/10.1103/PhysRevX.12.041032} {\bibfield  {journal} {\bibinfo
  {journal} {Phys. Rev. X}\ }\textbf {\bibinfo {volume} {12}},\ \bibinfo
  {pages} {041032} (\bibinfo {year} {2022})}\BibitemShut {NoStop}%
\bibitem [{\citenamefont {Schüttelkopf}\ \emph {et~al.}()\citenamefont
  {Schüttelkopf}, \citenamefont {Tajik}, \citenamefont {Bazhan}, \citenamefont
  {Cataldini}, \citenamefont {Ji}, \citenamefont {Schmiedmayer},\ and\
  \citenamefont {Møller}}]{Schuttelkopf2024}%
  \BibitemOpen
  \bibfield  {author} {\bibinfo {author} {\bibfnamefont {P.}~\bibnamefont
  {Schüttelkopf}}, \bibinfo {author} {\bibfnamefont {M.}~\bibnamefont
  {Tajik}}, \bibinfo {author} {\bibfnamefont {N.}~\bibnamefont {Bazhan}},
  \bibinfo {author} {\bibfnamefont {F.}~\bibnamefont {Cataldini}}, \bibinfo
  {author} {\bibfnamefont {S.-C.}\ \bibnamefont {Ji}}, \bibinfo {author}
  {\bibfnamefont {J.}~\bibnamefont {Schmiedmayer}},\ and\ \bibinfo {author}
  {\bibfnamefont {F.}~\bibnamefont {Møller}},\ }\href
  {https://doi.org/10.1126/science.ads8327} {\bibinfo {title} {Characterizing
  transport in a quantum gas by measuring drude weights}},\ \Eprint
  {https://arxiv.org/abs/https://www.science.org/doi/pdf/10.1126/science.ads8327}
  {https://www.science.org/doi/pdf/10.1126/science.ads8327} \BibitemShut
  {NoStop}%
\bibitem [{\citenamefont {Dubois}\ \emph {et~al.}(2024)\citenamefont {Dubois},
  \citenamefont {Th\'em\`eze}, \citenamefont {Nogrette}, \citenamefont
  {Dubail},\ and\ \citenamefont {Bouchoule}}]{dubois2024}%
  \BibitemOpen
  \bibfield  {author} {\bibinfo {author} {\bibfnamefont {L.}~\bibnamefont
  {Dubois}}, \bibinfo {author} {\bibfnamefont {G.}~\bibnamefont {Th\'em\`eze}},
  \bibinfo {author} {\bibfnamefont {F.}~\bibnamefont {Nogrette}}, \bibinfo
  {author} {\bibfnamefont {J.}~\bibnamefont {Dubail}},\ and\ \bibinfo {author}
  {\bibfnamefont {I.}~\bibnamefont {Bouchoule}},\ }\bibfield  {title} {\bibinfo
  {title} {Probing the local rapidity distribution of a one-dimensional bose
  gas},\ }\href {https://doi.org/10.1103/PhysRevLett.133.113402} {\bibfield
  {journal} {\bibinfo  {journal} {Phys. Rev. Lett.}\ }\textbf {\bibinfo
  {volume} {133}},\ \bibinfo {pages} {113402} (\bibinfo {year}
  {2024})}\BibitemShut {NoStop}%
\bibitem [{\citenamefont {Yang}\ \emph {et~al.}(2024)\citenamefont {Yang},
  \citenamefont {Zhang}, \citenamefont {Li}, \citenamefont {Lin}, \citenamefont
  {Gopalakrishnan}, \citenamefont {Rigol},\ and\ \citenamefont
  {Lev}}]{Yang2024Phantom}%
  \BibitemOpen
  \bibfield  {author} {\bibinfo {author} {\bibfnamefont {K.}~\bibnamefont
  {Yang}}, \bibinfo {author} {\bibfnamefont {Y.}~\bibnamefont {Zhang}},
  \bibinfo {author} {\bibfnamefont {K.-Y.}\ \bibnamefont {Li}}, \bibinfo
  {author} {\bibfnamefont {K.-Y.}\ \bibnamefont {Lin}}, \bibinfo {author}
  {\bibfnamefont {S.}~\bibnamefont {Gopalakrishnan}}, \bibinfo {author}
  {\bibfnamefont {M.}~\bibnamefont {Rigol}},\ and\ \bibinfo {author}
  {\bibfnamefont {B.~L.}\ \bibnamefont {Lev}},\ }\bibfield  {title} {\bibinfo
  {title} {Phantom energy in the nonlinear response of a quantum many-body scar
  state},\ }\href {https://doi.org/10.1126/science.adk8978} {\bibfield
  {journal} {\bibinfo  {journal} {Science}\ }\textbf {\bibinfo {volume}
  {385}},\ \bibinfo {pages} {1063} (\bibinfo {year} {2024})},\ \Eprint
  {https://arxiv.org/abs/https://www.science.org/doi/pdf/10.1126/science.adk8978}
  {https://www.science.org/doi/pdf/10.1126/science.adk8978} \BibitemShut
  {NoStop}%
\bibitem [{\citenamefont {Horvath}\ \emph {et~al.}(2025)\citenamefont
  {Horvath}, \citenamefont {Bastianello}, \citenamefont {Dhar}, \citenamefont
  {Koch}, \citenamefont {Guo}, \citenamefont {Caux}, \citenamefont {Landini},\
  and\ \citenamefont {Nägerl}}]{horvath2025}%
  \BibitemOpen
  \bibfield  {author} {\bibinfo {author} {\bibfnamefont {M.}~\bibnamefont
  {Horvath}}, \bibinfo {author} {\bibfnamefont {A.}~\bibnamefont
  {Bastianello}}, \bibinfo {author} {\bibfnamefont {S.}~\bibnamefont {Dhar}},
  \bibinfo {author} {\bibfnamefont {R.}~\bibnamefont {Koch}}, \bibinfo {author}
  {\bibfnamefont {Y.}~\bibnamefont {Guo}}, \bibinfo {author} {\bibfnamefont
  {J.-S.}\ \bibnamefont {Caux}}, \bibinfo {author} {\bibfnamefont
  {M.}~\bibnamefont {Landini}},\ and\ \bibinfo {author} {\bibfnamefont {H.-C.}\
  \bibnamefont {Nägerl}},\ }\href {https://arxiv.org/abs/2505.10550} {\bibinfo
  {title} {Observing bethe strings in an attractive bose gas far from
  equilibrium}} (\bibinfo {year} {2025}),\ \Eprint
  {https://arxiv.org/abs/2505.10550} {arXiv:2505.10550 [cond-mat.quant-gas]}
  \BibitemShut {NoStop}%
\bibitem [{\citenamefont {Zeng}\ \emph {et~al.}(2026)\citenamefont {Zeng},
  \citenamefont {Bastianello}, \citenamefont {Dhar}, \citenamefont {Wang},
  \citenamefont {Yu}, \citenamefont {Horvath}, \citenamefont {Astrakharchik},
  \citenamefont {Guo}, \citenamefont {Nägerl},\ and\ \citenamefont
  {Landini}}]{zeng2026}%
  \BibitemOpen
  \bibfield  {author} {\bibinfo {author} {\bibfnamefont {Y.}~\bibnamefont
  {Zeng}}, \bibinfo {author} {\bibfnamefont {A.}~\bibnamefont {Bastianello}},
  \bibinfo {author} {\bibfnamefont {S.}~\bibnamefont {Dhar}}, \bibinfo {author}
  {\bibfnamefont {Z.}~\bibnamefont {Wang}}, \bibinfo {author} {\bibfnamefont
  {X.}~\bibnamefont {Yu}}, \bibinfo {author} {\bibfnamefont {M.}~\bibnamefont
  {Horvath}}, \bibinfo {author} {\bibfnamefont {G.~E.}\ \bibnamefont
  {Astrakharchik}}, \bibinfo {author} {\bibfnamefont {Y.}~\bibnamefont {Guo}},
  \bibinfo {author} {\bibfnamefont {H.-C.}\ \bibnamefont {Nägerl}},\ and\
  \bibinfo {author} {\bibfnamefont {M.}~\bibnamefont {Landini}},\ }\href
  {https://arxiv.org/abs/2602.17657} {\bibinfo {title} {Realization of
  fractional fermi seas}} (\bibinfo {year} {2026}),\ \Eprint
  {https://arxiv.org/abs/2602.17657} {arXiv:2602.17657 [cond-mat.quant-gas]}
  \BibitemShut {NoStop}%
\bibitem [{\citenamefont {Suret}\ \emph {et~al.}(2024)\citenamefont {Suret},
  \citenamefont {Randoux}, \citenamefont {Gelash}, \citenamefont {Agafontsev},
  \citenamefont {Doyon},\ and\ \citenamefont {El}}]{suret2024soliton}%
  \BibitemOpen
  \bibfield  {author} {\bibinfo {author} {\bibfnamefont {P.}~\bibnamefont
  {Suret}}, \bibinfo {author} {\bibfnamefont {S.}~\bibnamefont {Randoux}},
  \bibinfo {author} {\bibfnamefont {A.}~\bibnamefont {Gelash}}, \bibinfo
  {author} {\bibfnamefont {D.}~\bibnamefont {Agafontsev}}, \bibinfo {author}
  {\bibfnamefont {B.}~\bibnamefont {Doyon}},\ and\ \bibinfo {author}
  {\bibfnamefont {G.}~\bibnamefont {El}},\ }\bibfield  {title} {\bibinfo
  {title} {Soliton gas: Theory, numerics, and experiments},\ }\href@noop {}
  {\bibfield  {journal} {\bibinfo  {journal} {Physical Review E}\ }\textbf
  {\bibinfo {volume} {109}},\ \bibinfo {pages} {061001} (\bibinfo {year}
  {2024})}\BibitemShut {NoStop}%
\bibitem [{\citenamefont {Shabat}\ and\ \citenamefont
  {Zakharov}(1972)}]{shabat1972exact}%
  \BibitemOpen
  \bibfield  {author} {\bibinfo {author} {\bibfnamefont {A.}~\bibnamefont
  {Shabat}}\ and\ \bibinfo {author} {\bibfnamefont {V.}~\bibnamefont
  {Zakharov}},\ }\bibfield  {title} {\bibinfo {title} {Exact theory of
  two-dimensional self-focusing and one-dimensional self-modulation of waves in
  nonlinear media},\ }\href@noop {} {\bibfield  {journal} {\bibinfo  {journal}
  {Sov. Phys. JETP}\ }\textbf {\bibinfo {volume} {34}},\ \bibinfo {pages} {62}
  (\bibinfo {year} {1972})}\BibitemShut {NoStop}%
\bibitem [{\citenamefont {Yang}(2010)}]{yang2010nonlinear}%
  \BibitemOpen
  \bibfield  {author} {\bibinfo {author} {\bibfnamefont {J.}~\bibnamefont
  {Yang}},\ }\href@noop {} {\emph {\bibinfo {title} {Nonlinear waves in
  integrable and nonintegrable systems}}}\ (\bibinfo  {publisher} {SIAM},\
  \bibinfo {year} {2010})\BibitemShut {NoStop}%
\bibitem [{\citenamefont {Koch}\ and\ \citenamefont
  {Bastianello}(2023)}]{Koch2023}%
  \BibitemOpen
  \bibfield  {author} {\bibinfo {author} {\bibfnamefont {R.}~\bibnamefont
  {Koch}}\ and\ \bibinfo {author} {\bibfnamefont {A.}~\bibnamefont
  {Bastianello}},\ }\bibfield  {title} {\bibinfo {title} {{Exact thermodynamics
  and transport in the classical sine-Gordon model}},\ }\href
  {https://doi.org/10.21468/SciPostPhys.15.4.140} {\bibfield  {journal}
  {\bibinfo  {journal} {SciPost Phys.}\ }\textbf {\bibinfo {volume} {15}},\
  \bibinfo {pages} {140} (\bibinfo {year} {2023})}\BibitemShut {NoStop}%
\bibitem [{\citenamefont {Bastianello}\ \emph {et~al.}(2024)\citenamefont
  {Bastianello}, \citenamefont {Krajnik},\ and\ \citenamefont
  {Ilievski}}]{Bastianello2024}%
  \BibitemOpen
  \bibfield  {author} {\bibinfo {author} {\bibfnamefont {A.}~\bibnamefont
  {Bastianello}}, \bibinfo {author} {\bibfnamefont {i.~c.~v.}\ \bibnamefont
  {Krajnik}},\ and\ \bibinfo {author} {\bibfnamefont {E.}~\bibnamefont
  {Ilievski}},\ }\bibfield  {title} {\bibinfo {title} {Landau-lifschitz
  magnets: Exact thermodynamics and transport},\ }\href
  {https://doi.org/10.1103/PhysRevLett.133.107102} {\bibfield  {journal}
  {\bibinfo  {journal} {Phys. Rev. Lett.}\ }\textbf {\bibinfo {volume} {133}},\
  \bibinfo {pages} {107102} (\bibinfo {year} {2024})}\BibitemShut {NoStop}%
\bibitem [{\citenamefont {Ilievski}\ \emph {et~al.}(2015)\citenamefont
  {Ilievski}, \citenamefont {De~Nardis}, \citenamefont {Wouters}, \citenamefont
  {Caux}, \citenamefont {Essler},\ and\ \citenamefont
  {Prosen}}]{ilievski2015complete}%
  \BibitemOpen
  \bibfield  {author} {\bibinfo {author} {\bibfnamefont {E.}~\bibnamefont
  {Ilievski}}, \bibinfo {author} {\bibfnamefont {J.}~\bibnamefont {De~Nardis}},
  \bibinfo {author} {\bibfnamefont {B.}~\bibnamefont {Wouters}}, \bibinfo
  {author} {\bibfnamefont {J.-S.}\ \bibnamefont {Caux}}, \bibinfo {author}
  {\bibfnamefont {F.~H.}\ \bibnamefont {Essler}},\ and\ \bibinfo {author}
  {\bibfnamefont {T.}~\bibnamefont {Prosen}},\ }\bibfield  {title} {\bibinfo
  {title} {Complete generalized gibbs ensembles in an interacting theory},\
  }\href@noop {} {\bibfield  {journal} {\bibinfo  {journal} {Physical review
  letters}\ }\textbf {\bibinfo {volume} {115}},\ \bibinfo {pages} {157201}
  (\bibinfo {year} {2015})}\BibitemShut {NoStop}%
\bibitem [{\citenamefont {Copie}\ \emph {et~al.}(2023)\citenamefont {Copie},
  \citenamefont {Suret},\ and\ \citenamefont {Randoux}}]{copie2023space}%
  \BibitemOpen
  \bibfield  {author} {\bibinfo {author} {\bibfnamefont {F.}~\bibnamefont
  {Copie}}, \bibinfo {author} {\bibfnamefont {P.}~\bibnamefont {Suret}},\ and\
  \bibinfo {author} {\bibfnamefont {S.}~\bibnamefont {Randoux}},\ }\bibfield
  {title} {\bibinfo {title} {Space–time observation of the dynamics of
  soliton collisions in a recirculating optical fiber loop},\ }\href
  {https://doi.org/https://doi.org/10.1016/j.optcom.2023.129647} {\bibfield
  {journal} {\bibinfo  {journal} {Optics Communications}\ }\textbf {\bibinfo
  {volume} {545}},\ \bibinfo {pages} {129647} (\bibinfo {year}
  {2023})}\BibitemShut {NoStop}%
\bibitem [{\citenamefont {Fache}\ \emph {et~al.}(2025)\citenamefont {Fache},
  \citenamefont {Copie}, \citenamefont {Suret},\ and\ \citenamefont
  {Randoux}}]{fache2025perturbed}%
  \BibitemOpen
  \bibfield  {author} {\bibinfo {author} {\bibfnamefont {L.}~\bibnamefont
  {Fache}}, \bibinfo {author} {\bibfnamefont {F.}~\bibnamefont {Copie}},
  \bibinfo {author} {\bibfnamefont {P.}~\bibnamefont {Suret}},\ and\ \bibinfo
  {author} {\bibfnamefont {S.}~\bibnamefont {Randoux}},\ }\bibfield  {title}
  {\bibinfo {title} {Perturbed nonlinear evolution of optical soliton gases:
  Growth and decay in integrable turbulence},\ }\href@noop {} {\bibfield
  {journal} {\bibinfo  {journal} {Physical Review Letters}\ }\textbf {\bibinfo
  {volume} {135}},\ \bibinfo {pages} {157201} (\bibinfo {year}
  {2025})}\BibitemShut {NoStop}%
\bibitem [{\citenamefont {Kraych}\ \emph {et~al.}(2019)\citenamefont {Kraych},
  \citenamefont {Agafontsev}, \citenamefont {Randoux},\ and\ \citenamefont
  {Suret}}]{kraych2019statistical}%
  \BibitemOpen
  \bibfield  {author} {\bibinfo {author} {\bibfnamefont {A.~E.}\ \bibnamefont
  {Kraych}}, \bibinfo {author} {\bibfnamefont {D.}~\bibnamefont {Agafontsev}},
  \bibinfo {author} {\bibfnamefont {S.}~\bibnamefont {Randoux}},\ and\ \bibinfo
  {author} {\bibfnamefont {P.}~\bibnamefont {Suret}},\ }\bibfield  {title}
  {\bibinfo {title} {Statistical properties of the nonlinear stage of
  modulation instability in fiber optics},\ }\href@noop {} {\bibfield
  {journal} {\bibinfo  {journal} {Physical review letters}\ }\textbf {\bibinfo
  {volume} {123}},\ \bibinfo {pages} {093902} (\bibinfo {year}
  {2019})}\BibitemShut {NoStop}%
\bibitem [{sup()}]{suppmat}%
  \BibitemOpen
  \href@noop {} {\bibinfo  {journal} {See Supplemental Material [url] for
  details on experiments and theory,, which includes Refs. [79-80]}\
  }\BibitemShut {NoStop}%
\bibitem [{\citenamefont {Agrawal}(2000)}]{agrawal2000nonlinear}%
  \BibitemOpen
\bibfield  {journal} {  }\bibfield  {author} {\bibinfo {author} {\bibfnamefont
  {G.~P.}\ \bibnamefont {Agrawal}},\ }\bibfield  {title} {\bibinfo {title}
  {Nonlinear fiber optics},\ }in\ \href@noop {} {\emph {\bibinfo {booktitle}
  {Nonlinear Science at the Dawn of the 21st Century}}}\ (\bibinfo  {publisher}
  {Springer},\ \bibinfo {year} {2000})\ pp.\ \bibinfo {pages}
  {195--211}\BibitemShut {NoStop}%
\bibitem [{\citenamefont {Gelash}\ \emph {et~al.}(2019)\citenamefont {Gelash},
  \citenamefont {Agafontsev}, \citenamefont {Zakharov}, \citenamefont {El},
  \citenamefont {Randoux},\ and\ \citenamefont {Suret}}]{gelash2019bound}%
  \BibitemOpen
  \bibfield  {author} {\bibinfo {author} {\bibfnamefont {A.}~\bibnamefont
  {Gelash}}, \bibinfo {author} {\bibfnamefont {D.}~\bibnamefont {Agafontsev}},
  \bibinfo {author} {\bibfnamefont {V.}~\bibnamefont {Zakharov}}, \bibinfo
  {author} {\bibfnamefont {G.}~\bibnamefont {El}}, \bibinfo {author}
  {\bibfnamefont {S.}~\bibnamefont {Randoux}},\ and\ \bibinfo {author}
  {\bibfnamefont {P.}~\bibnamefont {Suret}},\ }\bibfield  {title} {\bibinfo
  {title} {Bound state soliton gas dynamics underlying the spontaneous
  modulational instability},\ }\href@noop {} {\bibfield  {journal} {\bibinfo
  {journal} {Physical review letters}\ }\textbf {\bibinfo {volume} {123}},\
  \bibinfo {pages} {234102} (\bibinfo {year} {2019})}\BibitemShut {NoStop}%
\bibitem [{\citenamefont {Mandel}\ and\ \citenamefont
  {Wolf}(1995)}]{Mandel_Wolf_1995}%
  \BibitemOpen
  \bibfield  {author} {\bibinfo {author} {\bibfnamefont {L.}~\bibnamefont
  {Mandel}}\ and\ \bibinfo {author} {\bibfnamefont {E.}~\bibnamefont {Wolf}},\
  }\href@noop {} {\emph {\bibinfo {title} {Optical Coherence and Quantum
  Optics}}}\ (\bibinfo  {publisher} {Cambridge University Press},\ \bibinfo
  {year} {1995})\BibitemShut {NoStop}%
\bibitem [{\citenamefont {Walczak}\ \emph {et~al.}(2015)\citenamefont
  {Walczak}, \citenamefont {Randoux},\ and\ \citenamefont
  {Suret}}]{walczak_2015}%
  \BibitemOpen
  \bibfield  {author} {\bibinfo {author} {\bibfnamefont {P.}~\bibnamefont
  {Walczak}}, \bibinfo {author} {\bibfnamefont {S.}~\bibnamefont {Randoux}},\
  and\ \bibinfo {author} {\bibfnamefont {P.}~\bibnamefont {Suret}},\ }\bibfield
   {title} {\bibinfo {title} {Optical rogue waves in integrable turbulence},\
  }\href {https://doi.org/10.1103/PhysRevLett.114.143903} {\bibfield  {journal}
  {\bibinfo  {journal} {Phys. Rev. Lett.}\ }\textbf {\bibinfo {volume} {114}},\
  \bibinfo {pages} {143903} (\bibinfo {year} {2015})}\BibitemShut {NoStop}%
\bibitem [{\citenamefont {Agafontsev}\ \emph {et~al.}(2021)\citenamefont
  {Agafontsev}, \citenamefont {Randoux},\ and\ \citenamefont
  {Suret}}]{agafontsev2021extreme}%
  \BibitemOpen
  \bibfield  {author} {\bibinfo {author} {\bibfnamefont {D.}~\bibnamefont
  {Agafontsev}}, \bibinfo {author} {\bibfnamefont {S.}~\bibnamefont
  {Randoux}},\ and\ \bibinfo {author} {\bibfnamefont {P.}~\bibnamefont
  {Suret}},\ }\bibfield  {title} {\bibinfo {title} {Extreme rogue wave
  generation from narrowband partially coherent waves},\ }\href@noop {}
  {\bibfield  {journal} {\bibinfo  {journal} {Physical Review E}\ }\textbf
  {\bibinfo {volume} {103}},\ \bibinfo {pages} {032209} (\bibinfo {year}
  {2021})}\BibitemShut {NoStop}%
\bibitem [{\citenamefont {Faddeev}\ and\ \citenamefont
  {Takhtajan}(2007)}]{faddeev2007hamiltonian}%
  \BibitemOpen
  \bibfield  {author} {\bibinfo {author} {\bibfnamefont {L.}~\bibnamefont
  {Faddeev}}\ and\ \bibinfo {author} {\bibfnamefont {L.}~\bibnamefont
  {Takhtajan}},\ }\href@noop {} {\emph {\bibinfo {title} {Hamiltonian methods
  in the theory of solitons}}}\ (\bibinfo  {publisher} {Springer Science \&
  Business Media},\ \bibinfo {year} {2007})\BibitemShut {NoStop}%
\bibitem [{\citenamefont {Jenkins}\ and\ \citenamefont
  {Tovbis}(2024)}]{jenkins2024approximation}%
  \BibitemOpen
  \bibfield  {author} {\bibinfo {author} {\bibfnamefont {R.}~\bibnamefont
  {Jenkins}}\ and\ \bibinfo {author} {\bibfnamefont {A.}~\bibnamefont
  {Tovbis}},\ }\href {https://arxiv.org/abs/2408.13700} {\bibinfo {title}
  {Approximation of the thermodynamic limit of finite-gap solutions of the
  focusing nls hierarchy by multisoliton solutions}} (\bibinfo {year} {2024}),\
  \Eprint {https://arxiv.org/abs/2408.13700} {arXiv:2408.13700 [nlin.SI]}
  \BibitemShut {NoStop}%
\bibitem [{\citenamefont {Zakharov}\ and\ \citenamefont
  {Shabat}(1973)}]{zakharov1973interaction}%
  \BibitemOpen
  \bibfield  {author} {\bibinfo {author} {\bibfnamefont {V.~E.}\ \bibnamefont
  {Zakharov}}\ and\ \bibinfo {author} {\bibfnamefont {A.~B.}\ \bibnamefont
  {Shabat}},\ }\bibfield  {title} {\bibinfo {title} {Interaction between
  solitons in a stable medium},\ }\href@noop {} {\bibfield  {journal} {\bibinfo
   {journal} {Sov. Phys. JETP}\ }\textbf {\bibinfo {volume} {37}},\ \bibinfo
  {pages} {823} (\bibinfo {year} {1973})}\BibitemShut {NoStop}%
\bibitem [{\citenamefont {Congy}\ \emph {et~al.}(2021)\citenamefont {Congy},
  \citenamefont {El},\ and\ \citenamefont {Roberti}}]{congy2021soliton}%
  \BibitemOpen
  \bibfield  {author} {\bibinfo {author} {\bibfnamefont {T.}~\bibnamefont
  {Congy}}, \bibinfo {author} {\bibfnamefont {G.}~\bibnamefont {El}},\ and\
  \bibinfo {author} {\bibfnamefont {G.}~\bibnamefont {Roberti}},\ }\bibfield
  {title} {\bibinfo {title} {Soliton gas in bidirectional dispersive
  hydrodynamics},\ }\href@noop {} {\bibfield  {journal} {\bibinfo  {journal}
  {Physical Review E}\ }\textbf {\bibinfo {volume} {103}},\ \bibinfo {pages}
  {042201} (\bibinfo {year} {2021})}\BibitemShut {NoStop}%
\bibitem [{con(2025)}]{congy2025riemann}%
  \BibitemOpen
  \bibfield  {title} {\bibinfo {title} {Riemann problem for polychromatic
  soliton gases: a testbed for the spectral kinetic theory},\ }\href@noop {}
  {\bibfield  {journal} {\bibinfo  {journal} {Wave Motion}\ }\textbf {\bibinfo
  {volume} {134}},\ \bibinfo {pages} {103480} (\bibinfo {year}
  {2025})}\BibitemShut {NoStop}%
\bibitem [{\citenamefont {Suret}\ \emph {et~al.}(2023)\citenamefont {Suret},
  \citenamefont {Dufour}, \citenamefont {Roberti}, \citenamefont {El},
  \citenamefont {Copie},\ and\ \citenamefont {Randoux}}]{suret2023soliton}%
  \BibitemOpen
  \bibfield  {author} {\bibinfo {author} {\bibfnamefont {P.}~\bibnamefont
  {Suret}}, \bibinfo {author} {\bibfnamefont {M.}~\bibnamefont {Dufour}},
  \bibinfo {author} {\bibfnamefont {G.}~\bibnamefont {Roberti}}, \bibinfo
  {author} {\bibfnamefont {G.}~\bibnamefont {El}}, \bibinfo {author}
  {\bibfnamefont {F.}~\bibnamefont {Copie}},\ and\ \bibinfo {author}
  {\bibfnamefont {S.}~\bibnamefont {Randoux}},\ }\bibfield  {title} {\bibinfo
  {title} {Soliton refraction by an optical soliton gas},\ }\href@noop {}
  {\bibfield  {journal} {\bibinfo  {journal} {Physical Review Research}\
  }\textbf {\bibinfo {volume} {5}},\ \bibinfo {pages} {L042002} (\bibinfo
  {year} {2023})}\BibitemShut {NoStop}%
\bibitem [{\citenamefont {Fache}\ \emph {et~al.}(2024)\citenamefont {Fache},
  \citenamefont {Bonnefoy}, \citenamefont {Ducrozet}, \citenamefont {Copie},
  \citenamefont {Novkoski}, \citenamefont {Ricard}, \citenamefont {Roberti},
  \citenamefont {Falcon}, \citenamefont {Suret}, \citenamefont {El} \emph
  {et~al.}}]{fache2024interaction}%
  \BibitemOpen
  \bibfield  {author} {\bibinfo {author} {\bibfnamefont {L.}~\bibnamefont
  {Fache}}, \bibinfo {author} {\bibfnamefont {F.}~\bibnamefont {Bonnefoy}},
  \bibinfo {author} {\bibfnamefont {G.}~\bibnamefont {Ducrozet}}, \bibinfo
  {author} {\bibfnamefont {F.}~\bibnamefont {Copie}}, \bibinfo {author}
  {\bibfnamefont {F.}~\bibnamefont {Novkoski}}, \bibinfo {author}
  {\bibfnamefont {G.}~\bibnamefont {Ricard}}, \bibinfo {author} {\bibfnamefont
  {G.}~\bibnamefont {Roberti}}, \bibinfo {author} {\bibfnamefont
  {E.}~\bibnamefont {Falcon}}, \bibinfo {author} {\bibfnamefont
  {P.}~\bibnamefont {Suret}}, \bibinfo {author} {\bibfnamefont
  {G.}~\bibnamefont {El}}, \emph {et~al.},\ }\bibfield  {title} {\bibinfo
  {title} {Interaction of soliton gases in deep-water surface gravity waves},\
  }\href@noop {} {\bibfield  {journal} {\bibinfo  {journal} {Physical Review
  E}\ }\textbf {\bibinfo {volume} {109}},\ \bibinfo {pages} {034207} (\bibinfo
  {year} {2024})}\BibitemShut {NoStop}%
\bibitem [{\citenamefont {Huang}(2009)}]{huang2009}%
  \BibitemOpen
  \bibfield  {author} {\bibinfo {author} {\bibfnamefont {K.}~\bibnamefont
  {Huang}},\ }\href@noop {} {\emph {\bibinfo {title} {Introduction to
  statistical physics}}}\ (\bibinfo  {publisher} {Chapman and Hall/CRC},\
  \bibinfo {year} {2009})\BibitemShut {NoStop}%
\bibitem [{\citenamefont {Chung}(1990)}]{Chung1990}%
  \BibitemOpen
  \bibfield  {author} {\bibinfo {author} {\bibfnamefont {S.~G.}\ \bibnamefont
  {Chung}},\ }\bibfield  {title} {\bibinfo {title} {Breakdown of the
  soliton-gas phenomenology for the classical statistical mechanics of the
  sine-gordon model},\ }\href {https://doi.org/10.1088/0305-4470/23/23/010}
  {\bibfield  {journal} {\bibinfo  {journal} {Journal of Physics A:
  Mathematical and General}\ }\textbf {\bibinfo {volume} {23}},\ \bibinfo
  {pages} {L1241} (\bibinfo {year} {1990})}\BibitemShut {NoStop}%
\bibitem [{\citenamefont {Kovrizhin}(2001)}]{Kovrizhin2001}%
  \BibitemOpen
  \bibfield  {author} {\bibinfo {author} {\bibfnamefont {D.}~\bibnamefont
  {Kovrizhin}},\ }\bibfield  {title} {\bibinfo {title} {Exact form of the
  bogoliubov excitations in one-dimensional nonlinear schrödinger equation},\
  }\href {https://doi.org/https://doi.org/10.1016/S0375-9601(01)00503-5}
  {\bibfield  {journal} {\bibinfo  {journal} {Physics Letters A}\ }\textbf
  {\bibinfo {volume} {287}},\ \bibinfo {pages} {392} (\bibinfo {year}
  {2001})}\BibitemShut {NoStop}%
\bibitem [{\citenamefont {Kivshar}\ and\ \citenamefont
  {Agrawal}(2003)}]{kivshar2003optical}%
  \BibitemOpen
  \bibfield  {author} {\bibinfo {author} {\bibfnamefont {Y.~S.}\ \bibnamefont
  {Kivshar}}\ and\ \bibinfo {author} {\bibfnamefont {G.~P.}\ \bibnamefont
  {Agrawal}},\ }\href@noop {} {\emph {\bibinfo {title} {Optical solitons: from
  fibers to photonic crystals}}}\ (\bibinfo  {publisher} {Academic press},\
  \bibinfo {year} {2003})\BibitemShut {NoStop}%
\bibitem [{\citenamefont {De~Nardis}\ \emph {et~al.}(2018)\citenamefont
  {De~Nardis}, \citenamefont {Bernard},\ and\ \citenamefont
  {Doyon}}]{DeNardis2018}%
  \BibitemOpen
  \bibfield  {author} {\bibinfo {author} {\bibfnamefont {J.}~\bibnamefont
  {De~Nardis}}, \bibinfo {author} {\bibfnamefont {D.}~\bibnamefont {Bernard}},\
  and\ \bibinfo {author} {\bibfnamefont {B.}~\bibnamefont {Doyon}},\ }\bibfield
   {title} {\bibinfo {title} {Hydrodynamic diffusion in integrable systems},\
  }\href {https://doi.org/10.1103/PhysRevLett.121.160603} {\bibfield  {journal}
  {\bibinfo  {journal} {Phys. Rev. Lett.}\ }\textbf {\bibinfo {volume} {121}},\
  \bibinfo {pages} {160603} (\bibinfo {year} {2018})}\BibitemShut {NoStop}%
\bibitem [{\citenamefont {Doyon}\ and\ \citenamefont
  {Myers}(2020)}]{Doyon2020}%
  \BibitemOpen
  \bibfield  {author} {\bibinfo {author} {\bibfnamefont {B.}~\bibnamefont
  {Doyon}}\ and\ \bibinfo {author} {\bibfnamefont {J.}~\bibnamefont {Myers}},\
  }\bibfield  {title} {\bibinfo {title} {Fluctuations in ballistic transport
  from euler hydrodynamics},\ }\href
  {https://doi.org/10.1007/s00023-019-00860-w} {\bibfield  {journal} {\bibinfo
  {journal} {Annales Henri Poincar{\'e}}\ }\textbf {\bibinfo {volume} {21}},\
  \bibinfo {pages} {255} (\bibinfo {year} {2020})}\BibitemShut {NoStop}%
\bibitem [{\citenamefont {Doyon}\ \emph
  {et~al.}(2023{\natexlab{a}})\citenamefont {Doyon}, \citenamefont {Perfetto},
  \citenamefont {Sasamoto},\ and\ \citenamefont {Yoshimura}}]{Doyon2023}%
  \BibitemOpen
  \bibfield  {author} {\bibinfo {author} {\bibfnamefont {B.}~\bibnamefont
  {Doyon}}, \bibinfo {author} {\bibfnamefont {G.}~\bibnamefont {Perfetto}},
  \bibinfo {author} {\bibfnamefont {T.}~\bibnamefont {Sasamoto}},\ and\
  \bibinfo {author} {\bibfnamefont {T.}~\bibnamefont {Yoshimura}},\ }\bibfield
  {title} {\bibinfo {title} {{Ballistic macroscopic fluctuation theory}},\
  }\href {https://doi.org/10.21468/SciPostPhys.15.4.136} {\bibfield  {journal}
  {\bibinfo  {journal} {SciPost Phys.}\ }\textbf {\bibinfo {volume} {15}},\
  \bibinfo {pages} {136} (\bibinfo {year} {2023}{\natexlab{a}})}\BibitemShut
  {NoStop}%
\bibitem [{\citenamefont {Doyon}\ \emph
  {et~al.}(2023{\natexlab{b}})\citenamefont {Doyon}, \citenamefont {Perfetto},
  \citenamefont {Sasamoto},\ and\ \citenamefont {Yoshimura}}]{Doyon2023L}%
  \BibitemOpen
  \bibfield  {author} {\bibinfo {author} {\bibfnamefont {B.}~\bibnamefont
  {Doyon}}, \bibinfo {author} {\bibfnamefont {G.}~\bibnamefont {Perfetto}},
  \bibinfo {author} {\bibfnamefont {T.}~\bibnamefont {Sasamoto}},\ and\
  \bibinfo {author} {\bibfnamefont {T.}~\bibnamefont {Yoshimura}},\ }\bibfield
  {title} {\bibinfo {title} {Emergence of hydrodynamic spatial long-range
  correlations in nonequilibrium many-body systems},\ }\href
  {https://doi.org/10.1103/PhysRevLett.131.027101} {\bibfield  {journal}
  {\bibinfo  {journal} {Phys. Rev. Lett.}\ }\textbf {\bibinfo {volume} {131}},\
  \bibinfo {pages} {027101} (\bibinfo {year} {2023}{\natexlab{b}})}\BibitemShut
  {NoStop}%
\bibitem [{\citenamefont {H\"ubner}\ \emph {et~al.}(2025)\citenamefont
  {H\"ubner}, \citenamefont {Biagetti}, \citenamefont {De~Nardis},\ and\
  \citenamefont {Doyon}}]{hubner2025diffusive}%
  \BibitemOpen
  \bibfield  {author} {\bibinfo {author} {\bibfnamefont {F.}~\bibnamefont
  {H\"ubner}}, \bibinfo {author} {\bibfnamefont {L.}~\bibnamefont {Biagetti}},
  \bibinfo {author} {\bibfnamefont {J.}~\bibnamefont {De~Nardis}},\ and\
  \bibinfo {author} {\bibfnamefont {B.}~\bibnamefont {Doyon}},\ }\bibfield
  {title} {\bibinfo {title} {Diffusive hydrodynamics from long-range
  correlations},\ }\href {https://doi.org/10.1103/PhysRevLett.134.187101}
  {\bibfield  {journal} {\bibinfo  {journal} {Phys. Rev. Lett.}\ }\textbf
  {\bibinfo {volume} {134}},\ \bibinfo {pages} {187101} (\bibinfo {year}
  {2025})}\BibitemShut {NoStop}%
\bibitem [{\citenamefont {Prilepsky}\ \emph {et~al.}(2014)\citenamefont
  {Prilepsky}, \citenamefont {Derevyanko}, \citenamefont {Blow}, \citenamefont
  {Gabitov},\ and\ \citenamefont {Turitsyn}}]{prilepsky2014nonlinear}%
  \BibitemOpen
  \bibfield  {author} {\bibinfo {author} {\bibfnamefont {J.~E.}\ \bibnamefont
  {Prilepsky}}, \bibinfo {author} {\bibfnamefont {S.~A.}\ \bibnamefont
  {Derevyanko}}, \bibinfo {author} {\bibfnamefont {K.~J.}\ \bibnamefont
  {Blow}}, \bibinfo {author} {\bibfnamefont {I.}~\bibnamefont {Gabitov}},\ and\
  \bibinfo {author} {\bibfnamefont {S.~K.}\ \bibnamefont {Turitsyn}},\
  }\bibfield  {title} {\bibinfo {title} {Nonlinear inverse synthesis and
  eigenvalue division multiplexing in optical fiber channels},\ }\href@noop {}
  {\bibfield  {journal} {\bibinfo  {journal} {Physical review letters}\
  }\textbf {\bibinfo {volume} {113}},\ \bibinfo {pages} {013901} (\bibinfo
  {year} {2014})}\BibitemShut {NoStop}%
\bibitem [{\citenamefont {Takahashi}(2005)}]{takahashi2005}%
  \BibitemOpen
  \bibfield  {author} {\bibinfo {author} {\bibfnamefont {M.}~\bibnamefont
  {Takahashi}},\ }\href@noop {} {\emph {\bibinfo {title} {Thermodynamics of
  one-dimensional solvable models}}}\ (\bibinfo  {publisher} {Cambridge
  University Press},\ \bibinfo {year} {2005})\BibitemShut {NoStop}%
\bibitem [{\citenamefont {Congy}\ \emph {et~al.}(2024)\citenamefont {Congy},
  \citenamefont {El}, \citenamefont {Roberti}, \citenamefont {Tovbis},
  \citenamefont {Randoux},\ and\ \citenamefont {Suret}}]{congy2024}%
  \BibitemOpen
  \bibfield  {author} {\bibinfo {author} {\bibfnamefont {T.}~\bibnamefont
  {Congy}}, \bibinfo {author} {\bibfnamefont {G.~A.}\ \bibnamefont {El}},
  \bibinfo {author} {\bibfnamefont {G.}~\bibnamefont {Roberti}}, \bibinfo
  {author} {\bibfnamefont {A.}~\bibnamefont {Tovbis}}, \bibinfo {author}
  {\bibfnamefont {S.}~\bibnamefont {Randoux}},\ and\ \bibinfo {author}
  {\bibfnamefont {P.}~\bibnamefont {Suret}},\ }\bibfield  {title} {\bibinfo
  {title} {Statistics of extreme events in integrable turbulence},\ }\href
  {https://doi.org/10.1103/PhysRevLett.132.207201} {\bibfield  {journal}
  {\bibinfo  {journal} {Phys. Rev. Lett.}\ }\textbf {\bibinfo {volume} {132}},\
  \bibinfo {pages} {207201} (\bibinfo {year} {2024})}\BibitemShut {NoStop}%
\end{thebibliography}

\begin{thebibliography}{16}%
\makeatletter
\providecommand \@ifxundefined [1]{%
 \@ifx{#1\undefined}
}%
\providecommand \@ifnum [1]{%
 \ifnum #1\expandafter \@firstoftwo
 \else \expandafter \@secondoftwo
 \fi
}%
\providecommand \@ifx [1]{%
 \ifx #1\expandafter \@firstoftwo
 \else \expandafter \@secondoftwo
 \fi
}%
\providecommand \natexlab [1]{#1}%
\providecommand \enquote  [1]{``#1''}%
\providecommand \bibnamefont  [1]{#1}%
\providecommand \bibfnamefont [1]{#1}%
\providecommand \citenamefont [1]{#1}%
\providecommand \href@noop [0]{\@secondoftwo}%
\providecommand \href [0]{\begingroup \@sanitize@url \@href}%
\providecommand \@href[1]{\@@startlink{#1}\@@href}%
\providecommand \@@href[1]{\endgroup#1\@@endlink}%
\providecommand \@sanitize@url [0]{\catcode `\\12\catcode `\$12\catcode
  `\&12\catcode `\#12\catcode `\^12\catcode `\_12\catcode `\%12\relax}%
\providecommand \@@startlink[1]{}%
\providecommand \@@endlink[0]{}%
\providecommand \url  [0]{\begingroup\@sanitize@url \@url }%
\providecommand \@url [1]{\endgroup\@href {#1}{\urlprefix }}%
\providecommand \urlprefix  [0]{URL }%
\providecommand \Eprint [0]{\href }%
\providecommand \doibase [0]{https://doi.org/}%
\providecommand \selectlanguage [0]{\@gobble}%
\providecommand \bibinfo  [0]{\@secondoftwo}%
\providecommand \bibfield  [0]{\@secondoftwo}%
\providecommand \translation [1]{[#1]}%
\providecommand \BibitemOpen [0]{}%
\providecommand \bibitemStop [0]{}%
\providecommand \bibitemNoStop [0]{.\EOS\space}%
\providecommand \EOS [0]{\spacefactor3000\relax}%
\providecommand \BibitemShut  [1]{\csname bibitem#1\endcsname}%
\let\auto@bib@innerbib\@empty
\bibitem [{\citenamefont {Copie}\ \emph {et~al.}(2023)\citenamefont {Copie},
  \citenamefont {Suret},\ and\ \citenamefont {Randoux}}]{SMcopie2023space}%
  \BibitemOpen
  \bibfield  {author} {\bibinfo {author} {\bibfnamefont {F.}~\bibnamefont
  {Copie}}, \bibinfo {author} {\bibfnamefont {P.}~\bibnamefont {Suret}},\ and\
  \bibinfo {author} {\bibfnamefont {S.}~\bibnamefont {Randoux}},\ }\bibfield
  {title} {\bibinfo {title} {Space–time observation of the dynamics of
  soliton collisions in a recirculating optical fiber loop},\ }\href
  {https://doi.org/https://doi.org/10.1016/j.optcom.2023.129647} {\bibfield
  {journal} {\bibinfo  {journal} {Optics Communications}\ }\textbf {\bibinfo
  {volume} {545}},\ \bibinfo {pages} {129647} (\bibinfo {year}
  {2023})}\BibitemShut {NoStop}%
\bibitem [{\citenamefont {Fache}\ \emph {et~al.}(2025)\citenamefont {Fache},
  \citenamefont {Copie}, \citenamefont {Suret},\ and\ \citenamefont
  {Randoux}}]{SMfache2025perturbed}%
  \BibitemOpen
  \bibfield  {author} {\bibinfo {author} {\bibfnamefont {L.}~\bibnamefont
  {Fache}}, \bibinfo {author} {\bibfnamefont {F.}~\bibnamefont {Copie}},
  \bibinfo {author} {\bibfnamefont {P.}~\bibnamefont {Suret}},\ and\ \bibinfo
  {author} {\bibfnamefont {S.}~\bibnamefont {Randoux}},\ }\bibfield  {title}
  {\bibinfo {title} {Perturbed nonlinear evolution of optical soliton gases:
  Growth and decay in integrable turbulence},\ }\href@noop {} {\bibfield
  {journal} {\bibinfo  {journal} {Physical Review Letters}\ }\textbf {\bibinfo
  {volume} {135}},\ \bibinfo {pages} {157201} (\bibinfo {year}
  {2025})}\BibitemShut {NoStop}%
\bibitem [{\citenamefont {Walczak}\ \emph {et~al.}(2015)\citenamefont
  {Walczak}, \citenamefont {Randoux},\ and\ \citenamefont
  {Suret}}]{SMwalczak_2015}%
  \BibitemOpen
  \bibfield  {author} {\bibinfo {author} {\bibfnamefont {P.}~\bibnamefont
  {Walczak}}, \bibinfo {author} {\bibfnamefont {S.}~\bibnamefont {Randoux}},\
  and\ \bibinfo {author} {\bibfnamefont {P.}~\bibnamefont {Suret}},\ }\bibfield
   {title} {\bibinfo {title} {Optical rogue waves in integrable turbulence},\
  }\href {https://doi.org/10.1103/PhysRevLett.114.143903} {\bibfield  {journal}
  {\bibinfo  {journal} {Phys. Rev. Lett.}\ }\textbf {\bibinfo {volume} {114}},\
  \bibinfo {pages} {143903} (\bibinfo {year} {2015})}\BibitemShut {NoStop}%
\bibitem [{\citenamefont {Agafontsev}\ \emph {et~al.}(2021)\citenamefont
  {Agafontsev}, \citenamefont {Randoux},\ and\ \citenamefont
  {Suret}}]{SMagafontsev2021extreme}%
  \BibitemOpen
  \bibfield  {author} {\bibinfo {author} {\bibfnamefont {D.}~\bibnamefont
  {Agafontsev}}, \bibinfo {author} {\bibfnamefont {S.}~\bibnamefont
  {Randoux}},\ and\ \bibinfo {author} {\bibfnamefont {P.}~\bibnamefont
  {Suret}},\ }\bibfield  {title} {\bibinfo {title} {Extreme rogue wave
  generation from narrowband partially coherent waves},\ }\href@noop {}
  {\bibfield  {journal} {\bibinfo  {journal} {Physical Review E}\ }\textbf
  {\bibinfo {volume} {103}},\ \bibinfo {pages} {032209} (\bibinfo {year}
  {2021})}\BibitemShut {NoStop}%
\bibitem [{\citenamefont {Yang}(2010)}]{SMyang2010nonlinear}%
  \BibitemOpen
  \bibfield  {author} {\bibinfo {author} {\bibfnamefont {J.}~\bibnamefont
  {Yang}},\ }\href@noop {} {\emph {\bibinfo {title} {Nonlinear waves in
  integrable and nonintegrable systems}}}\ (\bibinfo  {publisher} {SIAM},\
  \bibinfo {year} {2010})\BibitemShut {NoStop}%
\bibitem [{\citenamefont {Gelash}\ \emph {et~al.}(2019)\citenamefont {Gelash},
  \citenamefont {Agafontsev}, \citenamefont {Zakharov}, \citenamefont {El},
  \citenamefont {Randoux},\ and\ \citenamefont {Suret}}]{SMgelash2019bound}%
  \BibitemOpen
  \bibfield  {author} {\bibinfo {author} {\bibfnamefont {A.}~\bibnamefont
  {Gelash}}, \bibinfo {author} {\bibfnamefont {D.}~\bibnamefont {Agafontsev}},
  \bibinfo {author} {\bibfnamefont {V.}~\bibnamefont {Zakharov}}, \bibinfo
  {author} {\bibfnamefont {G.}~\bibnamefont {El}}, \bibinfo {author}
  {\bibfnamefont {S.}~\bibnamefont {Randoux}},\ and\ \bibinfo {author}
  {\bibfnamefont {P.}~\bibnamefont {Suret}},\ }\bibfield  {title} {\bibinfo
  {title} {Bound state soliton gas dynamics underlying the spontaneous
  modulational instability},\ }\href@noop {} {\bibfield  {journal} {\bibinfo
  {journal} {Physical review letters}\ }\textbf {\bibinfo {volume} {123}},\
  \bibinfo {pages} {234102} (\bibinfo {year} {2019})}\BibitemShut {NoStop}%
\bibitem [{\citenamefont {Takahashi}(2005)}]{SMtakahashi2005}%
  \BibitemOpen
  \bibfield  {author} {\bibinfo {author} {\bibfnamefont {M.}~\bibnamefont
  {Takahashi}},\ }\href@noop {} {\emph {\bibinfo {title} {Thermodynamics of
  one-dimensional solvable models}}}\ (\bibinfo  {publisher} {Cambridge
  University Press},\ \bibinfo {year} {2005})\BibitemShut {NoStop}%
\bibitem [{\citenamefont {Doyon}(2018)}]{SMDoyon2018}%
  \BibitemOpen
  \bibfield  {author} {\bibinfo {author} {\bibfnamefont {B.}~\bibnamefont
  {Doyon}},\ }\bibfield  {title} {\bibinfo {title} {{Exact large-scale
  correlations in integrable systems out of equilibrium}},\ }\href
  {https://doi.org/10.21468/SciPostPhys.5.5.054} {\bibfield  {journal}
  {\bibinfo  {journal} {SciPost Phys.}\ }\textbf {\bibinfo {volume} {5}},\
  \bibinfo {pages} {054} (\bibinfo {year} {2018})}\BibitemShut {NoStop}%
\bibitem [{\citenamefont {Koch}\ \emph {et~al.}(2022)\citenamefont {Koch},
  \citenamefont {Caux},\ and\ \citenamefont {Bastianello}}]{SMKoch2022}%
  \BibitemOpen
  \bibfield  {author} {\bibinfo {author} {\bibfnamefont {R.}~\bibnamefont
  {Koch}}, \bibinfo {author} {\bibfnamefont {J.-S.}\ \bibnamefont {Caux}},\
  and\ \bibinfo {author} {\bibfnamefont {A.}~\bibnamefont {Bastianello}},\
  }\bibfield  {title} {\bibinfo {title} {Generalized hydrodynamics of the
  attractive non-linear schr\"odinger equation},\ }\href
  {https://doi.org/10.1088/1751-8121/ac53c3} {\bibfield  {journal} {\bibinfo
  {journal} {Journal of Physics A: Mathematical and Theoretical}\ }\textbf
  {\bibinfo {volume} {55}},\ \bibinfo {pages} {134001} (\bibinfo {year}
  {2022})}\BibitemShut {NoStop}%
\bibitem [{\citenamefont {Faddeev}\ and\ \citenamefont
  {Takhtajan}(2007)}]{SMfaddeev2007hamiltonian}%
  \BibitemOpen
  \bibfield  {author} {\bibinfo {author} {\bibfnamefont {L.}~\bibnamefont
  {Faddeev}}\ and\ \bibinfo {author} {\bibfnamefont {L.}~\bibnamefont
  {Takhtajan}},\ }\href@noop {} {\emph {\bibinfo {title} {Hamiltonian methods
  in the theory of solitons}}}\ (\bibinfo  {publisher} {Springer Science \&
  Business Media},\ \bibinfo {year} {2007})\BibitemShut {NoStop}%
\bibitem [{\citenamefont {Zakharov}(1971)}]{SMzakharov1971}%
  \BibitemOpen
  \bibfield  {author} {\bibinfo {author} {\bibfnamefont {V.}~\bibnamefont
  {Zakharov}},\ }\bibfield  {title} {\bibinfo {title} {Kinetic equation for
  solitons},\ }\href@noop {} {\bibfield  {journal} {\bibinfo  {journal} {Sov.
  Phys. JETP}\ }\textbf {\bibinfo {volume} {33}},\ \bibinfo {pages} {538}
  (\bibinfo {year} {1971})}\BibitemShut {NoStop}%
\bibitem [{\citenamefont {Bertini}\ \emph {et~al.}(2016)\citenamefont
  {Bertini}, \citenamefont {Collura}, \citenamefont {De~Nardis},\ and\
  \citenamefont {Fagotti}}]{SMBertini2016}%
  \BibitemOpen
  \bibfield  {author} {\bibinfo {author} {\bibfnamefont {B.}~\bibnamefont
  {Bertini}}, \bibinfo {author} {\bibfnamefont {M.}~\bibnamefont {Collura}},
  \bibinfo {author} {\bibfnamefont {J.}~\bibnamefont {De~Nardis}},\ and\
  \bibinfo {author} {\bibfnamefont {M.}~\bibnamefont {Fagotti}},\ }\bibfield
  {title} {\bibinfo {title} {Transport in out-of-equilibrium $xxz$ chains:
  Exact profiles of charges and currents},\ }\href
  {https://doi.org/10.1103/PhysRevLett.117.207201} {\bibfield  {journal}
  {\bibinfo  {journal} {Phys. Rev. Lett.}\ }\textbf {\bibinfo {volume} {117}},\
  \bibinfo {pages} {207201} (\bibinfo {year} {2016})}\BibitemShut {NoStop}%
\bibitem [{\citenamefont {Castro-Alvaredo}\ \emph {et~al.}(2016)\citenamefont
  {Castro-Alvaredo}, \citenamefont {Doyon},\ and\ \citenamefont
  {Yoshimura}}]{SMAlvaredo2016}%
  \BibitemOpen
  \bibfield  {author} {\bibinfo {author} {\bibfnamefont {O.~A.}\ \bibnamefont
  {Castro-Alvaredo}}, \bibinfo {author} {\bibfnamefont {B.}~\bibnamefont
  {Doyon}},\ and\ \bibinfo {author} {\bibfnamefont {T.}~\bibnamefont
  {Yoshimura}},\ }\bibfield  {title} {\bibinfo {title} {Emergent hydrodynamics
  in integrable quantum systems out of equilibrium},\ }\href
  {https://doi.org/10.1103/PhysRevX.6.041065} {\bibfield  {journal} {\bibinfo
  {journal} {Phys. Rev. X}\ }\textbf {\bibinfo {volume} {6}},\ \bibinfo {pages}
  {041065} (\bibinfo {year} {2016})}\BibitemShut {NoStop}%
\bibitem [{\citenamefont {Congy}\ \emph {et~al.}(2024)\citenamefont {Congy},
  \citenamefont {El}, \citenamefont {Roberti}, \citenamefont {Tovbis},
  \citenamefont {Randoux},\ and\ \citenamefont {Suret}}]{SMCongy2024}%
  \BibitemOpen
  \bibfield  {author} {\bibinfo {author} {\bibfnamefont {T.}~\bibnamefont
  {Congy}}, \bibinfo {author} {\bibfnamefont {G.~A.}\ \bibnamefont {El}},
  \bibinfo {author} {\bibfnamefont {G.}~\bibnamefont {Roberti}}, \bibinfo
  {author} {\bibfnamefont {A.}~\bibnamefont {Tovbis}}, \bibinfo {author}
  {\bibfnamefont {S.}~\bibnamefont {Randoux}},\ and\ \bibinfo {author}
  {\bibfnamefont {P.}~\bibnamefont {Suret}},\ }\bibfield  {title} {\bibinfo
  {title} {Statistics of extreme events in integrable turbulence},\ }\href
  {https://doi.org/10.1103/PhysRevLett.132.207201} {\bibfield  {journal}
  {\bibinfo  {journal} {Phys. Rev. Lett.}\ }\textbf {\bibinfo {volume} {132}},\
  \bibinfo {pages} {207201} (\bibinfo {year} {2024})}\BibitemShut {NoStop}%
\bibitem [{\citenamefont {Spohn}(2012)}]{SMspohn2012large}%
  \BibitemOpen
  \bibfield  {author} {\bibinfo {author} {\bibfnamefont {H.}~\bibnamefont
  {Spohn}},\ }\href@noop {} {\emph {\bibinfo {title} {Large scale dynamics of
  interacting particles}}}\ (\bibinfo  {publisher} {Springer Science \&
  Business Media},\ \bibinfo {year} {2012})\BibitemShut {NoStop}%
\bibitem [{\citenamefont {Doyon}(2022)}]{SMDoyon2022}%
  \BibitemOpen
  \bibfield  {author} {\bibinfo {author} {\bibfnamefont {B.}~\bibnamefont
  {Doyon}},\ }\bibfield  {title} {\bibinfo {title} {Hydrodynamic projections
  and the emergence of linearised euler equations in one-dimensional isolated
  systems},\ }\href {https://doi.org/10.1007/s00220-022-04310-3} {\bibfield
  {journal} {\bibinfo  {journal} {Communications in Mathematical Physics}\
  }\textbf {\bibinfo {volume} {391}},\ \bibinfo {pages} {293} (\bibinfo {year}
  {2022})}\BibitemShut {NoStop}%
\end{thebibliography}
\end{document}